\documentclass[10pt,journal,compsoc]{IEEEtran}

\ifCLASSOPTIONcompsoc
  \usepackage[nocompress]{cite}
\else
  \usepackage{cite}
\fi
\usepackage{enumitem}

\usepackage{graphicx}
\usepackage{verbatim} 
\usepackage{amssymb}
\usepackage{amsmath}
\usepackage{multirow}
\usepackage{longtable}
\usepackage[table,xcdraw]{xcolor}
\graphicspath{../Images/}
\usepackage{caption}
\usepackage[textsize=scriptsize]{todonotes}
\ifCLASSINFOpdf
\else
\fi
\usepackage{array}
\newcolumntype{C}[1]{>{\centering\let\newline\\\arraybackslash\hspace{0pt}}m{#1}}

\usepackage{physics}
\usepackage{multicol}
\usepackage{soul}

\usepackage{titlesec}
\usepackage{ragged2e}
\usepackage{rotating}
\usepackage{url}

\titleformat{\paragraph}
{\normalfont\normalsize\bfseries}{\theparagraph}{1em}{}
\titlespacing*{\paragraph}
{0pt}{3.25ex plus 1ex minus .2ex}{1.5ex plus .2ex}

\begin{document}
%
\title{A Survey on Adversarial Attacks \\ for Malware Analysis}
%
%
%
%

\author{Kshitiz Aryal, Maanak Gupta,~\IEEEmembership{Member,~IEEE,}
        and Mahmoud Abdelsalam,~\IEEEmembership{Member,~IEEE}
\IEEEcompsocitemizethanks{\IEEEcompsocthanksitem Kshitiz Aryal is a Doctoral student in the Department of Computer Science, Tennessee Tech University, Cookeville, TN, 38501, USA. \protect\\
E-mail: karyal42@tntech.edu}
\IEEEcompsocitemizethanks{\IEEEcompsocthanksitem Maanak Gupta is an Assistant Professor in the Department of Computer Science, Tennessee Tech University, Cookeville, TN, 38501, USA. \protect\\
E-mail: mgupta@tntech.edu}
\IEEEcompsocitemizethanks{\IEEEcompsocthanksitem Mahmoud Abdelsalam is an Assistant Professor in the Department of Computer Science, North Carolina A\&T State University, NC, USA. \protect\\
E-mail: mabdelsalam1@ncat.edu}}
\IEEEtitleabstractindextext{%
\begin{abstract}
\justifying
The last decade has experienced an exponential growth in research and adoption of Machine Learning (ML) and Artificial Intelligence (AI), and its application in different use-cases. Traditional machine learning algorithms have evolved into data intensive deep learning  architectures, which have fostered cutting edge and unprecedented technological advancements revolutionizing today's world. The capability of these ML algorithms to uncover knowledge and patterns from semi- or unstructured data to support automation in decision making has led to the revamping of domains such as medicine, e-commerce, autonomous cars, and cybersecurity. The growth and adoption of machine learning solutions have been lately slowed down with the advent of adversarial attacks. Adversaries are able to modify data at training and testing time, maximizing the classification error of the ML models. Minor intentional perturbations in test samples are crafted by the adversaries to exploit the discovered blind spots in trained models. The increased data dependency 
of these algorithms have offered a way for high incentives to disguise ML models. In order to survive against possible catastrophic implications, continuous research is required to find vulnerabilities in form of adversarial and design resilient autonomous systems.

Machine learning-based malware analysis approaches are widely researched and deployed in critical infrastructures for detecting and classifying evasive and growing malware threats. However, minor perturbations or few ineffectual bytes insertion can easily 'fool' these trained ML classifiers, essentially making them ineffective against these crafted and smart malicious software.  
This survey aims at providing an encyclopedic introduction to adversarial evasion attacks that are carried out specifically against malware detection and classification systems. Since most of the research in the adversarial malware domain is new and has been performed in the last couple of years, our survey will cover the relevant literature published on the malware adversarial evasion attacks between the year 2013 to 2021. The paper will begin by introducing various machine learning techniques used to generate adversarial in malware analysis and explaining the structures of target files. The survey will model the threat posed by adversaries followed by brief descriptions of widely used adversarial algorithms. We will also provide a taxonomy of adversarial evasion attacks with respect to the attack domains and adversarial generation techniques that are widely used in malware detection and classification. Adversarial evasion attacks carried out against malware detectors will be discussed under each taxonomical headings and compared with related literature. 
The survey will conclude by highlighting the open problems, challenges and future research directions.
\end{abstract}

\begin{IEEEkeywords}
Adversarial Evasion attack, Adversary Modeling, Data Poisoning, Malware Analysis, Machine Learning, Deep Learning, Security, Windows Malware, Android Malware, PDF Malware
\end{IEEEkeywords}}

\maketitle

\IEEEdisplaynontitleabstractindextext

%
\IEEEpeerreviewmaketitle

\ifCLASSOPTIONcompsoc
\IEEEraisesectionheading{\section{Introduction}\label{sec:introduction}}
\else
\section{Introduction}
\label{sec:introduction}
\fi

\IEEEPARstart{M}{achine} Learning  has revolutionized the modern world due to its ubiquity and generalization power over the humongous volume of data. With Zettabytes of data hovering around the cloud~\cite{statista_data}, modern technology's power resides in extracting knowledge from these unstructured raw data. Machine learning (ML) has provided the unprecedented power to automate the decision-making process, outperforming humans by far margin. ML has powered more robust and representative feature set in comparison to hand-crafted features. Transformation of ML approaches from classical algorithms to modern deep learning technologies are providing the major breakthroughs in state-of-art research problems. Further, deep learning (DL) has excelled in areas where traditional ML approaches were infeasible (or unsuccessful) to apply. The evolving deep learning techniques have furnished the fields of natural language processing~\cite{9075398,8416973,gardner2018allennlp}, image classification~\cite{hemanth2017deep,razzak2018deep,perez2017effectiveness,chan2015pcanet}, autonomous driving~\cite{al2017deep,fujiyoshi2019deep,sallab2017deep,huval2015empirical}, neuro science~\cite{marblestone2016toward,richards2019deep,tanaka2019deep,mathis2020deep} and many other wide range of domains. Society is experiencing high-end amazing products like Apple Siri\footnote{https://www.apple.com/siri/}, Amazon Alexa\footnote{https://alexa.amazon.com/} and Microsoft Cortana\footnote{https://www.microsoft.com/en-us/cortana} due to recent advances in machine learning and artificial intelligence (AI).
Needless to say that machine learning has started shaping our daily life habits; connecting to people on social media, ordering food and groceries from online stores, listening to music on Spotify\footnote{https://www.spotify.com/us/}, watching movies on Netflix\footnote{https://www.netflix.com/}, reading online news and books, are all examples of systems built around the recommendation engines powered by deep learning based models. Machine learning based solutions not only control our lifestyle but it has also revolutionized cyber security critical operations in different domains including malware analysis~\cite{abdelsalam2018malware,sahs2012machine,mcdole2021deep, mcdole2020analyzing, kimmel2021recurrent,kimmell2021analyzing}, spam filtering~\cite{guzella2009review,tretyakov2004machine,lai2007empirical,dada2019machine}, fraud detection~\cite{awoyemi2017credit, varmedja2019credit, adewumi2017survey,roy2018deep}, medical analysis~\cite{shen2017deep, wernick2010machine,kononenko2001machine,erickson2017machine,dhakalartificial}, access control~\cite{martin2006inferring,gupta2017poster, karimi2018unsupervised, gupta2021towards}, among others.

Malware analysis is one of the most critical fields where ML is being significantly employed.
Traditional malware detection approaches~\cite{bazrafshan2013survey,zheng2013droid,venugopal2008efficient,sahoo2014signature,abbas2017low} rely on signatures where unique identifiers of malware files are maintained in a database and are compared to extracted signatures from newly encountered suspicious files.
However, several techniques are used to rapidly evolve the malware to avoid detection (more details in Section~\ref{sec:Malware_Detection}).
With security researchers looking for detection techniques addressing such sophisticated zero-day and evasive malware, ML based approaches came to their rescue~\cite{kim2018zero}. Most of the modern anti-malware engines, such as Windows Defender\footnote{https://www.microsoft.com/en-us/windows/comprehensive-security}, Avast\footnote{https://www.avast.com/}, Deep Instinct D-Client\footnote{https://www.deepinstinct.com/endpoint-security} and Cylance Smart Antivirus\footnote{https://shop.cylance.com/us}, are powered by machine learning\cite{phillips_gavin}, making them robust against emerging variants and polymorphic malware~\cite{rieck2011automatic}. As per some estimates \cite{iot_analytics_2021}, around 12.3 billion devices are connected worldwide and spread of malware in this scale can result in catastrophic consequences. As such, it is evident that economies worth billions of dollars are directly or indirectly relying on machine learning's performance and growth to be protect from this rapidly evolving menace of malware.
Despite the existence of numerous malware detection approaches, including ones that leverage ML, recent ransomware attacks, like the Colonial Pipeline attack where operators had to pay around \$5 million for recovering 5,500-mile long pipeline\cite{morrison_2021}, the MediaMarkt attack worth around \$50M bitcoin payment~\cite{yahoo_mediamarkt} and the computer giant Acer attack~\cite{abrams_2021}, 
highlight the vulnerabilities and limitations of current security approaches, and necessitates more robust, real-time, adaptable and autonomous defense mechanisms powered by AI and ML.

\begin{figure}[!t]
    \centering
    \includegraphics[scale = 0.9]{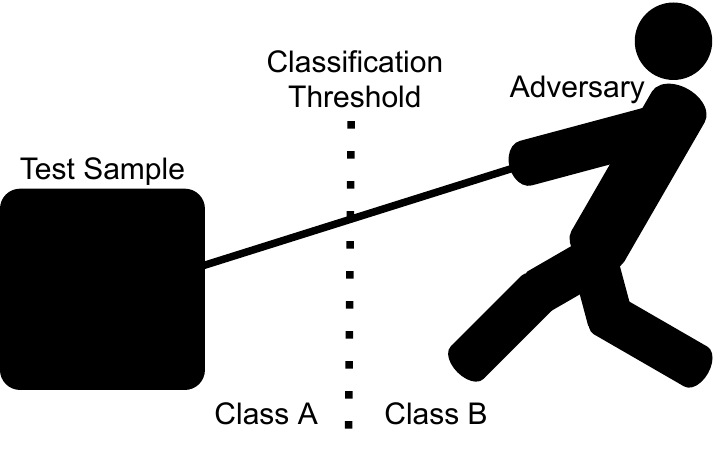}
    \caption{Evasion attack drags test sample from class A to B.}
    \label{fig:adversarial_pull}
\end{figure} 
The performance of ML models relies on the basic assumption that training and testing are carried out under similar settings and that samples from training and testing datasets follow independent and identical distribution. This assumption is overly simplified and, in many cases, does not hold true for real world use-cases where adversaries deceive the ML models into performing wrong predictions (i.e. adversarial attacks). In addition to traditional threats like malware attack~\cite{khouzani2012maximum,hu2009wifi,schwarz2019practical}, phishing~\cite{gupta2016literature,hong2012state,halevi2015spear}, man-in-the-middle attack~\cite{meyer2004man,callegati2009man,eberz2012practical}, denial-of-service~\cite{mirkovic2004internet,schuba1997analysis,jamal2018denial} and SQL injection~\cite{halfond2006classification,kieyzun2009automatic,anley2002advanced}, adversarial attacks has now emerged as a serious concern, threatening to dismantle and undermine all the progress made in the machine learning domain. 

Adversarial attacks are carried out either by poisoning the training data or manipulating the test data (evasion attacks). Data poisoning attacks~\cite{shafahi2018poison,liu2019unified,Cao_pois_feder,tensorclog_shen} have been prevalent for some time but are less scrutinized as access to training data by the attackers is considered unlikely. In contrast, evasion attacks, first introduced by Szegedy et al.~\cite{szegedy2014intriguing} against deep learning architectures, are carried out by carefully crafting imperceptible perturbation in test samples, forcing models to mis-classify as illustrated in Figure \ref{fig:adversarial_pull}. Here, the attacker's effort is to drag a test sample across the ML's decision boundary through the addition of minimal perturbation to that sample. Considering the availability of research works and higher-risk in practicality, this survey will entirely focus on adversarial evasion attacks that are carried out against the malware detectors.

\begin{figure}[!t]
    \centering
    \includegraphics[width=\columnwidth]{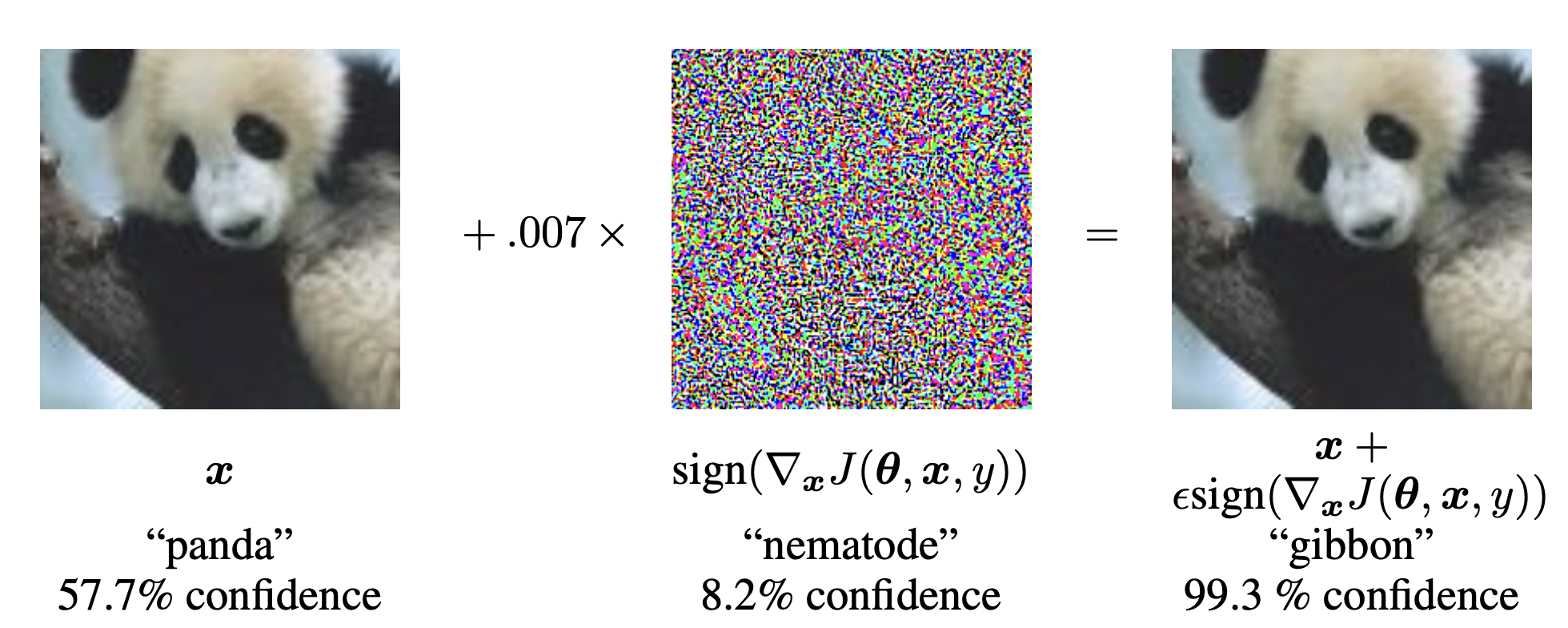}
    \caption{ An adversarial example against GoogLeNet \cite{szegedy2014going} on ImageNet \cite{5206848}, demonstrated by Goodfellow et al.~\cite{goodfellow2015explaining}.}
    \label{fig:goodfellow_adv}
\end{figure}

Adversarial evasion attacks were initially crafted on images as the only requirement for perturbation in an image is that it should be imperceptible to the human eye~\cite{kurakin2017adversarial,yilmaz2020practical}. A very common example for adversarial attack in images, shown in Figure \ref{fig:goodfellow_adv}, is performed by Goodfellow et al.~\cite{goodfellow2015explaining} where GoogLeNet~\cite{szegedy2014going} trained on ImageNet\cite{5206848} classifies panda as gibbon with addition of very small perturbations. This threat is not limited to experimental research labs but have already been successfully demonstrated in real world environments. For instance, Eykholt et al. performed sticker attacks to road signs forcing the image recognition system to detect 'STOP' sign as a speed limit. Researchers from the Chinese technology company Tencent\footnote{https://www.tencent.com/} tricked Tesla's\footnote{https://www.tesla.com/} Autopilot in Model S and forced it to switch lanes by adding few stickers on the road~\cite{jr_2019}. Such adversarial attacks on real world applications force us to rethink the increasing reliability over smart technologies like Tesla Autopilot\footnote{https://www.tesla.com/autopilot}. 

\begin{table*}[!t]
    \centering
    \def\arraystretch{1.5}
    \rowcolors{2}{}{lightgray}
    \caption{{Surveys focusing on security of machine learning.}}
    \begin{tabular}{|c|c|c|c|c|c|}
    \rowcolor{gray!25} 
    \hline
         \textbf{Paper}&\textbf{Year}&\textbf{Application  Domain}&\textbf{Taxonomy}&
         \rotatebox[origin=c]{90}{\textbf{Threat Modeling}}&
         \rotatebox[origin=c]{90}{\textbf{Adversarial Example}} \\
         \hline
         Barreno et al.~\cite{BARRENO_SECURITY}& 2008&Security&Attack Nature&&\\
         \hline
         Gardiner et al.~\cite{gardiner}&2016&Security&Attack Type/Algorithm&$\surd$&$\surd$\\
         \hline
         Kumar et al.~\cite{kumar2017survey}&2017&General&Attack Type&&\\
         \hline
         Yuan et al.~\cite{yuan2018adversarial}&2017&Image&Algorithm&$\surd$&$\surd$\\
         \hline
         Chakraborty et al.~\cite{chakraborty2018adversarial}&2018&Image/Intrusion&Attack Phase&$\surd$&$\surd$\\
         \hline
         Akhtar et al.~\cite{akhtar2018threat}&2018&Image&Image domains&&$\surd$\\
         \hline
         Duddu et al.~\cite{Duddu2018ASO}&2018&Security&Attack Type&$\surd$&\\
         \hline
         Li et al.~\cite{li2018security}&2018&General&Algorithm&&\\
         \hline
         Liu et al.~\cite{Liu2018ASO}&2018&General&Target Phase&$\surd$&\\
         \hline
         Biggio et al.~\cite{2018}&2018&Image&Attack Type&$\surd$&$\surd$\\
         \hline
         Sun et al.~\cite{Sun2018ASO}&2018&Image&Image Type&$\surd$&$\surd$\\
         \hline
         Pitropakis et al.~\cite{Pitropakis2019ATA}&2019&Image/Intrusion/Spam&Algorithm&$\surd$&\\
         \hline
         Wang et al.~\cite{Wang2019TheSO}&2019&Image&Algorithm&$\surd$&$\surd$\\
         \hline
         Qiu et al.~\cite{Qiu2019ReviewOA}&2019&Image&Knowledge&$\surd$&$\surd$\\
         \hline
         Xu et al.~\cite{xu2019adversarial}&2019&Image/Graph/Text&Attack Type&$\surd$&\\
         \hline
         Zhang et al.~\cite{zhang2019adversarial}&2019&Natural Language Processing&Knowledge/Algorithm&$\surd$&
         $\surd$\\
         \hline
         Martins et al.~\cite{IntrusionMartins}&2019&Intrusion/Malware&Approach&&$\surd$\\
         \hline
         Moisejevs ~\cite{Moisejevs2019AdversarialAA}&2019&Malware Classification&Attack Phase&&$\surd$\\
         \hline
         Ibitoye et al.~\cite{ibitoye2020threat}&2020&Network Security&Approach/Algorithm&$\surd$&$\surd$\\
         \hline
         Our Work&2021&Malware Analysis&Domain/Algorithm&$\surd$&$\surd$\\
         \hline
    \end{tabular}\\
    \vspace{2mm}
    \footnotesize\textit{{\textbf{Year}: Published Year,
    \textbf{Application Domain}: Dataset domain on which adversarial is crafted, \textbf{Taxonomy}: Basis on which attack taxonomy is made, \textbf{Threat Modeling}: Presence of threat modeling, \textbf{Adversarial Example}: Discuss actual adversarial attacks crafted in literature}}
    \label{tab:literature}
\end{table*}
However, adversarial generation is a completely different game in the malware domain, in comparison to computer vision, due to the increased number of constraints. Perturbations in malware files should be generated in a way that it should not affect both their functionality and executability. Adversarial evasion attacks on malware are carried out by manipulating or inserting few ineffectual bytes in the malware executables in a way that does not tamper with its original state, but change the classification decision by the ML model.
For instance, one early demonstrated attack against anti-malware engine was carried out by Anderson et al.~\cite{Anderson2017EvadingML} using reinforcement learning. This black-box attack was able to bypass Random forest and gradient boosted decision trees (GBDT) detectors by modifying few bytes of Windows PE malware files. Kolosnjaji et al.~\cite{8553214} carried out evasion attack using gradient based approach against convolutional neural network (CNN) based malware detector. Since then, there has been numerous works trying to optimize the attacks, discovering better approaches to attack wide domains of malware detectors. Demetrio et al.~\cite{Demetrio2019} success in crafting adversarial from few header byte modification and Suciu et al.~\cite{Suciu2019Malware} experiment on inserting perturbations in different file locations, further magnified the interest towards improving the standard of attacks. The fear of evolving adversarial attack is growing among the cyber security research community and has provoked the everlasting war between adversarial attackers and defenders. \textit{To help researchers better understand the current situation of adversarial attacks in the malware domain and infer vulnerabilities on current approaches, this paper will provide a comprehensive survey of ongoing adversarial evasion attack researches against Windows, Android, PDF, Linux and Hardware-based malware.}


\subsection{Motivation and Contribution}

\subsubsection{Prior Surveys and Limitations}
The surveys on adversarial attacks crafted in different domains have been summarized in Table~\ref{tab:literature}. Majority of surveys on adversarial attacks are focused on computer vision for images mis-classification. Yuan et al.~\cite{yuan2018adversarial} summarized major adversarial generation methods for images. Chakraborty et al.~\cite{chakraborty2018adversarial} surveyed adversarial in form of evasion and poisoning in image and anomaly detection. Akhtar et al.'s~\cite{akhtar2018threat} work was restricted on the computer vision domain like most of the works. Biggio et al.~\cite{2018} presented a historical timeline of evasion attacks along with works carried out on security of deep neural networks.  Sun et al.~\cite{Sun2018ASO} surveyed practical adversarial examples on graph data. Many of the surveys did not only focused on a single domain but covered generalized field across multiple domains including image, text, graph, intrusion, spam and malware. Kumar et al.~\cite{kumar2017survey} classified adversarial attacks into four overlapping classes.  Li et al.~\cite{li2020sok} explains adversarial generation and defense mechanism through formal representation. Liu et al.~\cite{Liu2018ASO} reviewed some general security threats and associated defensive techniques.  Pitropakis et al.~\cite{Pitropakis2019ATA} surveyed adversarial in intrusion detection, spam filtering and image domain. Xu et al.~\cite{xu2019adversarial} surveyed vulnerabilities, analysed reasons behind it and also proposed ways to detect adversarial examples. There have been a few works focusing on the security related domains like intrusion, malware, and network security.  Barreno et al.~\cite{BARRENO_SECURITY} worked on one of the very first surveys done on security of machine learning where different categories of attacks and defenses against ML systems are discussed. Gardiner et al.~\cite{gardiner} focused on reviewing call and control detection techniques. They identified vulnerabilities and also pointed limitations of malware detection systems. Duddu et al.~\cite{Duddu2018ASO} discussed the concern of privacy being leaked by information handled by machine learning algorithms. They also presented cyber-warfare testbed for the effectiveness of attack and defense strategies. Martins et al.~\cite{IntrusionMartins} performed generalized survey on attacks focusing on cloud security, malware detection and intrusion detection.  Ibitoye et al.~\cite{ibitoye2020threat} surveyed adversarial attacks in network domain using risk grid map.

With the discussed surveys, we can make certain conclusions reflecting the growing attention and concerns in the community as the world moves toward automation. 
First, the interest of people in adversarial domain has surged in last 3 or 4 years. Second, very few of the survey papers is solely focused on adversarial malware analysis, which is a growing menace. Majority of the surveys conducted on adversarial domain is built around computer vision attacks. Recent flux of works are spread in wide domains including network, natural language processing, security, and intrusion detection. There has been limited research on adversarial attacks in malware analysis, being a relatively new domain. The few existing surveys on malware domain are not focused on malware analysis but spread around multiple domains. The current surveys also does not cover entire attacks carried out on malware detection domain, but focuses on small subset of attacks. The outpouring interests in adversarial and lack of surveys justifying entire adversarial attacks on malware domain, motivates us to extensively survey adversarial evasion attack on malware. 

\begin{table}[]
    \centering
    \caption{{List of acronyms in alphabetical order.}}
    \rowcolors{2}{gray!25}{white}
    \def\arraystretch{1.21}
    \begin{tabular}{|p{1.25cm}|p{6.25cm}|}
    \rowcolor{gray!25}
    \hline
         \textbf{Acronym} & \textbf{Full Form} \\
    \hline
     ACER&Actor Critic model with Experience Replay\\
\hline
AE&Adversarial Example\\
\hline
AI&Artificial Intelligence\\
\hline
AMAO&Adversarial Malware Alignment Obfuscation\\
\hline
API & Application Programming Interface\\
\hline
ATMPA & Adversarial Texture Malware Perturbation Attack\\
\hline
BFA&Benign Features Append\\
\hline
BRN&Benign Random Noise\\
\hline
CFG & Control Flow Graph\\
\hline
CNN & Convolutional Neural Network\\
\hline
CRT&Cross Reference Table\\
\hline
CW&Carlini-Wagner\\
\hline
DCGAN&Deep Convolutional GAN\\
\hline
DE & Differential Evolution\\
\hline
DL&Deep Learning\\
\hline
DNN&Deep Neural Network\\
\hline
DRL&Deep Reinforcement Learning\\
\hline
FGM & Fast Gradient Method \\
\hline
FGSM&Fast Gradient Sign Method\\
\hline
GADGET&
Generative API Adversarial Generic Example by Transferability\\
\hline
GAN&Generative Adversarial Network\\
\hline
GAP&Global Average Pooling\\
\hline
GBDT&Gradient Boosted Decision Trees\\
\hline
GD-KDE&Gradient Descent and Kernel Density Estimation\\
\hline
GEA& Graph Embedding and Augmentation\\
\hline
GRU&Gated Recurrent Units\\
\hline
HDL & Hardware Description Language \\
\hline
HMD & Hardware Malware Detectors\\
\hline
IoT&Internet of Things\\
\hline
KNN&K-Nearest Neighbors\\
\hline
LLC & Logical Link Control\\
\hline
LR & Logistic Regression \\
\hline
LRP & Layer-wise Relevance Propagation\\
\hline
LSTM&Long Short Term Memory\\
\hline
MEV & Modification Evaluating Value\\
\hline
MIM & Momentum Iterative Method\\
\hline
ML&Machine Learning\\
\hline
MLP&Multi Layer Perceptron\\
\hline
MRV&Malware Recomposition Variation\\
\hline
OPA & One Pixel Attack \\
\hline
PDF&Portable Document Format\\
\hline
PE&Portable Executable\\
\hline
PGD & Projected Gradient Descent\\
\hline
ReLU&Rectified Linear Unit\\
\hline
RF & Random Forest\\
\hline
RL&Reinforcement Learning\\
\hline
RNN&Recurrent Neural Network\\
\hline
RTLD&Resource Temporal Locale Dependency\\
\hline
SDG&System Dependency Graph\\
\hline
SHAP & SHapley Additive exPlanation \\
\hline
SR&Success Rate\\
\hline
SVM&Support Vector Machine\\
\hline
TCD & Trojan-net Concealment Degree \\
\hline
TLAMD&
{A Testing framework for Learning based Android Malware Detection systems for IoT Devices}\\
\hline
TPR&True Positive Rate\\
\hline
VOTE & VOTing based Ensemble\\
\hline
ZOO & Zeroth Order Optimization\\
\hline
\end{tabular}
\label{tab:Acros}
\end{table}

\begin{figure*}[!t]
    \centering
    \captionsetup{justification=centering}
    \includegraphics[scale=0.855]{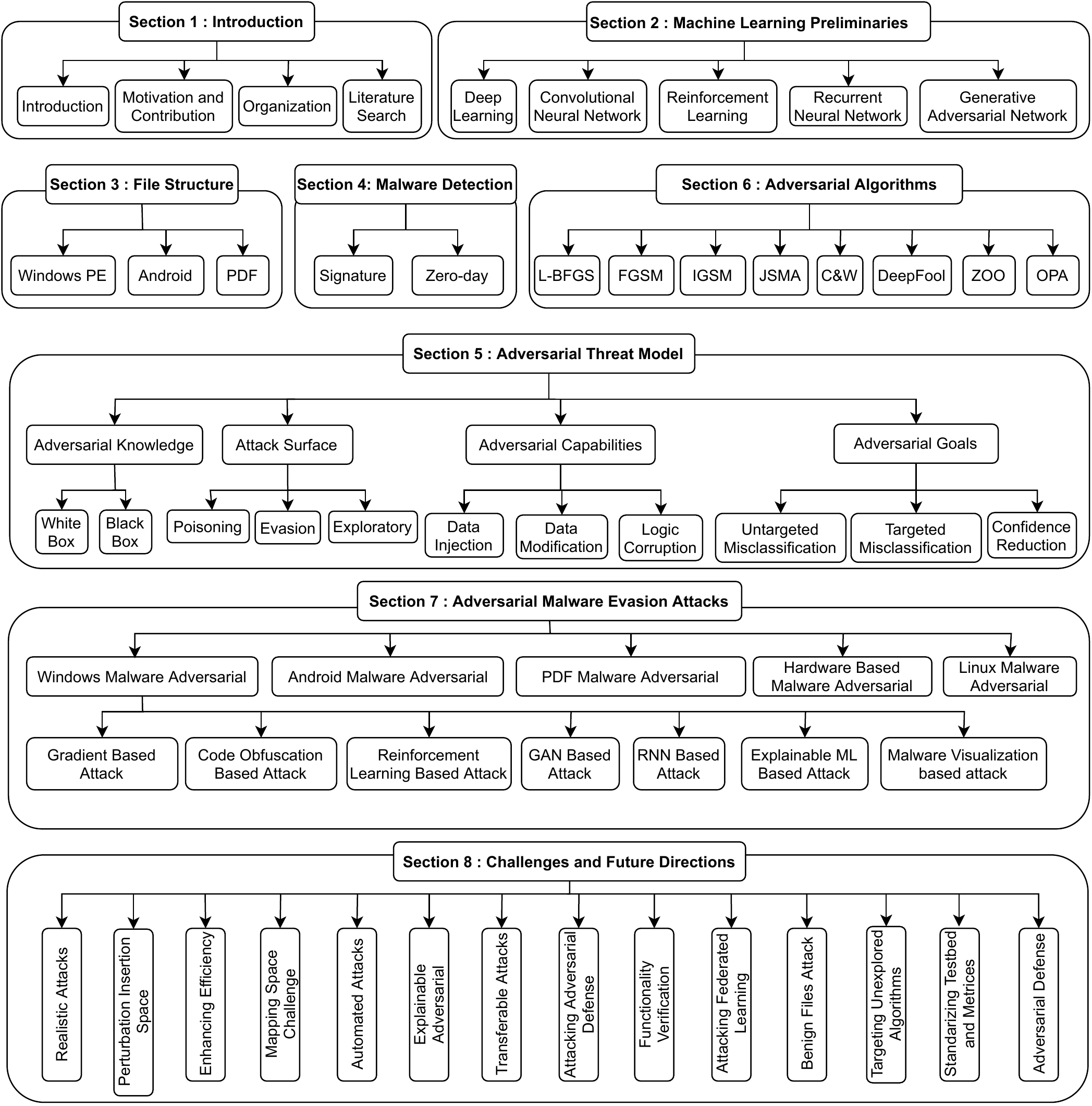}
     \caption{Diagrammatic view of the paper organization (Position of Section 6 is changed to save space)}
    \label{fig:organization}
\end{figure*}

\subsubsection{Our Contributions}
This work will contribute in understanding the arms race between attacker and defender by discussing adversarial evasion attacks in different folds of the malware domain. We aim to provide completely self-contained survey on adversarial attacks carried out against malware detection techniques. Based on our knowledge, this work is one among the first to solely focus on adversarial attacks on malware detection systems.
In this work, our contributions cover the following dimensions: 
\begin{itemize}
    \item As our goal is to make the survey as comprehensive as possible, we provide all the related information required to completely comprehend the contents of the survey. We discuss the machine learning approaches used, the adversarial generation algorithms used by attackers, the malware detection methods attacked and the structure of files that has been exploited to insert adversarial perturbations.
    
    \item We provide the threat modeling to adversarial evasion attacks carried out in malware domain. The threat model helps in quantify and analyze the attack-specific risk associated to particular target of malware. The threat is modeled in terms of attack surface of the malware detector, attacker's knowledge about the malware detector, attacker's capabilities on malware, and adversarial goals that is to be achieved through the malware files. The proper threat modeling also helps to well understand the behaviors of malware, allowing the adversarial attacker to craft effective perturbations.
    
    \item We systematically analyze different adversarial generation algorithms proposed in different domains, which have been attempted to be used in the malware domain. We then discuss the basics of standard adversarial algorithms and taxonomize adversarial evasion attacks in the malware domain with respect to various attack domains. As Windows malware are the most abundant and also the most exploited area, we further taxonomize attacks on Windows malware based on the attack algorithms used. We also discuss attacks carried out in the less frequent file structures like Android and PDF. 
    
    \item We discuss real evasion attacks carried out against anti-malware engines by the researchers, under each taxonomical headings. We also cover the attack strategies used by researchers to generate adversarial attacks, showing how the attacks evolved with time. Further, we compare the motivation and limitations of each research in tabular forms for each taxonomy-class. 
    
    \item We discuss the challenges and limitations on existing adversarial attacks while carrying out in real world environment. We also highlight the future research directions to carry out more practical, robust, efficient and generalized adversarial attacks on malware classifiers. 
\end{itemize}

\subsection{Survey Organization}
In this paper, we structure our survey in a hierarchical manner as shown in Figure \ref{fig:organization}. Section \ref{sec:Adv_Alg} is placed before section \ref{sec:Modeling} in the Figure \ref{fig:organization} just to manage space but actual order in the paper is in incremental order of section number. We begin our survey, as discussed in Section \ref{sec:introduction}, by introducing the field of adversarial machine learning along with motivation for the need to study adversarial attacks in the malware analysis domain. Note that Table \ref{tab:Acros} provides acronyms that are used frequently in the survey. Section \ref{sec: ML} discusses different machine and deep learning algorithms that are used in state-of-art adversarial research. Understanding key concepts of machine learning prerequisites provides the readers the appropriate background to grasp adversarial generation techniques discussed later in the survey. Section \ref{sec:File_struct} explains the structures of Windows, Android and PDF files. The structure of files plays a key role in apprehending adversarial attacks as perturbation depends on flow and robustness of files' structure. Section \ref{sec:Malware_Detection} provides an introduction to malware detection approaches against which adversarial attacks are designed. Section \ref{sec:Modeling} models the adversarial threat from different dimensions. This section briefly elaborates on attack surface, attacker's knowledge, attacker's capabilities and adversarial goals. Section \ref{sec:Adv_Alg} discusses various adversarial algorithms that are considered as standard techniques for perturbation generation across different domains. Section \ref{sec:Adv_Alg} taxonomizes existing real adversarial attacks based on the execution domains (Windows, PDF, Android, Hardware, Linux) and algorithms maneuvered to carry out attack. This section discusses real attacks carried out against malware detection approaches in detail and provides comparisons among related works. Section \ref{sec:Future} highlights challenges of current adversarial generation approaches and sheds the light on open research areas and future directions for adversarial generation in malware analysis. Finally, Section \ref{sec:conclusion} concludes our survey.


\subsection{Literature Search Resources}
To discover the relevant state-of-art works and publications in adversarial attacks on malware analysis, we relied on different digital libraries for computer science scholarly articles. Our major sources are IEEE Xplore\footnote{https://www.ieee.org/}, ACM digital library\footnote{https://dl.acm.org/}, DBLP\footnote{https://dblp.uni-trier.de/}, Semantic Scholar\footnote{https://www.semanticscholar.org/} and arXiv\footnote{https://arxiv.org/}. Apart from these digital libraries, we also searched directly through Google\footnote{https://www.google.com/} and Google Scholar\footnote{https://scholar.google.com/} to get impactful papers in domain that were somehow missed in other libraries. Among numerous keywords used to fetch the papers from public libraries, "Adversarial Malware" and "Adversarial attacks in malware" gave us the most number of relevant papers. After listing all the published works in adversarial generation between year 2013 to 2021, we filtered out papers with good impact, relevance and prepared the final list to conduct our detailed survey. 


\section{Machine Learning Preliminaries}
\label{sec: ML}
\begin{figure}[!t]
    \centering
    \includegraphics[width = \columnwidth]{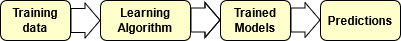}
    \caption{A basic machine learning workflow.}
    \label{fig:ML_workflow}
\end{figure}

\begin{figure}[!t]
    \centering
    \includegraphics[width = \columnwidth]{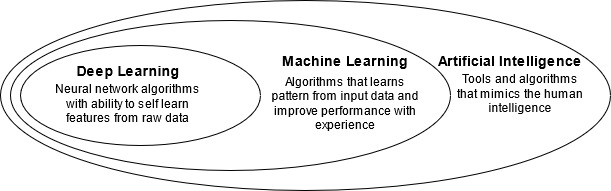}
    \caption{The venn-diagram showing a relation between artificial intelligence, machine learning, and deep learning.}
    \label{fig:Relation_AI_ML_DL}
\end{figure}

We are in the era of Big Data~\cite{hurwitz2013big,sagiroglu2013big,george2014big,gupta2017object,gupta2017multi,gupta2018attribute}, and an unprecedented amount of digital information is generated and flowing around us. With more than 2.5 Quintilian data bytes every day, around 200 million active web pages, 500 million tweets every day and a few million years of videos in YouTube, we can imagine the magnanimity of data around~\cite{Vish_2020,Total_data_vstatista,websites_exist_globally,Twitter_Stats_and_Statistics}. Manual extraction of valuable information from raw data is a cumbersome, tedious, and infeasible task given the volume of data. Machine learning, due to its intrinsic capability to process this humongous amount of data which can learn from raw data, discover patterns and give decisions with least human interference. ML allows automatic detection of patterns in data and use the learned model to predict future data. Prediction on unseen data helps in probabilistic decision-making under uncertainty. Tom M. Mitchell, chair of Machine Learning at Carnegie Mellon University in his book Machine Learning (McGraw-Hill1997)~\cite{ML_MITCHELL} defines machine learning as "\textit{A computer program is said to learn from experience E with respect to some class of tasks T and performance measure P, if its performance at tasks in T, as measured by P, improves with the experience E}". Simply stating, machine learning is a branch of artificial intelligence that enables learning from data used for training. Modern literature often confuses the term 'Artificial Intelligence' with 'Machine Learning' and they are used interchangeably. Machine learning are the subset of artificial intelligence as shown in Figure \ref{fig:Relation_AI_ML_DL} that focuses on learning the patterns and improving the predictions as experience grows. Another term 'Deep Learning' is the current hot-topic inside machine learning which we will discuss later in this section. A trained system should learn and improve with experience, being able to make predictions based on previous learning. The normal workflow of machine learning is shown in Figure \ref{fig:ML_workflow}, where training data is passed through learning algorithm and trained models are used to make predictions. 

Interests in computational approaches to learning can be seen starting back in mid-1950s~\cite{Elements_OF_ML} and since then there has been continuous growth in the development of learning systems. It was only after the 1980s that ML was observed as real-world potential, and today it continues to foster its growth towards increased intelligence in the form of deep learning. The unprecedented power of making predictions, advancement of machine learning techniques and broadening of its application areas have increased exponentially. Originating from data analysis and statistics, it has already gained a pioneering position in fields of text recognition, Natural Language Processing (NLP), speech processing, computer vision application, computational biology, fraud detection, and many security-related critical applications. Classification\cite{kotsiantis2007supervised,bost2014machine,pennacchiotti2011machine}, regression\cite{rong2018research,babii2021machine,steyerberg2014risk}, clustering~\cite{mcgregor2004flow,bijuraj2013clustering,pham2007clustering}, dimensionality reduction~\cite{reddy2020analysis,mladenic2005feature}, and ranking~\cite{richardson2006beyond,cossock2006subset,agarwal2010ranking} are some examples of the major machine learning tasks applied in different applications. Starting from search engines, online product recommendations to high-end self-driving cars and space missions, the growth of human civilization has already started to be driven by the progress of machine learning. 

Classical machine learning is classified based on the way of interaction between the learner and the environment~\cite{Understanding_ML}. The most basic approaches include supervised \cite{hastie2009overview,schuster1999supervised}, unsupervised \cite{barlow1989unsupervised,hastie2009unsupervised}, semi-supervised \cite{zhu2009introduction,kingma2014semi} and reinforcement learning\cite{wiering2012reinforcement,mnih2013playing}. Supervised learning deals with training from a set of labeled training data while unsupervised learning trains on unlabeled data to find any meaningful patterns. Having the capability of finding associations among data, machine learning was able to provide tailored product development based on customer demands. The normal workflow of traditional machine learning algorithms is shown in Figure \ref{fig:tradmlvsdl}. The raw data presented as image in the figure are first passed to feature extraction phase which outputs feature vectors in a form that is suitable to be fed to machine learning models. The feature vectors are then used for either training or testing machine learning algorithms.  In this section, we will discuss the core machine learning terminologies that will be used frequently later in the survey. This section will not be covering traditional machine learning approaches as they are rarely used for generating adversarial these days in comparison to modern deep-learning based algorithms. The subsections of this section follows no particular hierarchical order, with each subsection being a stand-alone topic. 

\subsection{Deep Learning}

Deep Learning (DL) is a sub-field of machine learning that uses supervised and unsupervised techniques to learn multiple level of representations and features in hierarchical architectures. The ability of conventional machine learning was very limited while processing a raw data. Deep learning has been able to make significant breakthroughs for challenges faced by ML practitioners by showcasing its ability to find patterns in very high dimensional data. Deep learning has enabled researchers to reach unparalleled success in fields of image recognition \cite{Imagesucc1,He_2016_CVPR,Qi_2017_CVPR}, speech recognition \cite{Speechrecogn1,speechrecog2,hannun2014deep}, neuro-science integration \cite{neurosience}, malware detection \cite{droidsec,droiddetect} and most of the ML powered research areas. Since the start, structuring conventional machine learning algorithm required careful feature engineering and high level domain expertise to extract meaningful features from raw data. The effectiveness of machine learning is largely dependent on the representation ability of feature vectors. 
Representation learning is an approach for ML which allows models to be fed with raw data and automatically learns the representation required to make decisions.


\begin{figure}[!t]
    \centering
    \includegraphics[scale=0.8]{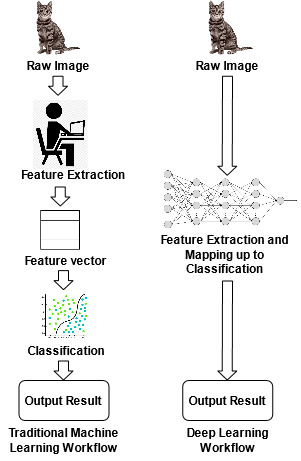}
    \caption{Traditional machine learning vs. deep learning workflow.}
    \label{fig:tradmlvsdl}
\end{figure}

\begin{figure*}
    \centering
    \includegraphics[scale=0.45]{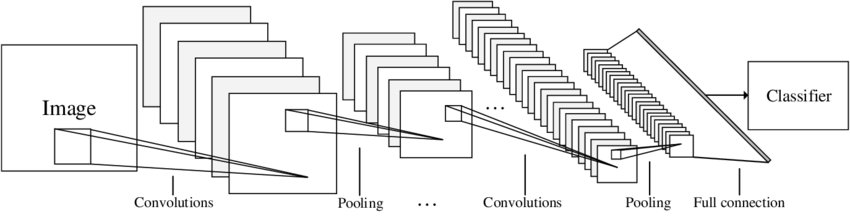}
    \caption{LeNet-5 architecture}
    \label{fig:LeNet_5}
\end{figure*}
Deep-learning methods are representation learning approach with multiple level of representation, obtained by non-linear transformations from lower to higher abstraction level. By combining such simple, non-linear transformation, machine finally learns complex function. Taking representation to higher level in each step signifies amplifying aspects of input which are important for discrimination while suppressing irrelevant features. It can be observed in Figure \ref{fig:tradmlvsdl} that all the feature extraction overhead of traditional learning is replaced by neural nets in deep learning. A standard neural network architecture are made up of connected neurons which are the processors and each neuron outputs a sequence of real-valued activation. Environmental input obtained by sensors activate the input neurons while deeper neurons get activated through weighted connections from previously active neurons. Deep learning operations are usually composed of weighted combination of a group of hidden units having a non-linear activation function, based on a model structure. The architecture of neural network resembles to the perception process of human brain, where a specific sets of unit get activated if it has a role in influencing the output of neural network model. Mathematically, the deep neural network architecture are usually differentiable, so that the optimal weights of the model are learned by minimizing a loss function using variants of stochastic gradient descent through back propagation. For the example mentioned in Figure \ref{fig:tradmlvsdl}, we can consider an image classification example where image input comes in the form of an array of pixel value \cite{Deep_Learning_Lecun}. During the first layer of representation, deep learning models learns the presence or absence of edges at particular orientation. Second layer tries to detect some arrangements in detected edges discarding the minute variations in the position of edges. These arrangement of edges are combined into larger combinations, corresponding to the sections of familiar objects and subsequent layer capable of giving the detection results.

\subsection{Convolutional Neural Network}

Convolutional Neural Network (CNN) is one of the most popular deep learning architecture inspired by natural visual perception of the living beings\cite{gu2017recent}. It takes its name from mathematical linear operation between the matrixes called convolution.  One of the first multi layer artificial neural network (ANN), LeNet-5 \cite{LeCun1989HandwrittenDR,Lecun2017document_recognition}, as shown in Figure \ref{fig:LeNet_5}, is considered to have established the modern framework of CNN architecture. CNN has received ground breaking success in recent years in the field of image processing \cite{Krizhevsky2012ImageNetCW} which has been replicated to many other fields. One of the biggest aspects behind the success of CNN is its ability of reducing the parameters in ANN. 

CNN is mainly composed of 3 layers as shown in Figure \ref{fig:LeNet_5}: convolutional layer, pooling layer and fully connected layer. 
Convolutional layer aims to learn feature representation from the input raw data. Feature maps are computed using convolutional kernels, with each neuron of a feature map connected to a region of neighbouring neurons in the previous layer. New feature map is received by convolving around the inputs with a learned kernel and applying element-wise nonlinear activation function on the convolved results. During feature map generation, kernels are shared by all the spatial locations of the input. The role of pooling layer is to reduce the number of connections among convolutional layers which in turn helps in reducing the complexity of computation. Pooling layer hovers over each activation map and scales the dimensionality using appropriate functions like max, average and so on. Stride and filter size of pooling defines the scaling. Fully connected layers in CNN have same role as that of standard ANN, producing class scores from the activation. Other common CNN architectures include AlexNet~\cite{Krizhevsky2012ImageNetCW}, VGG 16~\cite{simonyan2014very}, Inception ResNet~\cite{szegedy2016inceptionv4}, ResNeXt~\cite{xie2017aggregated}, DenseNet~\cite{huang2018densely}.

\subsection{Reinforcement Learning}
\begin{figure}[!t]
    \centering
    \includegraphics[scale=0.9]{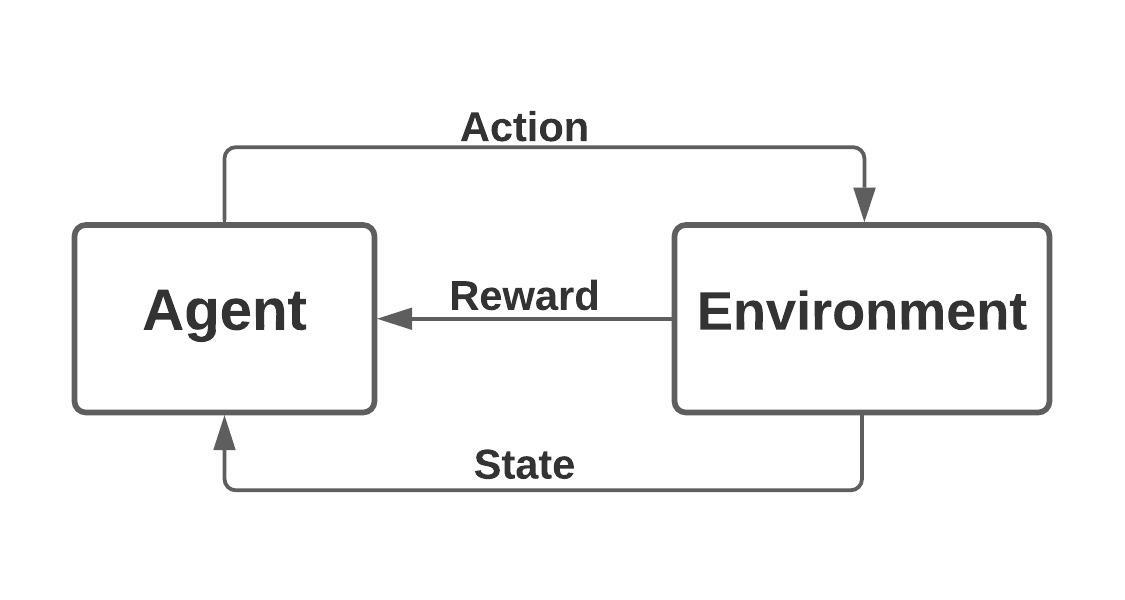}
    \caption{A basic structure of reinforcement learning.}
    \label{fig:RL_Struct}
\end{figure}

Reinforcement learning (RL) can be viewed as a learning problem and a sub-field of machine learning~\cite{algo_for_rl}. 
Basically, its about using past experience to enhance the future manipulation of a dynamic system and learning by maximizing some numerical value which helps to meet long term goals. A supervised learning model learns from data and its labels whereas a RL model completely relies on its experience. 
In RL, a model is trained to make sequence of decisions through the action of agent in a game-like environment. Diagrammatic representation of reinforcement learning is shown in Figure \ref{fig:RL_Struct} where it is shown as the combination of four elements: an agent capable of learning, the current environment state, an action space from which an agent can choose an action and the reward value that an agent is provided in response to each action. Program is deployed to go through a trial and error process to reach the solution of a problem. Agent acting on an environment gets either a reward or penalties for an action it performs and the goal of learning is to maximize the total reward. The programmer sets up the action space, environment and reward policy required for learning and the model figures out the way of performing tasks and maximizing the reward. An agent learning starts with random trials and errors leading to highly sophisticated tactics and superhuman decision making. In a formal definition, a system governed by machine learning algorithm observes a state \(s_t\) from its environment at time step \(t\). The agent performs action \(a_t\) in state \(s_t\) to make transition to a new state \(s_{t+1}\). The state is basically the information about environment which is sufficient for an agent to take best possible actions. The best sequence of actions are defined by the rewards provided by the environment while executing the actions. After completion of each action and transition of environment to new state, environment provides a scalar reward \(r_{t+1}\) to the agent in form of feedback. Rewards could be positive to increase the strength and frequency of the action or negative to stop the occurrence of action. The goal of an agent is to learn a policy \(\pi\) that maximizes the reward. Reinforcement learning faces the challenge of requiring extensive experience before reaching optimal policy.

Exploration and exploitation through all possible directions in high dimensional state spaces leads the learning process to an overwhelming number of states and negatively impact the performance. This had limited the previous success of reinforcement learning \cite{RL_AUTO_HELI,RL_PRIM1,RL_Singh_2002} to lower-dimensional problems. We have discussed the rise of deep learning in last decade by providing low dimensional representation in previous sections. In its way to solve the curse of dimensionality, deep learning also enabled reinforcement learning to scale to very high-dimensional states problem, which were previously considered impractical. Mnih et al. \cite{mnih2013playing} work to play Atari game using deep reinforcement learning and beating human level experts, easily elevated the application of reinforcement learning in combination with deep learning. An actor-critic model with experience replay was used to reach such performance on the Atari game. In deep reinforcement learning framework, agent acting on end-to-end way, takes raw pixels as an input and outputs the associated rewards for each actions. The learned reward function is the basis for deep Q-learning which keeps refining over the experience. Deep reinforcement learning has already been very successful in fields such as robotics \cite{gu2017deep,lillicrap2015continuous,tai2017virtual,vecerik2017leveraging} and game playing \cite{lample2017playing,mnih2015human,heinrich2016deep} where learning from experience is very effective, replacing hand-engineered low-dimensional states.

\subsection{Recurrent Neural Network}
\begin{figure}[!t]
    \centering
    \includegraphics[scale=0.6]{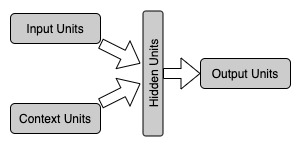}
    \caption{An early schema of recurrent unit.}
    \label{fig:RNN_SCHEMA}
\end{figure}

Neural networks has already been established as a very powerful tool to perform in many supervised and unsupervised machine learning problems. Their ability to learn from underlying raw features which are not individually decipherable has been unparalleled. Despite their significant power to learn from hierarchical representations, they rely on assumption of independence among the training and test sets \cite{lipton2015critical}. Despite of neural net's ability to function perfectly with independent test cases, their assumption of independence fails while data points are correlated in time or space. Recurrent Neural Network (RNN) being a connectionist model, are able to pass information across the sequence steps and processes single sequential data at a time. We can relate this to understanding meaning of a word in text by understanding the previous contexts. RNN is a adaptation of the standard feed-forward neural network allowing it to model sequential data. The basic schema of RNN is shown in Figure \ref{fig:RNN_SCHEMA} where hidden unit takes input of current unit as well as contextual units to provide output. Different from the feed-forward neural networks, the decision for current input depends on activation from previous time steps\cite{sak2014long}. The activation values from previous state are stored inside the hidden layers of a network which provides all the temporal contextual information in place of fixed contextual Windows used for feed forward neural networks (FFNN). Hence, dynamically changing contextual window helps RNN better suited for sequence modeling tasks. 

The gradients of the RNN are very easily computed using back-propagation through time \cite{back_prop_1988}, and gradient descent is a suitable option to train RNN. However, dynamics of RNN makes effectiveness of gradient highly unstable, resulting to exponential gradient decays or gradient blows up. To resolve this issue, enhanced RNN architecture, Long Short-Term Memory (LSTM) is designed \cite{LSTM_1997}. The architecture of LSTM are made up of special units called memory blocks inside the hidden layer of RNN. Memory cells are made up of memory blocks storing the temporary state of network and gates controlling the information flow. A forget gate prevents LSTM models from processing continuous input streams by resetting the cell states. 
Today RNN are being extended towards deep RNNs, bidirectional RNNs and recursive neural nets. Among many application areas, language modeling~\cite{kombrink2011recurrent,chien2015bayesian,sundermeyer2012lstm}, text generation ~\cite{pawade2018story,lu2018neural,sha2018order}, speech recognition~\cite{miao2015eesen,graves2013speech,li2019improving}, text summarization~\cite{nallapati2016sequence,ma2017improving,nallapati2016abstractive} are the major areas transformed by the use of RNN models. 

\subsection{Generative Adversarial Network}
\begin{figure}[!t]
    \centering
    \includegraphics[scale=0.6]{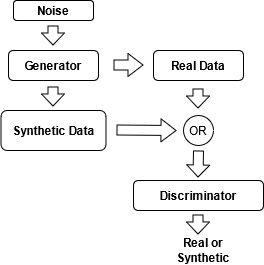}
    \caption{An architecture of two training models of GAN: discriminator and generator.}
    \label{fig:GAN_archi}
\end{figure}
Generative Adversarial Network (GANs) are the generative modeling approach using deep learning methods. Goodfellow et al.~\cite{goodfellow2014generative} proposed GAN as a technique for unsupervised and semi-supervised learning. In a GAN model, two pairs of networks namely: Generator and Discriminator are trained in combination to reach the goal as presented in Figure \ref{fig:GAN_archi}. Creswell et al.~\cite{GAN_creswell} define the generator as an art forger and the discriminator as an art expert. The forger create forgeries, with the aim of making realistic images whereas discriminator tries to distinguish between forgeries and real image. The generator's goal is to mimic a model distribution and the discriminator separates the model distribution from the target~\cite{miyato2018spectral}. The concept here is to consecutively train the generator and the discriminator in turn, with goal of reducing difference between the model distribution and the target distribution. During the training of GANs, discriminator learns its parameters in such a way that its classification accuracy is maximized and generator learns its parameters is such a way that it maximally forges the discriminator. The generator and the discriminator must be differentiable, while not necessarily being invertible.

GAN's ability to train a flexible generator functions, without absolutely computing likelihood has made GAN successful in image generation~\cite{radford2016unsupervised,salimans2016improved} and image super resolution~\cite{bulat2018learn,shamsolmoali2019g}. The flexibility of the GAN models has allowed them to be extended to structured prediction~\cite{hong2018conditional,simo2018mastering}, training energy based models~\cite{zhao2016energy,finn2016connection}, generating adversarial examples for malware~\cite{papernot2017practical,li2020feature}, and robust malware detection~\cite{kim2018zero,kim2017malware}. GAN models suffers from issues of oscillation during training process~\cite{liang2018generative}, depriving them from converging to a fixed point. Approaches that has been taken to stabilize the learning process still rely on heuristics which are very sensitive to modifications \cite{Arjovsky2017TowardsPM}. Recent research work \cite{pan2019recent,karnewar2019msg} is being carried out to address the stability issues of GANs. 
 
 \begin{figure}
    \centering
    \includegraphics[scale=0.9]{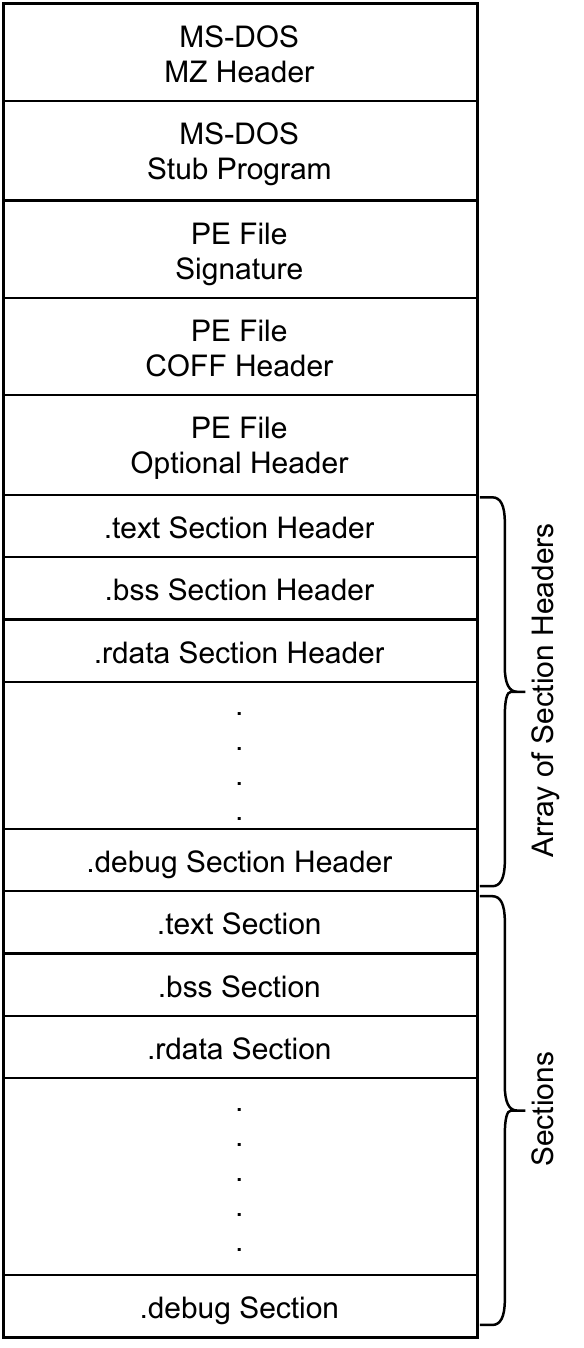}
    \caption{{A structure of Windows PE file}}
    \label{fig:WIN_PE_STRUCT}
\end{figure}
\section{File Structure} 
\label{sec:File_struct}
Executable files are structured differently based on the target/host OS. In this survey, we briefly cover the adversarial attacks across Windows portable executable (PE) file, PDF file and Android files. Although detailed discussions on file structure is out-of-scope for survey, a good understanding of file structure is essential for successful generation of adversarial examples. Different sections of a file are classified into two groups, mutable and immutable. Mutable sections are those which can be modified for adversarial generation without altering the functionality of file whereas immutable sections either breaks the file or alters the functionality on modification. This section will provide brief overview of three kinds of file's structure that are discussed in later parts of survey.

\subsection{Windows PE File Structure}
Windows PE file format is an executable file format based on the Common Object File Format (COFF) specification. The PE file is composed of linear streams of data. The structure of Windows PE file as shown in Figure \ref{fig:WIN_PE_STRUCT} is derived and confirmed from \cite{kexugit,krzysztof_kowalczyk,pe_file_structure}. The header section consists of MS-DOS MZ header, MS-DOS stub program, PE file signature, the COFF file header and an optional header. File headers are followed by body sections, before closing the file with debug information. First 64 bytes of PE file are occupied by MS-DOS header. This header is required to maintain the compatibility with files created on Windows version 3.1 or earlier. In absence of MZ header, the operating system will fail to load the incompatible file~\cite{krzysztof_kowalczyk}. The Magic number used in the header determines if the file is of compatible type. Stub-program is run by MS-DOS after loading the executable and is responsible for giving output messages which include errors and warnings.

PE file header is searched by indexing the \texttt{e\_lfanew} field to get the offset of file which is the actual memory-mapped address. This section of the PE file is one of the target areas to perform modification by using these locations as macros in order to create adversarial examples. The macro returns the offset of file signature location without any dependency on the type of executable file. At offset \texttt{0x3c}, 4-byte signature is placed which helps to identify the file as a PE image. The next 224 bytes is taken by optional header. Even though it may be absent in few types of file, it is not an optional segment for PE files. It contains information like initial stack size, program entry point location, preferred base address, operating system version, section alignment information and few other~\cite{krzysztof_kowalczyk}.
Section headers are of 40 bytes without any padding in between. The number of entries in the section portion is given by the NumberofSections field in the file header~\cite{karl_bridge_microsoft}. Section header contains fields like Name, PhysicalAddress or VirtualSize, VirtualAddress, SizeOfRawData, PointerToRawData and few more pointers with characteristics.

Data is located in data directories inside data section. Information from both the section header as well as optional header are required to retrieve data directories. The .text section contains all the executable code sections along with the entry point. An uninitialized data for the applications are stored in the .bss section which includes all declared static variables and .rdata section represents all the read only data like constants, strings and debug directory information. The .rsrc section contains resorce information for the module and export data for an application are present in .edata section. Section data are the major area where perturbation takes place to make a file adversarial. Debug information is placed on .debug section but the actual debug directories resides in the .rdata section. 

\subsection{Android File Structure}
\begin{figure}
    \centering
    \includegraphics[scale=0.6]{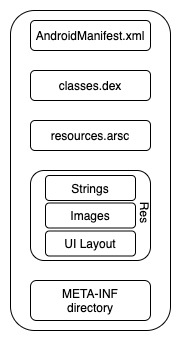}
    \caption{{A structure of Android APK file.}}
    \label{fig:android_apk}
\end{figure}
Android APK file has been recently victimized as a tool for adversarial attacks ~\cite{li2019adversarial,rosenberg2018generic,android_adversarial_Shahpasand,pierazzi2020intriguing}. 
APK file is basically a ZIP files containing different entries as shown in Figure \ref{fig:android_apk}.  Different sections of APK files are described below:
\begin{itemize}
    \item \textbf{Androidmanifest.xml:} AndroidManifest.xml contains the information to describe the application. It contains the information like application's package name, components of application, permissions required and compatibility features~\cite{android_developers}. Due to presence of large amount of information, AndroidManifest.xml is one of the majorly exploited section in APK file for adversarial attack.
    
    \item \textbf{classes.dex:} As Android applications are written in Java, source code will be with extension .java. These source code are optimized and packed into this classes.dex file.
    \item \textbf{resources.arsc:} This file is an archive  of compiled resources. Resources include the design part of apps like layout, strings and images. This file form the optimized package of these resources.
    \item \textbf{res:} Resources of app which is not compiled to store in resources.arsc stays in \textit{res} folder. The XML files present inside this folder are compiled to binary XML to boost performance \cite{ajin_asokan_2016}. Each sub-folder inside \textit{res} store different types of resources. 
    
    \item \textbf{Meta-INF:} This section is only present in signed APKs and has all the files in APK along with their signatures. Signature verification is done by comparing the signature with the uncompressed file in archive~\cite{kalicnsk_2016}.
\end{itemize}

\subsection{PDF File Structure}
\begin{figure}
    \centering
    \includegraphics[scale=0.6]{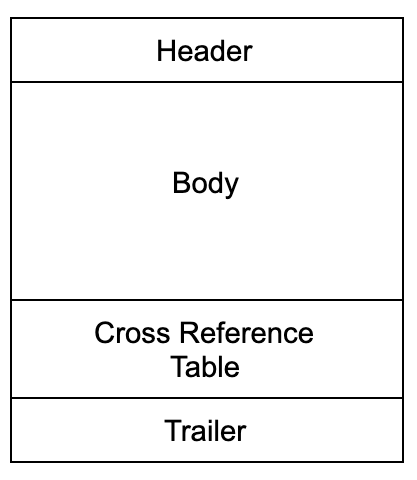}
    \caption{{A PDF file structure.}}
    \label{fig:pdf_struct}
\end{figure}
In this section we will look into the internal structure of PDF file format. PDF is a portable document with wide range of features, capable of representing documents which includes text, images, multimedia and many others. The basic structure of a PDF file is shown in Figure \ref{fig:pdf_struct} and are discussed below:
\begin{itemize}
    \item \textbf{PDF header: }PDF header is the first line of PDF which specifies the version of a PDF file format. 
    \item \textbf{PDF Body: }The body of a PDF file consists of objects present in the document. The objects include image, data, fonts, annotations, text streams, etc.~\cite{simp_pdf_toolscreate}. Interactive features like animation and graphics can also be embedded in the document. This section provides the possibility of injecting contents and files within it, which makes it the most favourable avenue for adversarial attackers. 
    \item \textbf{Cross-reference table: }The cross-reference table stores the links of all the objects or elements in a file. Table helps on navigating to other pages and contents of a document. Cross-reference table automatically gets updated on updating the PDF file.
    \item \textbf{The Trailer: }The trailer denoted end of PDF file and contain a links to cross-reference table. The last line of trailer contains the end-of-file marker, \%\%EOF.
\end{itemize}

\section{Malware Detection}
\label{sec:Malware_Detection}

In globally networked world, malware has posed a serious threat to data, devices and users on internet. From data theft to disrupting the computer operation, with increasing reliability over internet, malware is a growing menace. Malware is being used as a weapon on digital world carrying malicious intentions throughout the internet. Malware attacker tries to take advantage from legitimate users and accomplishing financial or  other goals. Malware can be in any forms like viruses, trojan, ransomware, rootkits, spyware and so on. Global cybercrime cost is projected to be around \$10.5 trillion in 2025~\cite{cybercrimemag_2021} which shows the required urgency to mitigate or limit the damage from these malicious software. Security researchers all around the world are working to combat with these malware files via antivirus software, firewalls and numerous other approaches. However, with big incentives driving malware production, millions of new malware\footnote{https://www.av-test.org/en/statistics/malware/} are  introduced to cyber world every year. These exponentially growing malware number comes with highly equipped tools and techniques, thus requiring continuous work on  effective and efficient malware detection technologies.

\begin{figure}
    \centering
    \includegraphics[scale=0.75]{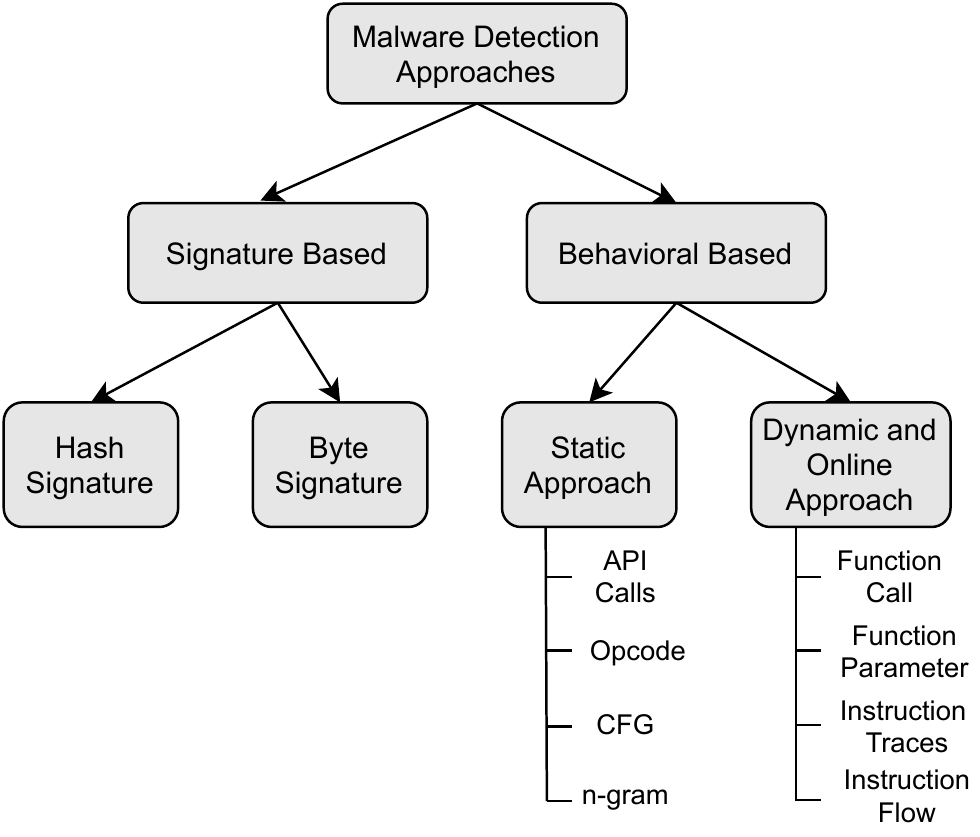}
    \caption{{The malware detection approaches.}}
    \label{fig:mal_det}
\end{figure}
Current malware detection techniques are broadly classified into signature based and behavioral based approaches as shown in Figure \ref{fig:mal_det}.
Traditionally, signature based approaches were used to detect malware. However, due to inability of this approach to detect zero-day attacks, the much focused has been moved into behavioral based approaches (dynamic and online). In modern day anti-virus, hybrid approaches are considered by combining signature based approaches with behavioral based techniques. We will now discuss different types of malware detection approaches.
\subsection{Signature Based Malware Detection}
\begin{figure}
    \centering
    \includegraphics[scale=0.55]{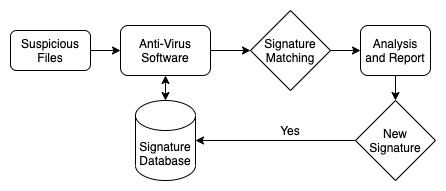}
    \caption{A signature based malware detection approach.}
    \label{fig:sign_based_mal}
\end{figure}
Signature is a short sequence of bytes unique to each malware and helps in identifying malware from rest of the files. Since this approach works by maintaining malware signature database, there are very low false positives rate. Signature based detection has been very effective and fast for malware detection but it is not able to capture the `unseen' malware. Figure \ref{fig:sign_based_mal} shows the malware detection process using signature. As shown in figure, signature database is predefined list of all the possible malware signatures and is solely responsible for entire malware detection process. The anti-malware engine if detects the malicious objects, malware signature is updated in signature database for future detection. A good malware detector's signature database has a huge number of signature that can detect malware~\cite{Souri2018ASS}. Signature based malware detection are good at speed of detection, efficiency to run and broad accessibility~\cite{cloudeyes}. 
However, the inability to detect zero-day malware whose signature is not available in database of anti-malware engine led to question the reliability of signature based approaches. Digital signature patterns can be extracted easily by attacker and implemented to confuse the signature of malware. Current malware comes with polymorphic and metamorphic properties \cite{WANG20151012,Fraley2016PolymorphicMD} which can easily change their behavior enough to change the signature of file. With complete dependence over known malware, signature based detection can neither detect zero day attacks nor the variations of existing attacks. In addition to it, signature database grows exponentially with malware family growing at a rapid pace~\cite{5487493}. 

\subsection{Zero-day Malware Detection}
\begin{figure}[!t]
        \centering
        \includegraphics[scale=0.6]{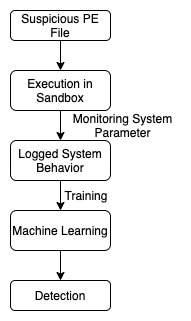}
        \caption{A basic workflow for dynamic malware detection.}
        \label{fig:dynamics}
    \end{figure}

To overcome the limitations posed by signature based approach, zero-day malware detection techniques are focused to capture the unseen malware. In modern zero-day detection approaches, suspicious objects are identified based on behavior or potential behavior of the file~\cite{kimmell2021analyzing,kimmel2021recurrent}. An object's potential behavior is first analyzed for suspicious activities before deploying in a real-time production environment. Those behavior which are anomalous to benign file actions, indicates the presence of malware. Most of the zero-day detection approaches are built around machine learning systems, with state-of-art works using modern deep learning architectures. The captured behavior of the file under inspection is generalized using machine learning models, which is later used to detect unseen malware family. Here, we discuss three different types of zero-day malware detection in this section below:
\begin{figure}
    \centering
    \includegraphics[scale=0.6]{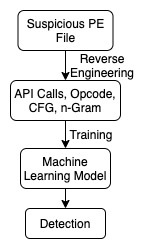}
    \caption{A basic workflow for static malware detection.}
    \label{fig:static_det}
\end{figure}
\begin{itemize}
    \item \textbf{Static Approach: } Static malware detection is the closest approach to signature based system as detection is carried out without running the file. Execution of unknown file may not be always possible in system due to security risks and this is where static detection comes into play. Anti-malware system captures static attributes like hashes, header information, file type, file size, presence of API calls, n-grams etc. from binary code of executable using reverse engineering tools. Once the features are extracted, they are pre-processed to keep only non-redundant and important features. Among numerous available features, n-Grams~\cite{10.5555/1248547.1248646,10.1007/978-3-540-73547-2_48} for byte sequence analysis and Opcode~\cite{10.1504/IJESDF.2007.016865} used to analyze the frequency of 'Operation Code' appearance are the most widely used ones. As shown in Figure \ref{fig:static_det}, the extracted features are fed to different machine learning algorithms ranging from classical to deep learning architectures to train the detection model. The trained model is then used to carry out detection of malware from static features. However, static detection alone is not sufficient to detect more sophisticated attacks \cite{han2019maldae,fleshman2018static,moser2007limits} as the static features can not reflect the exact behavior of malware on run time, which limits its applications~\cite{miller2016reviewer} in real world. 
    
    \item \textbf{Dynamic and Online Approach:} Dynamic approaches~\cite{anderson2011graph,wu2014droiddolphin,feng2018novel} are constructed by executing a suspicious file inside the isolated virtual environments like a sandbox~\cite{4140988} and detecting malware based on a run-time behavior of a program. The use of closed environment is to prevent malware from escaping and attacking the system where analysis is being conducted. Malware on execution can change the registry key maliciously and obtain the privileged mode of operating system~\cite{8667136}. During the execution of malware, properties of operating system changes which is logged by agent in controlled environments. Dynamic analysis enables system to capture dynamic indicators like application programming interface (API) calls, registry keys, domain names, file locations and other system metrices. These features are pre-processed and fed to machine learning model to train malware detector in the flow as shown in Figure \ref{fig:dynamics}. Dynamic analysis are considered more powerful than static due to ability to capture more number of system features. Code obfuscation approaches and polymorphic malware are considered ineffective against dynamic malware detection~\cite{5432531} reflecting its resilience from such sophisticated malware.
    
    Dynamic approach, though overcoming some limitations of the static detection, have its own challenges. Every suspicious file needs to be executed in an isolated environment for specific time frame which results in expense of significant time and resources~\cite{5665796}. The malware file does not guarantee to exhibit a same behavior both in a sandbox and live environment~\cite{lindorfer2011detecting}. Modern smart malware comes with the ability of detecting the presence of sandbox and stay dormant till they reach live systems. In most of the current applications, both static and dynamic analysis are combined to detect the presence of malware in the file~\cite{tzermias2011combining,shijo2015integrated,spreitzenbarth2015mobile}. To combat the issue of polymorphic and metamorphic malware which are evasive to control malicious functionality only during some particular events, online malware detection approaches are performed~\cite{7299317,ozsoy2015malware,demme2013feasibility,mcdole2020analyzing,mcdole2021deep}. Online malware analysis continuously monitors the system for the presence of maliciousness in any file~\cite{abdelsalam2019online}. Continuous monitoring helps to capture the malware at any time in live environment. However, online detection also demands for continuous monitoring overhead to the system. As most of the adversarial attacks performed so far in the literature are on static malware detection approaches, this survey will primarily focus on evasion attacks carried out against static malware detection. 
\end{itemize}

\section{Adversarial Threat Model}
\label{sec:Modeling}

Security threats are defined in terms of their goals and capabilities. 
In this section, we defined the adversarial threat model, tailored to evasion attacks in malware, into four parts: adversarial knowledge, attack surface, adversarial capabilities and adversarial goals. This section aims to provide the readers with explanations to the major components of adversarial attacks.

\subsection{Adversarial Knowledge}
The adversary's knowledge is the amount of information about a model under attack that the attacker has, or is assumed to have, to carry out adversarial attacks against the model. An adversarial attack can be classified into two groups based on the attacker's knowledge:
\begin{itemize}
    \item \textbf{White box attack:} In a white box approach, an attacker has full knowledge about the underlying model. Such knowledge might include, but not limited to, the name of the algorithm, training data, tuned hyper-parameter, gradient information, among others. It is relatively easy to carry out attacks in white box model due to large amount of available knowledge. Current state-of-art works on white box environment have achieved near perfect adversarial attacks~\cite{Suciu2019Malware}.
    \item \textbf{Black box attack:} In a black box approach, an attacker only have access to inputs and outputs of the model. There is no information provided about the internal structure of the model. Generally in black box attack, surrogate model is created by making guess on internal structure of target model using input and output~\cite{hu2017blackbox,yuan2018adversarial}. In addition, in a gray box attack~\cite{vivek2018gray}, a type of black box attacks, the attacker knows the output performance of the model in the form of accuracy, confusion matrix or some other performance metrics.
\end{itemize}
There is large variation on the amount of adversarial knowledge starting from complete access to actual source codes to receiving only output of models. In general, it is assumed that black box adversarial attacks are difficult to orchestrate compared to white box, primarily due to the information available regarding the underlying target model. However, black box attacks reflect more real world use-cases where, in practical sense, an attacker will not likely have any knowledge of models or other parameters.

\subsection{Attack Surface}
Attack surface includes different vulnerable points by which an attacker attacks the target model. Machine learning algorithms pass through a pipeline of different stages before deployment. The flow of data through this data pipeline introduces vulnerabilities in each stage~\cite{abs-1810-00069}. Starting from collection of data, transformation and processing to output generation, an attackers have different attack entry points. Attack surface comprise all those points in machine learning models (malware defender models in our case), where adversaries can carry out their attacks. Based on different approaches to carry out attacks, attack surface has been classified into following broad categories \cite{Biggio_2014}:
\begin{itemize}
    \item \textbf{Poisoning Attack:} This attack is carried out by contaminating training data during the training process of models~\cite{wang2018data,chen2017targeted,koh2018stronger}. Training data is poisoned with faulty data, making machine learning models learn on wrong dataset. As a result of poisoned training data, the entire training process is compromised.
    \item \textbf{Evasion Attack:} This attack is performed by trying to evade a trained system through adjusting malicious input samples at test time~\cite{Anderson2017EvadingML,autom_evad_clasf,Suciu2019Malware}. Evasion attacks do not require any access to training data but requires some level of access to the target model.
    \item \textbf{Exploratory Attack:} This attack is carried out against a model with blackbox access~\cite{hu2017blackbox,Yuan2020BlackBoxAA}. Attackers try to maximize their knowledge without direct access to the underlying algorithm and attempt to reflect the similar input data patterns.
\end{itemize}

\subsection{Adversarial Capabilities}
Adversarial capabilities denote the abilities of adversaries and are dependent on their knowledge of the target model. Some adversaries have access to training data, some have access to gradient information of the model, while others do not have any access to the model at all. The capabilities of attacker vary depending on the information and phase (i.e. training or testing phase) of the model they are attacking. The most straightforward attack approach is attacker having access to full or partial training data. For adversarial attacks carried out on malware files, adversarial capabilities can be classified into following categories:
\begin{itemize}
    \item \textbf{Data Injection: } Data injection is the ability of attackers to inject a new data. There are multiple types of data injection that might take place. One type of injection can be done on training data before training process. Another type of data injection is carried out by inserting a perturbations which forms a new section or replaces original section within an existing file. Injected data can corrupt the original model or cause the data injected file to evade detection.
    \item \textbf{Data Modification: } Data modification can also be performed both for training data and evading file. If attacker has access to training data, data can be modified to cause model learn on modified data. Attacker can also modify input data to cause perturbation and leading to evasion.
    \item \textbf{Logic Corruption: } Logic corruption is the most dangerous ability to be possessed by attacker and also the most improbable. Whenever an attacker has complete access over a model, they can modify the learning parameters and other hyper-parameters related to model. Logic corruption can go undetected which makes it hard to design any remedies.
\end{itemize}

\begin{figure}
    \centering
    \includegraphics[scale=0.75]{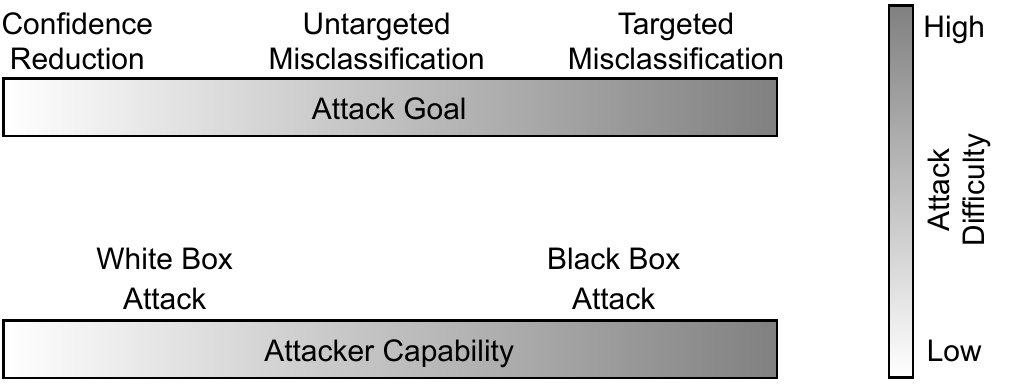}
    \caption{{An attack difficulty with adversarial knowledge and adversarial goals.}}
    \label{fig:adversarial_difficulty}
\end{figure}
\subsection{Adversarial Goals}
An attacker tries to fool the target model, causing it to produce misclassifications. Details of algorithms used to successfully attack and achieve the adversaries goals are discussed in section~\ref{sec:Adv_Alg}. Typically, the adversarial goals of attacker's are categorized as follows:
\begin{itemize}
    \item \textbf{Untargeted Misclassification:} An attacker tries to change the output of model to a value different than original prediction. For a malware classification problem, if a ML model is predicting a malware file as family A, the goal is to force the model to misclassify it as a family other than A.
    \item \textbf{Targeted Misclassification:} An attacker tries to change the output of the model to a target value. For example, if a ML model is predicting a malware file as family A, the goal is to force the model to misclassify it as a family B.
    \item \textbf{Confidence Reduction:} An attacker's goal is to reduce the confidence of a ML model's prediction. It is not necessary to change the prediction value but a reduction of confidence is enough to meet the goal.
\end{itemize}
To summarize, Figure \ref{fig:adversarial_difficulty} gives an overview of the adversarial attack difficulty with respect to the attacker's knowledge, capabilities and goals. While moving in the direction of increasing attack complexity from confidence reduction to targeted misclassification, attack difficulty also increases for the attacker. However, whitebox attacks with higher attacker's capability has least attack difficulty. 

\section{Adversarial Algorithms}
\label{sec:Adv_Alg}
In this section, we will explore the most distinguished adversarial attack algorithms that have been discovered in different domains and are applied to generate adversarial malware samples. Different algorithms are developed in numerous time frames battling the trade off in terms of application domain, performance, computational efficiency and complexity~\cite{9001114}. We will discuss the architecture, implementation and challenges of each algorithm. Most of the attack algorithms are gradient based approaches where perturbations are obtained by optimizing some distance metrics between original and perturbed samples.

\subsection{Limited-memory Broyden - Fletcher - Goldfarb - Shanno (L-BFGS)}

Szegedy et al.~\cite{szegedy2014intriguing} proposed one of the first gradient based approaches for adversarial example generation in the imaging domain using the box constrained Limited-Memory Broyden-Fletcher-Goldfarb-Shanno optimization technique. The authors studied counter-intuitive properties of deep neural networks which allow small perturbations in the images to fool deep learning models for misclassification. Adversarial examples trained for particular neural network are also able to evade other neural networks trained on completely different hyper-parameters. These results are attributed to non-intuitive characteristics and intrinsic blind spots of deep learning models learned by back propagation, with structure connected to data distribution in a non-obvious way. Traditionally, for small enough radius \(\epsilon\)$>$0 around the given training sample \(x\), \(x+r\) satisfying \(||r||<\epsilon\) will be classified correctly by a model with very high probability. However, many underlying kernels are found not holding to this kind of smoothness. Simple optimization procedure is able to find adversarial sample using imperceptibly small perturbations, leading to incorrect classifications by classifier.  While adding noise to an original image, the goal is to minimize perturbation \(r\) added to the original image under \(L_2\) distance.
  A classifier mapping pixel value vectors to a discrete label set is denoted as \(f:R^{m}\xrightarrow {}\{1...k\}\) and the loss function associated is given by \(loss_f : R_{m}* \{1...k\}\xrightarrow {}R^{+}\). For a given image \(x\in R^{m}\) with a target label \(l\in \{1...k\}\), box-constrained optimization problem is defined as :
\begin{equation}
    min ||r||^{2} \;subject \;to: f(x+r)=l, \; x+r\in [0,1]^{m}
\end{equation}
where x is the original image, r is the added perturbation, f is the loss function of the classifier and l is the label of incorrect prediction by the classifier. Perturbed \(x+r\) is arbitrarily chosen using distance minimizer. The computation of distance \(D(x,l)\) is done by approximation using box-constrained L-BFGS. After this early proposal of L-BFGS for adversarial examples generation, plenty of research were triggered to dive into flaws of deep learning. 

\subsection{Fast Gradient Sign Method (FGSM)}
Considering gradient-based optimization technique as a workhorse of modern AI, Goodfellow et al.~\cite{goodfellow2015explaining} proposed an efficient approach for generation of adversarial perturbation in image domain. In contrast to earlier works which explained adversarial phenomena to non-linearity and overfitting, the authors argued the linear nature of neural networks leading to their vulnerability. Linear behaviour in high dimensional space are found sufficient to cause adversarial samples. Linearity is the result of trade off while designing models that are easy to train. LSTMs~\cite{LSTM_1997}, ReLUs and maxout networks~\cite{goodfellow2013maxout} are all found to be intentionally designed to behave linearly for ease of optimization. To define the approach formally, let's consider \(\theta\) as a parameter of model, \(x\) as input to the model, \(y\) as target associated with \(x\) and \(J(\theta,x,y)\) be the cost function for training neural network. On linearizing the cost function around the current parameter values \(\theta\), perturbation can be obtained by
\begin{equation}
    \eta=\epsilon sign(\nabla_x J(\theta,x,y))
\end{equation}
where required gradient can be computed using backpropagation and the approach is called as Fast Gradient Sign Method.

Conversion of features from problem space to feature space affects the precision. Commonly images are represented by 8 bits per pixel and all other information below 1/255 of continuous range are discarded. With limited precision, classifier may not be able to respond to all perturbations whose size is smaller precision of feature. Classifier having well-separated decision boundary for for classes are expected to assign same class for original sample \(x\) and perturbed sample \(x^{'}\) until \(||\eta||_\infty < \epsilon\) where \(\epsilon\) is small enough to be discarded. Taking the dot product and weight vector w and an adversarial example \(x^{'}\):
\begin{equation}
    w^{T}x^{'}=w^{T}x+w^{T}\eta
\end{equation}
The adversarial perturbation increases the activation by \(w^{T}\eta\). The amount of perturbation can be controlled by keeping max norm constraint on \(\eta\) and assigning \(\eta=sign(w)\). Taking \(w\) with \(n\) dimensions and having average magnitude weight vector \(m\), the activation grows by \(\epsilon mn\).  Even though \(||\eta||_{\infty}\) does not grow with increasing dimensionality of the problem but for high-dimensional problems, the activation change caused by the adversarial perturbation can grow linearly. In presence of sufficient dimensions, even simple linear models are seen to have adversarial examples. Adversarial examples are found to occur in contiguous regions of  1-D subspace defined by the fast gradient sign method, where traditional belief was in fine pockets. This allows adversarial examples to be abundant and generalizable across different machine learning models. FGSM being one of the most efficient techniques for adversarial with fast generation of samples, is among the most used technique in this field.

\subsection{Iterative Gradient Sign Method (IGSM)}
Different from the one step perturbation approach where single large step in direction of increasing loss of classifier, Iterative Gradient Sign Method takes iterative small steps while adjusting the direction after each step~\cite{kurakin2017adversarial}. Basic iterative method extends FGSM approach by applying it multiple times with small step size and clipping the pixel values after each iteration to ensure the perturbation within \(\epsilon\) neighbourhood of original image. 
\begin{equation}
    X_{N+1}^{adv}=Clip_{X,\epsilon}{X_N^{adv}+\alpha sign (\nabla_X J(X_N^{adv},y_{true}))}
\end{equation}
where \(X_{N+1}^{adv}\) is the perturbed image at \(N^{th}\) iteration and \(Clip_{X,\epsilon}\{X^{'}\}\) function performs pixel wise clipping on image \(X^{'}\) in order to keep perturbation inside \(L_{\infty} \epsilon\)-neighbourhood of source image X. Kurakin et al.~\cite{kurakin2017adversarial} extended basic iterative method to iteratively least likely class method to produce adversarial for targeted misclassification. Desired class for this version of iterative approach is chosen based on the prediction of the trained network, given as:
\begin{equation}
    y_{LL}=arg_y min{p(y|X)}
\end{equation}
To make adversarial classified as \(y_{LL}, \;log p(y_{LL}|X)\) is maximized taking iterative steps in direction given by \(sign{-\nabla_x log p(y_{LL}|X)}\). Now the adversarial generation cost function can be viewed as:
\begin{equation}
    X_{N+1}^{adv}=Clip_{X,\epsilon}{X_N^{adv}-\alpha sign (\nabla_X J(X_N^{adv},y_{LL}))}
\end{equation}
This iterative algorithm helps to add finer perturbations without damaging the original sample even with higher \(\epsilon\). 

\subsection{Jacobian Saliency Map Attack (JSMA)}
Most of the adversarial generation techniques are based on observing output variations to generate input perturbations, while Papernot et al.~\cite{papernot2015limitations} crafted adversarial samples by constructing a mapping of input perturbations with output variations. The approach is based on limiting the \(l_0\)-norm of the perturbation which deals with minimal number of pixel modification. The proposed adversarial generation algorithm against feed forward DNN modifies small portion of input features by applying heuristic search approaches. Adversarial sample \(X^{*}\) is constructed by adding perturbation \(\delta_X\) to benign sample \(X\) through following optimization problem:
\begin{equation}
    arg  min_{\delta_X} ||\delta_X|| \; s.t.\; F(X+\delta_X)=Y^{*}
\end{equation}
where \(X^{*}=X+\delta_X\) is the adversarial sample and \(Y^{*}\) is the desired adversarial output. Forward derivative is used to evaluate the changes on output due to corresponding modifications in input and these changes are presented on matrix form called as Jacobian of the function. Replacing gradient descent techniques with forward derivative allows attacker to generalize attack for both supervised and unsupervised architecture for broad families of adversaries. Forward derivative of Jacobian matrix of function F is learnt by neural network during training process. For a function with single dimensional output, Jacobian matrix is given as:
\begin{equation}
    \nabla F(X)=\bigg[\frac{\partial F(X}{\partial x_1}, \frac{\partial F(X}{\partial x_2}\bigg]
\end{equation}
Forward derivative helps to distinguish the region which are unlikely to generate adversarial sample and focus on features with high forward derivative for efficient search and smaller distortions. JSMA is a black-box attack with only assumption of DNN architecture using  differentiable activation function. Algorithms take a benign sample \(X\), a target output \(Y^{*}\), a feedforward DNN F, a maximum distortion parameter \(\gamma\), feature variation parameter \(\theta\) and undergoes following steps to give adversarial sample \(X^{*}\) such that \(F(X^{*})=Y^{*}\).
\begin{itemize}
    \item Compute forward derivative \(\nabla F(X^{*})\) 
    \item Construct a Saliency map S based on the derivative
    \item Modify an input feature \(i_{max}\) by \(\theta\)
\end{itemize}
The forward derivative calculate gradients similar to those computed for back-propagation, taking derivative of network directly in place of its cost function and differentiating with respect to input features in place of the network parameters. Consequently, gradients are propagated forward which helps in determining input components leading to significant changes in network outputs. Authors extended application of saliency maps~\cite{simonyan2014deep} to construct adversarial saliency maps which gives features having significant impact on output and thus is a very versatile tools to generate wide range of adversarial examples. Once saliency map gives the input feature that needs to be perturbed, benign samples are perturbed using distortions limited by parameter \(\gamma\). The limiting parameter \(\gamma\) depends on human perception of adversarial sample. The experiment is carried out on LeNet architecture using MNIST dataset. Adversarial crafting is done by increasing or decreasing the pixel intensities of images. Before wrapping up JSMA, we discuss briefly about Saliency Vector. Saliency Vector contains the features in input blocks of data and their significance for machine learning model. Importance of input feature given by saliency vector can be thought of as a function of network's sensitivity to changes in the input feature~\cite{STEPPE1997109}. The regions of element in original files corresponds to the position of elements in the vector and value of that element gives the measure of importance of that feature region. Zhou et al.~\cite{zhou2015learning} proposed Class Activation Mapping to produce visual interpretation for CNN-based model. Authors used global average pooling to indicate discriminative image regions used by CNN to make the decision. Due to difficulty of modifying and retraining the original model to obtain CAM, Selvaraju et al.~\cite{Selvaraju_2019} proposed Grad-CAM method. Gradient-weighted CAM uses the gradient information flowing into the final convolutional layer to produce a localization map highlighting the important regions of image. Saliency vector allows to observe database bias and improve the models based on training data.  

\subsection{Carlini \& Wagner attack (C\&W)}
Carlini \& Wagner \cite{carlini2017evaluating} proposed an adversarial generation approach to overcome the defensive distillation. Defensive distillation has been recent discovery to harden any feed-forward neural network against adversarial examples by performing only a single retraining. Proposed approach is able to perform three types of attacks: $L_{0}$ attack, $L_{2}$ attack and $L_{\infty}$ attack to evade defensively distilled and undistilled networks. These attacks are based on different distance metrics which are:
\begin{itemize}
    \item $L_0$ distance, measuring the number of pixels modified in an image
    \item $L_2$ distance, measuring the standard Euclidean distance between original sample and perturbed sample
    \item $L_{\infty}$ distance, measuring the highest change among any of the perturbed coordinates
    \end{itemize}
The optimization problem for adversarial generation of input image $x$ is given as:
\begin{equation}
    min D(x,x+\delta)
    \; such\; that\; C(x+\delta)=t
              x+\delta \in [0,1]^{n}
\end{equation}
where input $x$ is fixed and goal is to reach $\delta$ that minimizes $D(x,x+\delta)$. $D$ could be any of distance metric among $L_0,\;L_2 \;or\; L_{\infty}$. Different approaches are taken to limit the modification and generate valid perturbations:
\begin{itemize}
    \item Projected gradient descent is allowed to perform only one standard gradient descent, clipping all other coordinates
    \item Clipping gradient descent does not clips input perturbation on each iteration, but clips into objective function to be minimized
\end{itemize}
From their experiments~\cite{carlini2017evaluating}, it is observed that $L_2$ attack has low distortion while $L_{\infty}$ and $L_0$ is not fully differentiable as well as bad suited for gradient descent. 

\subsection{DeepFool}
\begin{figure}
    \centering
    \includegraphics[scale=0.6]{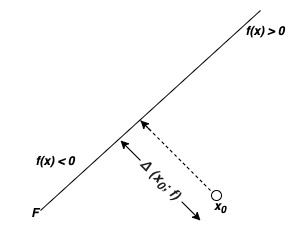}
    \caption{{DeepFool - perturbation is added to original sample $x_0$ in direction perpendicular to hyperplane.}}
    \label{fig:deepfool}
\end{figure}
Dezfooli et al. \cite{moosavidezfooli2016deepfool} proposed an untargeted white-box adversarial generation technique called as DeepFool. DeepFool works by minimizing the euclidean distance between perturbed sample and original samples. Attack begins by generating linear decision boundary to separate the given classes and accompanied by addition of perturbation perpendicular to the decision boundary that separates classes as demonstrated in Figure \ref{fig:deepfool}. Attacker projects the perturbation into a separating line called hyper-plane and tries to push it beyond for mis-classification. In high dimensional space, decision boundaries are usually non-linear, so the perturbation are added iteratively by performing multiple attacks till evasion. Attack for such multiclass finds the closest hyperplane and projects input towards that hyperplane and then proceeds to other. The minimal perturbation required to misclassifiy classifier is the orthogonal projection of $x_0$ onto $f$ and is given by closed loop formula in Equation~\ref{eq:deepfool}.
\begin{multline} \label{eq:deepfool}
    r_{*}(x_0) := argmin||r||_{2} \; \; subject \; to \;  \\sign (f(x_0+r)) \neq sign(f(x_0))
    =-\frac{f(x_0)}{||w||_2^{2}}w
\end{multline}
where $r$ is perturbation, $f$ is classifier function. $w$ is gradient and $x_0$ the original sample.

\subsection{Zeroth Order Optimization (ZOO)}
All of the previously discussed adversarial generation algorithms are dependent on gradient of detector module which limits the adversarial attack space within white-box attack. Chen et al.~\cite{Chen_ZOO_2017} proposed a black-box adversarial generation approach by estimating the gradients of targeted DNN with only access to input and output of a target. Zeroth order methods are gradient-free optimization approach requiring only the zeroth order oracle for optimization process. The objective function is analyzed at every two close points $f(x+hv)$ and $f(x-hv)$ with a very small h, so that a gradient along the direction of vector $v$ can be estimated. Gradient estimation is followed by an optimization algorithms like gradient descent. While attacking black-box DNN with large input size, use of a single minute step of gradient descent can be very inefficient as large number of gradient needs to be estimated. To resolve this optimization issue, coordinate descent method is used by optimizing each coordinate iteratively.

Zeroth Order Optimization attack is inspired by formulation of the C\&W attack. The loss function of C\&W attack is modified in such a way that it is only dependent to output of DNN and desired class label. A new hinge like loss function based on the output $F$ of DNN is defined in Equation \ref{eq:hinge}.

\begin{equation}\label{eq:hinge}
    f(x,t)=max\{\max_{i\neq t} log[F(x)]_i-log[F(x)]_t,-K\} 
\end{equation}
 The proposed adversarial generation technique does not require gradient to be estimated accurately. To accelerate the zeroth order methods, attack-space dimension reduction approach is carried out which reduces the number of gradient to be estimated. Attack-space dimension reduction might lead to insufficient search space to find adversarial. So, hierarchical attack scheme is used where dimension is gradually increased during process of optimization. Using coordinate descent and importance sampling, attacker can also update pixels on a selective basis. 
 
 \subsection{One Pixel Attack (OPA)}
Another gradient free adversarial generation approach is proposed by Su et al. \cite{Su_2019} by generating one pixel adversarial perturbations based on differential evolution (DE). Differential evolution is a population based optimization algorithm which has ability to find higher quality solutions than gradient based approaches~\cite{yina_de}. Since gradient information is not required for DE, the need of differentiable objective functions is also omitted. One pixel attack perturbs single pixel using only probability labels. Single pixel modification allows attackers to hide the adversarial modifications making it imperceptible. To carry out the attack, each image is represented as a vector where each scalar element represents one pixel. With $f$ as the target function, $x=(x_1,...,x_n)$ representing n-dimensional inputs, $t$ being the original class, $e(x)=(e_1,...,e_n)$ denoting the perturbation to be added to the input with maximum modification limited to $L$ , the optimized solution is given by Equation \ref{Eq:OPA}.
\begin{equation}
\label{Eq:OPA}
    \max_{e(x)^{*}} f_{adv}(x+e(x)) \; \;
    subject \: to \: ||e(x)||_0 \leq d,
\end{equation}
where \textit{d} is a small number. This approach deals with determining two values: dimension to be perturbed and the required corresponding magnitude of modification for each dimension. Unlike other attack strategies, OPA focus modifications on only one pixel without limiting the strength of modification.

\section{Adversarial Malware Evasion Attacks}
\label{sec:Adv_attacks}
Adversarial generation methods that originated in the image domain did not take long to migrate into the malware field. Among different adversarial threats, evasion attack has been the most worrisome approach that has already been exploited in different ways. Adversarial malware started with PDF and Windows files due to their abundance and then proliferated into other file formats. There have been significant work on adversarial generation for Android, PDF, Windows and Linux files. This section deals with adversarial examples generated to evade malware detection systems by making minor perturbations on input malware files. These subtle modifications on malware files during test time are able to sneak through blind-spots of machine learning models without breaking the functionality of malware. The following sections will briefly explain different adversarial generation works carried out by researcher on the malware domain. Adversarial work has been divided based on the attack domain which includes Windows, Android, PDF Hardware Based and Linux malware files. The following subsection discusses adversarial attacks in Windows files.

\subsection{Windows Malware Adversarial}
Microsoft Windows is a dominant operating system on PCs with more than 70\% market share and 1.5 billion users worldwide~\cite{liu_2021}. 
Gartner research~\cite{gartnerinc} predicts 30\% of cyberattacks by 2022 will be carried out in form of adversarial. Abundant availability has placed Windows malware at the core of adversarial threats. Even the continuously evolving machine learning based malware detectors are not able to withstand adversarial attacks. In this section, we will cover different adversarial attacks carried out on Windows malware detectors.

Machine learning based models being data hungry, feature engineering is a critical task to feed important features as input. However, the advent of deep neural network has allowed models to learn features themselves from complex raw data. Deep neural networks have shown impressive performance in malware detection by providing whole binary file as a input without any hand crafted feature engineering effort. Different malware detectors have been previously discussed but we want to include Raff et al. work \cite{raff2017malware} in this section (referred as MalConv) which has been a industry standard in the field by making detection considering whole executable. Most of the attacks discussed in the survey are carried out against MalConv.
\begin{figure}
    \centering
    \includegraphics[scale=0.65]{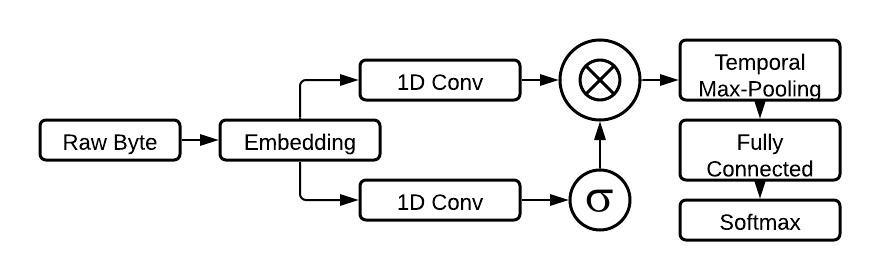}
    \caption{MalConv architecture}
    \label{fig:malconv}
\end{figure}
Detection using raw bytes comes with the sequence problem with millions of time steps and batch normalization hindering the learning process. Raff et al. tried to replicate the neural net's success in learning features from raw inputs, successfully performed in image \cite{szegedy2014going}, speech \cite{graves2013speech} and text \cite{zhang2016text} domain. With bytes in malware having multiple modalities of information and contents exhibiting multiple spatial correlation, MalConv becomes the first ever architecture with ability to process a raw byte sequences of more than two million steps. MalConv addresses high amount of positional variation present in executable files. It's architecture as shown in Figure \ref{fig:malconv}, combines the convolutional activation with a global max-pooling before going to fully connected layers allowing the model to produce its activation regardless of the locations of the detected features. MalConv as one of the only robust static detector, has been considered a baseline for most of the static adversarial attacks. 

\begin{table*}[t]
    \centering
    \def\arraystretch{1.5}
    \caption{{A gradient based approaches.}}
    \setlength{\tabcolsep}{0.8\tabcolsep}
    \rowcolors{2}{gray!25}{white}
    \begin{tabular}{|p{1.5cm}|p{2.25cm}|p{1.25cm}|p{5.75cm}|p{1.75cm}|p{3.5cm}|}
    \rowcolor{gray!25}
    \hline
     \textbf{Paper/Year}  &\textbf{Key Motivation}&\textbf{Target Model} & \textbf{Approach} & \textbf{Modification} & \textbf{Preserving Functionality} \\
     \hline
     Kolosnjaji et al. 2018 ~\cite{8553214} &
     Adversarial attack on malware detection using raw bytes &
     MalConv ~\cite{raff2017malware} &
     \begin{minipage}[t]{\linewidth}
     \begin{itemize}[leftmargin=*]
         \item Optimizing one byte at a time using gradient descent
         \item Embedded layer to tackle non-differentiable MalConv architecture
         \item Gradient calculation of objective function with respect to embedded representation
     \end{itemize}
     \vspace{1mm}
     \end{minipage}& Bytes are padded only at the end of file & 
     \begin{minipage}[t]{\linewidth}
     \begin{itemize}[leftmargin=*]
         \item Byte padding only at the end
         \item Padding byte closest to embedded byte is chosen
     \end{itemize}
     \vspace{1mm}
     \end{minipage}\\
     \hline
     Kreuk et al. 2018 ~\cite{Kreuk2018AdversarialEO} &
     Gradient based attack with better reconstruction &
     MalConv ~\cite{raff2017malware}&
     \begin{minipage}[t]{\linewidth}
     \begin{itemize}[leftmargin=*]
         \item Perturbation generation in embedded space
         \item Calculation of weighted distance between generated adversarial embedding from actual embedding
         \item Weighted gradient similar to iterative FGSM
     \end{itemize}
     \vspace{1mm}
     \end{minipage}&
     Padding bytes at the end of file &
     \begin{minipage}[t]{\linewidth}
     \begin{itemize}[leftmargin=*]
         \item Loss function enforcing perturbation close to embedding matrix row
         \item Payload bytes are inserted into marked and flagged region
     \end{itemize}
     \vspace{1mm}
     \end{minipage}\\
     \hline
     Demetrio et al. 2019 ~\cite{Demetrio2019} &
     Explainable technique for efficient adversarial generation &
     MalConv ~\cite{raff2017malware} &
     \begin{minipage}[t]{\linewidth}
     \begin{itemize}[leftmargin=*]
         \item Feature attribution to determine most influential feature
         \item Perturbation generation using gradient of classification function with respect to embedding layer
         \item Bytes modification in file header
     \end{itemize}
     \vspace{1mm}
     \end{minipage}&
     Changing bytes of file header & 
     \begin{minipage}[t]{\linewidth}
     \begin{itemize}[leftmargin=*]
         \item MZ magic number and offset at \texttt{0x3C} are not modified
     \end{itemize}
     \vspace{1mm}
     \end{minipage}\\
     \hline
     Suciu et al. 2018 ~\cite{Suciu2019Malware} &
      Test existing methods on production-scale dataset and 
      compare effectiveness of different strategies &
      MalConv ~\cite{raff2017malware} &
      \begin{minipage}[t]{\linewidth}
      \begin{itemize}[leftmargin=*]
          \item Random, gradient based and fast gradient perturbation
          \item End of file append and slack region insertion
          \item Transferability test across full, EMBER~\cite{anderson2018ember} and mini dataset
      \end{itemize}
      \vspace{1mm}
      \end{minipage}&
      Padding bytes at the end and in the slack regions &
      \begin{minipage}[t]{\linewidth}
      \begin{itemize}[leftmargin=*]
          \item Slackindexes calculated before adversarial payload insertion
          \item Updates only at the end or in slack regions
      \end{itemize}
      \vspace{1mm}
      \end{minipage}\\
      \hline
      Chen et al. 2019 ~\cite{CNNATTACK_SALIEN_VECT} &
      Enhancing effectiveness of gradient based approach through benign perturbation initialization &
      MalConv ~\cite{raff2017malware} &
      \begin{minipage}[t]{\linewidth}
      \begin{itemize}[leftmargin=*]
          \item Saliency vector generation using Grad-CAM method
          \item Random benign block initialization and enhanced benign file append, followed by FGSM for white box attack
          \item Experience based attack by summarizing the successful trajectories of random benign attacks for black box attack 
      \end{itemize}
      \vspace{1mm}
      \end{minipage}&
      Bytes are appended only at the end of file & 
      \begin{minipage}[t]{\linewidth}
      \begin{itemize}[leftmargin=*]
          \item No alteration of existing section
          \item Appending only at the end
      \end{itemize}
      \vspace{1mm}
      \end{minipage}\\
      \hline
    \end{tabular}\\
    \vspace{1mm}
    \footnotesize\textit{\textbf{Key Motivation}: The major motive behind the published work, \textbf{Target Model}: Target defense for adversarial attack, \textbf{Approach}: Key procedures to carry out adversarial attack, \textbf{Modification}: Changes on file to craft the adversarial perturbation, \textbf{Preserving Functionality}: Works towards safeguarding the functionality of a malware}
    \label{tab:Grad_approach}
    
\end{table*}

\subsubsection{Gradient Based Attack}
\begin{figure}
    \centering
    \includegraphics[scale=0.6]{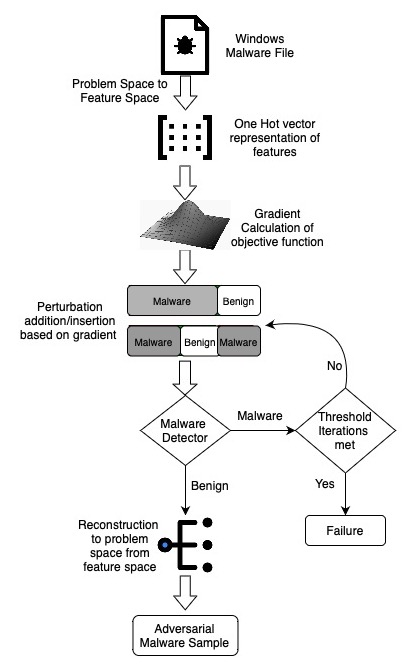}
    \caption{A gradient based adversarial attacks on Windows malware.}
    \label{fig:Grad_Attack}
\end{figure}
\begin{figure}[!t]
    \centering
    \includegraphics[scale=0.7]{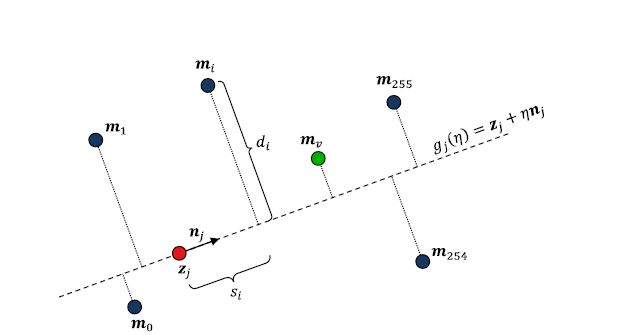}
    \caption{A representation of an exemplary two dimensional byte embedding space from ~\cite{8553214}.}
    \label{fig:kolosnjaji}
\end{figure}

Table \ref{tab:Grad_approach} presents comparative study of adversarial attacks using gradient of cost function against Windows malware. Since Anderson et al.\cite{Anderson2017EvadingML} proposed possibility of manipulating different sections of Windows PE malware to form adversarial, various researches have been conducted to bypass malware detector using malware PE files. Authors \cite{Anderson2017EvadingML} used random actions from action space to modify PE files resulting in very low accuracy. To reduce the randomness of payloads, Kolosnjaji et al. \cite{8553214} proposed appending optimized padding bytes using gradient descent, originally proposed by Biggio et al. \cite{DBLP:journals/corr/abs-1708-06131}. Gradient based approaches are carried out using either append or insertion method for perturbation generated using gradient of cost function as shown in Figure \ref{fig:Grad_Attack}. One-hot represented vector of malware is combined with gradient generated perturbation to bypass the malware detector. Making complex changes in architecture of file requires precise knowledge and could destroy file integrity as well. Kolosnjaji et al.~\cite{8553214} choose to append bytes only at the end of file by optimizing one padding byte at a time. In this case, attacker's goal is to minimize the confidence of malicious class, limiting the maximum perturbation to \(q_{max}\). Appending \(q_{max}\) bytes to input size \(x_0\) should satisfy \(k+q_{max}<=d\), where k is the number of informative bytes and d is fixed input size to detector. Due to non-differentiability issue of embedded layer in MalConv, negative gradient of loss function is calculated with respect to embedded representation. Embedding layer is a lookup table that contains mapping for each input \(x_j\) to a 8-dimensional vector \(z_j=\phi(x_j)\) while optimizing one padding bytes at a time. Byte \(x_j\) is replaced by optimal padding byte corresponding to embedded byte \(m_i\) closest to line \(g(\eta)=z+\eta n\), defined parallel to negative gradient descent \(w\) passing through embedded representation \(z\) as shown in Figure \ref{fig:kolosnjaji}.  Authors were able to achieve evasion rate up to 60\% by only modifying 1\% of bytes in PE file. 

Kreuk et al.~\cite{Kreuk2018AdversarialEO,kreuk2019deceiving} proposed the enhanced attack method against MalConv \cite{raff2017malware} using iterative FGSM \cite{goodfellow2015explaining}.  Authors focused this approach on enhancing reconstruction by introducing new surrogate loss function. Representation of binary files as a sequence of bytes are arbitrary and neural network is unable to work in this space. Generating adversarial examples deals with adding perturbations to original sample as given by increasing or decreasing gradient. However, this process is not that simple as perturbation in one-hot vector results to a vector that is no longer in one-hot vector space. This approach proceeds by generating perturbation in embedded space. In many cases, the perturbed embedding loses resemblance to embeddings in lookup table which contains mapping between bytes to embeddings. In absence of resemblance, reconstruction is not possible. Kreuk et al.  introduced new term to loss function which causes perturbations to be close to embedding matrix. Introduced term is the weighted distance of generated adversarial embeddings from actual embeddings, with a goal of minimizing the distance. In order to preserve the functionality, payload bytes and flag is appended and only the flagged appended bytes is perturbed. The proposed white-box attack could obtain very high evasion rate of around 95\%. The new loss function proposed by authors is given as:
\begin{equation}
    \Bar{l}^*(z,y;\theta)=\alpha.\Bar{l}(z,y;\theta)+(1-\alpha)\bigg[\sum_{i=1}^L\sum_{j=1}^Nd(z_i,M_j)\bigg]
\end{equation} 
where first part is the categorical loss called as the negative log-likelihood loss and second term gives the distance of generated adversarial embedding with the actual embedding in \textit{M}. Second term is responsible to steer the direction towards reconstructible adversarial embeddings.

To interpret the blackbox decisions of malware detection model, Demetrio et al. ~\cite{Demetrio2019} proposed a technique called integrated gradients initially proposed by Sundararajan et al. \cite{DBLP:journals/corr/SundararajanTY17}. With input model $f$, a point $x$ and baseline $x^{'}$, the attribution of \(i_{th}\) feature is computed as:
\begin{equation} \label{eq: IG}
    IG_i(x)=(x_i-x_i^{'}) \int_0^1 \frac{\partial f(x^{'} + \alpha(x-x^{'}))}{\partial{x_i}}d\alpha
\end{equation}
Equation (\ref{eq: IG}) is the integral of the gradient computed on all points on line passing through \(x\) and \(x^{'}\). Feature attribution is used to determine the most influential feature leading to meaningful explanations behind classifications of malware binaries. Feature attribution is extended in a way that information about relevant features are obtained from higher semantic level. Referencing the findings of research, authors are also able to generate adversarial malware samples by efficiently modifying few bytes in file header. Integrated gradient method satisfies both sensitivity and implementation invariance based on the concept of baseline. Findings of research showed that MalConv \cite{raff2017malware} learns to differentiate between malware and benign based on header bytes, ignoring the bytes present in other sections of file. Data sections and text sections being the major malicious section in file, considering only few header bytes for detection exposes serious concern in MalConv model. This approach is more efficient as it requires very few manipulations to bypass the detector. Authors were able to evade almost all malware by generating small perturbations on file header sections other than MZ magic number and value at offset \texttt{0x3C}. Perturbation generation using gradient of classification function with respect to embedding layer is same as implemented by Kolosnjaji et al. \cite{8553214}. Along with success in efficient adversarial attack from perturbations in file header, research also introduces to new challenges of perturbation being easily detected and patched. This study has directed further research towards hiding modifications from detection.

Suciu et al.~\cite{Suciu2019Malware} trained existing models to study their behaviour on a production-scale dataset. Author comparatively evaluated effectiveness of adversarial generation strategies at different scales and observed their transferability property. Full dataset with 16.3M PE files, EMBER dataset\cite{anderson2018ember} with 1.1M PE files and mini dataset with 8598 files were used to train different attack strategies. Existing adversarial attacks are constrained on appending adversarial noise at the end of binary file. However, appended bytes are found to be less influential and offsetted by bytes in original malware. Inability of byte appending strategies while using size constrained detector like MalConv(Only first 2MB are considered for detection), led authors to use slack attacks. Slack attacks are performed by discovering the region in executable files that are not mapped to memory and will not affect the functionality on modification. These slack gaps formed due to mis-alignments between physical addresses in disk and virtual addresses are combined across all the slack regions and returned as slackindexes indicating modifiable regions. Research also pointed MalConv to take only 128 most discriminative features from 4195 possible features to make classification decision. Attacking most influential feature will amplify attack effectiveness and sufficiently appended bytes can replace legitimate features.

Random append adds values sampled from uniform distribution where gradient append uses gradient value of output with respect to input layer, giving the direction towards benign class. Random append and benign append fails to perform in all data size whereas Fast Gradient Method (FGM) append could reach 71\% success rate in full dataset. Due to linearly increasing convergence time of gradient based approach with increase in number of appended bytes, FGM append attack is adapted as a one-shot gradient. Slack attacks are found to be more effective in both ember and full dataset with greater possibility for modifications in full dataset. FGM append achieved higher SR in comparison to slack attack but in expense of larger bytes modification. It is observed that models trained on larger data are more vulnerable as model encodes more sequential information allowing gradient evaluation be more meaningful. Need of appending bytes in each iteration for gradient append may lead to divergence from oscillating pattern. Single step gradient samples are found non-transferable against MalConv architecture. Authors were able to make comparative analysis over models in dataset of different scales and point out challenges of models.

\begin{table*}[t]
    \centering
    \setlength{\tabcolsep}{0.8\tabcolsep}
    \def\arraystretch{1.5}
    \caption{{A code obfuscation based approaches.}}
    \def\arraystretch{1.5}
    \rowcolors{2}{gray!25}{white}
    \begin{tabular}{|p{1.5cm}|p{2cm}|p{1.5cm}|p{5.5cm}|p{2cm}|p{3.5cm}|}
    \rowcolor{gray!25}
    \hline
     \textbf{Paper/Year}  & \textbf{Key Motivation}&\textbf{Target Model} & \textbf{Approach} & \textbf{Modification} & \textbf{Preserving Functionality} \\
     \hline
     Park et al. 2019 ~\cite{Park2019Mal_Obfus} &
      Generative model by obfuscation in raw binaries &
      Inception V3~\cite{Inception_v3}, MalConv ~\cite{raff2017malware}&
      \begin{minipage}[t]{\linewidth}
      \begin{itemize}[leftmargin=*]
          \item Dummy code insertion using Adversarial Malware Alignment Obfuscation
          \item Semantic nops insertion to match original malware to standard adversarial
          \item Optimization in closed loop till evasion
      \end{itemize}
      \vspace{1mm}
      \end{minipage}&
      Semantic nops insertion & 
      \begin{minipage}[t]{\linewidth}
      \begin{itemize}[leftmargin=*]
          \item Modification with executable adversarial
          \item Dummy code insertion in form of semantic nops
      \end{itemize}
      \vspace{1mm}
      \end{minipage}\\      
      \hline
      Song et al. 2020 ~\cite{song2020automatic} &
      Practical adversarial generation and evaluation against real world anti-virus system &
      Signature based and machine learning based detectors &
      \begin{minipage}[t]{\linewidth}
      \begin{itemize}[leftmargin=*]
          \item Selection and application of macro actions from action space
          \item Action sequence minimization, traversing through actions and removing unnecessary actions
          \item Entangling macro actions to micro actions to evaluate feature essence
      \end{itemize}
      \vspace{1mm}
      \end{minipage}& Through sequence of macro and micro action &
    \begin{minipage}[t]{\linewidth}
      \begin{itemize}[leftmargin=*]
          \item Functionality preserving actions
          \item Cuckoo sandbox verification
      \end{itemize}
      \vspace{1mm}
    \end{minipage}\\
     \hline
     \end{tabular}\\
     \vspace{1mm}
     \footnotesize\textit{\textbf{Key Motivation}: The major motive behind the published work, \textbf{Target Model}: Target defense for adversarial attack, \textbf{Approach}: Key procedures to carry out adversarial attack, \textbf{Modification}: Changes on file to craft the adversarial perturbation, \textbf{Preserving Functionality}: Works towards safeguarding the functionality of a malware}
    \label{tab:Code_obfus}
\end{table*}

All previous works based on gradient initialized perturbations with random noises and then iteratively updated using gradient of the model. Role of initializing perturbation in the success rate of adversarial generation can not be disregarded and Chen et al.~\cite{CNNATTACK_SALIEN_VECT} proposed use of saliency vector to select initializing perturbations from benign files. Researchers consider issue of accuracy and inefficiency in the work of Kreuk et al.~\cite{Kreuk2018AdversarialEO} and Suciu et al.~\cite{Suciu2019Malware} as a result of random perturbation initialization before gradient driven iterative modification. Benign feature append method was carried out by debugging victim model once for generating saliency vectors whereas continuous debugging of model is required while incorporating FGSM algorithm. Saliency vector assigns values to benign and malicious regions of file with higher values linked to more significant features. In white box environment, BFA attack appends perturbation at the end of malicious files, selected using saliency vectors while enhanced-BFA attack uses most significant benign blocks for initialization. Avoiding random initialization helps model to obtain back propagation gradients and, gradient based algorithms can be implemented more effectively. Benign bytes form saliency vector also helps in mapping between adversarial from continuous space to discrete space avoiding random perturbations which can not be accurately mapped back to corresponding raw-byte perturbations. Authors also performed black box attack to malware detectors using random benign append and perturbations obtained by summarizing the successful trajectories of random attacks. This work was successful to increase the accuracy of gradient based adversarial generation techniques just by replacing random initialization. 

\subsubsection{Code Obfuscation Based Attack}

\begin{figure}
    \centering
    \includegraphics[scale=0.6]{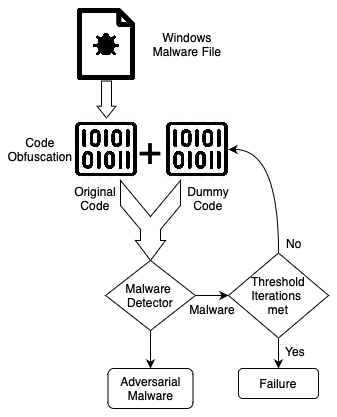}
    \caption{An adversarial generation workflow using code obfuscation.}
    \label{fig:code_obfus}
\end{figure}

Code obfuscation changes the pattern of program without any damage to program logic. Adversarial attacks using obfuscation deals with modifying code section without changing the functionality and flow of program, as shown in Figure \ref{fig:code_obfus}. Table \ref{tab:Code_obfus} discusses the code obfuscation attacks done against Windows malware detectors. Park et al.~\cite{Park2019Mal_Obfus} proposed a generative model for generating adversarial through obfuscation in raw binaries. The proposed approach minimally modifies malicious raw binaries using a dynamic programming based insertion algorithm, obfuscating the .txt section of a binary in executable byte sequence. Windows malware binaries are initially converted into grayscale images from byte code for obfuscation. Obfuscation technique called dummy code insertion is used to insert semantic \textit{nops} (no operation) into a program. From the study, authors proposed Adversarial Malware Alignment Obfuscation (AMAO) algorithm which resembles string matching algorithm. Algorithm takes two inputs along with insertion points for semantic \textit{nops}. One of the input is binary representation of non-executable adversarial example generated by standard adversarial generation algorithm (FGSM and C\&W) with another input being binary representation of original malware. AMAO inserts semantic \textit{nops} in original malware such that original malware matches to first input as closely as possible. At each iteration, algorithm chooses between inserting a semantic \textit{nop} or not inserting anything based on distance metric between binary strings. Adding semantic \textit{nops} is easier if the source code is given but without it, patching techniques are required~\cite{7174804,10.1145/2931037.2931047}. Algorithm outputs adversarial malware with original program's logic after operating in a closed loop model until the classifier gets fooled.  Authors tested AMAO algorithm against different classification models including simple CNN, LeNet5~\cite{szegedy2015rethinking}, Inception V3 and MalConv~\cite{raff2017malware} in a whitebox as well as blackbox environment. Proposed algorithm is found to be effective against classifiers employing both static and dynamic analysis with features such as API, system calls and n-grams.

\begin{figure}
  \centering
   \includegraphics[scale=0.8]{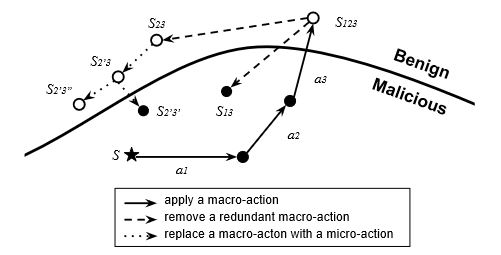}
   \caption{The workflow of action sequence minimizer \cite{song2020automatic}.}
    \label{fig:Songwork}
\end{figure}

Most of the attacks in adversarial domain are carried out in feature space and mapping features back to problem space is always not feasible. Attacks done in unrealistic scenarios are also not verified in terms of functionality. Song et al.~\cite{song2020automatic} proposed a open-source systematic framework for adversarial malware attack using code randomization and binary manipulation to evaluate against real world antivirus systems. Authors collected large categories of features from open source malware detectors namely: hash-based signatures, rule-based signatures and data distribution based features. To carry modifications in these features, generic action set is prepared as micro and macro actions which are given in Table \ref{tab:action_set_safe}. Micro actions are a relative concept, which only changes subset of actions inside macro-actions. Proposed workflow begins by selecting and applying macro-actions to original samples, till the original sample crosses decision boundary. Those macro-actions which do not have any roles are removed from action sequence to reach the most efficient evasive form. And finally, to get the detailed knowledge about the reason behind evasion, macro actions are broken into micro-action. From the modular point of view, binary rewriter module of framework generates different versions of original malware applying macro-actions randomly until the malware is able to evade static classifier. Action sequence minimization is carried out by traversing through actions and removing unnecessary actions. Cuckoo\footnote{https://cuckoosandbox.org/} sandbox is used to verify the functionality of malware. To provide reasoning for evasion, every actions are entangled into several micro-actions and each macro-actions is replaced with one micro-actions at a time as shown in Figure \ref{fig:Songwork}. This process helps in evaluation of essential features change responsible for classification decision. Evasion rate of adversarial from framework is found to be highest in EMBER classifier while lowest in ClamAV\footnote{https://www.clamav.net/}. Signature based antivirus were evaded by only one byte of perturbation where machine learning based antivirus required more perturbations for evasion.  This research directs future exploration towards generation of adversarial which can evade both static as well as dynamic detectors and also recommends antivirus systems to provide offline dynamic detection.

\begin{table}[]
\centering
\caption{{An action set for safe randomization method.}}
\begin{tabular}{|
>{\columncolor[HTML]{FFFFFF}}c |
>{\columncolor[HTML]{FFFFFF}}l |}
\hline
\cellcolor[HTML]{FFFFFF}                                              & Overlay Append             \\ \cline{2-2} 
\cellcolor[HTML]{FFFFFF}                                              & Section Append             \\ \cline{2-2} 
\cellcolor[HTML]{FFFFFF}                                              & Section Add                \\ \cline{2-2} 
\cellcolor[HTML]{FFFFFF}                                              & Section Rename             \\ \cline{2-2} 
\cellcolor[HTML]{FFFFFF}                                              & Remove Certificate         \\ \cline{2-2} 
\cellcolor[HTML]{FFFFFF}                                              & Remove Debug               \\ \cline{2-2} 
\cellcolor[HTML]{FFFFFF}                                              & Break Checksum             \\ \cline{2-2} 
\multirow{-8}{*}{\cellcolor[HTML]{FFFFFF}Macro}                       & Code Randomization         \\ \hline
\multicolumn{1}{|l|}{\cellcolor[HTML]{FFFFFF}}                        & Overlay Append 1 Byte      \\ \cline{2-2} 
\multicolumn{1}{|l|}{\cellcolor[HTML]{FFFFFF}}                        & Section Append 1 Byte      \\ \cline{2-2} 
\multicolumn{1}{|l|}{\cellcolor[HTML]{FFFFFF}}                        & Section Add 1 Byte         \\ \cline{2-2} 
\multicolumn{1}{|l|}{\cellcolor[HTML]{FFFFFF}}                        & Section Rename 1 Byte      \\ \cline{2-2} 
\multicolumn{1}{|l|}{\multirow{-5}{*}{\cellcolor[HTML]{FFFFFF}Micro}} & Code Section Append 1 Byte\\ \hline
\end{tabular}
\label{tab:action_set_safe}
\end{table}

\subsubsection{Reinforcement Learning Based Attack}

\begin{table*}[]
    \centering
    \setlength{\tabcolsep}{0.8\tabcolsep}
    \def\arraystretch{1.5}
    \caption{{A reinforcement learning adversarial attacks.}}
    \rowcolors{2}{gray!25}{white}
    \begin{tabular}{|p{2cm}|p{1.5cm}|p{2.5cm}|p{1.5cm}|p{4.5cm}|p{2cm}|p{1cm}|}
    \rowcolor{gray!25}
    \hline
         \textbf{Paper/Year} & \textbf{Target} & \textbf{Features} & \textbf{Action Space} & \textbf{Approach} & \textbf{Reward} & \textbf{SR}  \\
    \hline
     Anderson et al. 2018 ~\cite{Reinforce_Ander_sec} & Gradient Boosted Decision Trees (GBDT) & 
     \begin{minipage}[t]{\linewidth}
     \begin{itemize}[leftmargin=*]
         \item 2350-Dimensional feature vector
         \item Hashing trick to collapse into a vector of fixed size
     \end{itemize}
     \vspace{1mm}
     \end{minipage}
     & 
     10 stochastic actions for simplicity &
     \begin{minipage}[t]{\linewidth}
     \begin{itemize}[leftmargin=*]
         \item ACER with DQN learns both a policy model and a Q-function
         \item Boltzman exploration and exploitation where mutation are proportional to expected Q-value
         \item Mutations till evasion or 10 rounds
     \end{itemize}
     \vspace{1mm}
     \end{minipage}&
     Positive:10, Negative:0 & (12-24)\% \\
     \hline
     Fang et al. 2019 ~\cite{evading_anti_fang} &
     GBDT & 
     \begin{minipage}[t]{\linewidth}
     \begin{itemize}[leftmargin=*]
         \item Instability reduction using lower dimensional features
         \item Feature vector of 513-D
     \end{itemize}
     \vspace{1mm}
     \end{minipage}&
     4 stochastic actions, choosen after assessing malware&
     \begin{minipage}[t]{\linewidth}
     \begin{itemize}[leftmargin=*]
         \item DQN with prioritized version of experience replay
         \item Virtual address correction after modification
         \item Integrity verification using Cuckoo Sandbox
     \end{itemize}
     \vspace{1mm}
     \end{minipage}&
     TURN and discount factor based function&
     75\%\\
     \hline
     Chen et al. 2020 ~\cite{Chen_2020} &
     GBDT & 
     \begin{minipage}[t]{\linewidth}
     \begin{itemize}[leftmargin=*]
         \item Features similar to Anderson et al.'s work ~\cite{Reinforce_Ander_sec} 
     \end{itemize}
     \vspace{1mm}
     \end{minipage}&  10 deterministic actions &
     \begin{minipage}[t]{\linewidth}
     \begin{itemize}[leftmargin=*]
         \item DQN and A2C based approach called as gym-malware-mini
         \item Modification on original work of gym-malware
     \end{itemize}
     \vspace{1mm}
     \end{minipage}&
     Positive:10, Negative:-1 & 83\%\\
     \hline
     Fang et al. 2020 ~\cite{Fang2020DeepDetectNetVR} &
     Neural network based DeepDetectNet with AUC score upto 0.989 &
     \begin{minipage}[t]{\linewidth}
     \begin{itemize}[leftmargin=*]
         \item Import function feature
         \item General information feature
         \item Byte entropy features
         \item 2478-D feature vector
     \end{itemize}
     \vspace{1mm}
     \end{minipage} &
     200 deterministic actions &
     \begin{minipage}[t]{\linewidth}
     \begin{itemize}[leftmargin=*]
         \item Novel static feature extraction
         \item RLAttackNet using DQN and optimized using double and dueling DQN
         \item Different Q-network for choosing best action and Q-value
     \end{itemize}
     \vspace{1mm}
     \end{minipage}&
         r= k*MAXTURN / TURN & 19.13\% \\
    \hline
    \end{tabular}\\
    \vspace{1mm}
    \footnotesize\textit{\textbf{Target}: Target defense for adversarial attack, \textbf{Features}: Properties of features considered for processing, \textbf{Action Space}: Nature of actions in action space \textbf{Approach}: Key procedures to carry out adversarial attack, \textbf{Reward}: Reward values used for learning, \textbf{SR}: Success Rate of evasion}
    \label{tab:Reinforcement}
\end{table*}

\begin{figure}
  \centering
   \includegraphics[scale=0.6]{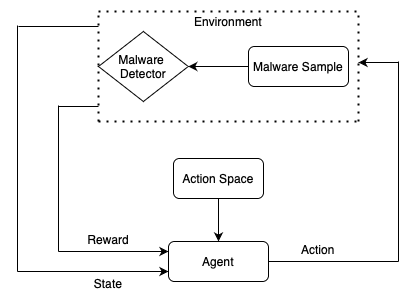}
   \caption{The Markov decision process formulation of the malware evasion reinforcement learning problem.}
    \label{fig:reinforcement_learn}
\end{figure}

To counter the need of differentiable model for gradient based approaches, reinforcement learning agent has been proposed to generate adversarial against malware detection. RL agent is provided with a set of operations to modify PE files while also preserving the functionality of malware. Goal of RL agent will be to perform sequence of operations on malware to evade detection. Reinforcement learning enables complete blackbox attacks to detector, creating real world attack scenario where attacker is completely unknown about detector. This reinforcement learning process is built around Markov decision process as shown in Figure \ref{fig:reinforcement_learn}. Table \ref{tab:Reinforcement} provides comparison of all RL approaches on adversarial evasion attack for Windows malware. Anderson et al. ~\cite{Anderson2017EvadingML} proposed a whitepaper on evading malware detection by modifying Windows PE bytes for the first time. Anderson et al. \cite{Reinforce_Ander_sec} extended results of work done in \cite{Anderson2017EvadingML} to perform generic black box attacks on static PE malware detection without assuming any knowledge of detector model's structure and features, retrieving only malicious/benign label. Actor Critic Model with Experience Replay (ACER) is used to learn both policy model \(\pi\) and a Q-function to estimate the state-action value. Reward of 0 is provided for malware samples which are detected by anti-malware engine and 10 for ones that can evade detection. Reward and state after each action are provided to an agent as a feedback. Feature vector is prepared such that it summarizes the state of the environment.  2350- dimensional feature vector is extracted from Windows PE malware consisting of features as:
\begin{itemize}
    \item Metadata of PE header
    \item Section metadata: section name, size and characteristics
    \item Metadata of Import and Export table 
    \item Counts of human readable strings
    \item Byte histogram
    \item 2D byte-entropy histogram
\end{itemize}

Countably infinite features are collapsed into a vector of fixed size using hashing trick. The obtained feature vector provides the complete view of malware files. 

Modifications to PE malware byte should not break file format and it's functionality. So, functionality preserving actions used in action space includes following:
\begin{itemize}
    \item New functions are added to unused import address table
    \item Existing section names can be manipulated
    \item New sections can be created
    \item Bytes can be appended to a space at the end of sections
    \item A new entry point can be defined which can direct to an original entry point
    \item A signer information can be removed
    \item Debug info can be manipulated
    \item Packing and unpacking operations can be carried out on file
    \item Header checksum can be modified
    \item Bytes can be appended in overlay
\end{itemize}
Functions in action space are stochastic in nature for simplicity. Appending bytes and compression level used by packer are chosen at random by the user to reduce the exponentially large number of mutations. During the construction of framework, authors considered feature representation used by agent to represent malware state, significantly overlaps with features of model under attack. Another important condition for success of reinforcement learning is that the agent's actions need to be fully observable by state representation. Agent performs mutation using a Boltzman exploration/exploitation strategy where mutation are proportional to their expected value. The game of exploration and exploitation continues with mutation in each round till evasion or permitted mutation. Anderson et al. showed cross evasion attack by training models with different data sizes. Authors also successfully demonstrated model hardening by adversarial training using the evading samples during training process. Regardless of few shortcomings, \cite{Anderson2017EvadingML} is able to direct the research towards adversarial world by just modifying binary bytes of Windows PE malware files in a blackbox environment. 

To reduce the instability and increase the convergence speed of Gym-Malware~\cite{Reinforce_Ander_sec}, Fang et al.~\cite{evading_anti_fang} proposed a Deep Q-network to Evade Antimalware engines (DQEAF) framework to evade anti-malware engines. DQEAF is able to reduce instability caused by higher dimensions, taking binary stream features of only 513 dimensions. It also takes only 4 functionality preserving actions in its action space to increase convergence and reports higher evasion rate. Action space is prepared only after assessing the malware to make actions more specific to the case. Actions proposed for deep Q-network training are as follows:
\begin{itemize}
    \item ARBE: Append random bytes at the end
    \item ARI: Append random library with random function to import address table
    \item ARS: Append a randomly named section to section table of PE data
    \item RS: Remove Signature
\end{itemize}

An agent feeds the features and rewards to two different neural networks to determine action value \(Q\) and target action value, which undergoes global optimization during training process. At the end of training process, agent gives the optimal parameter values of neural networks. The deep Q-network in DQEAF is extended version of convolutional neural network, performed by adding new features with action value and target action value. Rewards are provided based on number of training 'TURN' required to evade malware detection along with discount factor to consider future rewards, given in Equation \ref{eq:Fangreward}.
\begin{equation} \label{eq:Fangreward}
    r_t=20^{-(TURN-1)/MAXTURN}*100
\end{equation}
 where 'MAXTURN' is defined to claim failure if the 'MAXTURN' steps of modifications will still keep reward to 0. DQEAF also uses experience replay which allows reinforcement learning to remember and reuse experience from the past. Prioritized version of experience replay used, considers priority of transitions and important transitions are replayed more frequently. After number of transitions, action that leads to a maximum reward in state $s$ can be reached which is approximated in Equation \ref{eq:DQNFANG}. 
 \begin{equation} \label{eq:DQNFANG}
     a_t=arg\; {max}_a Q(s_t,a;\theta)
 \end{equation}
 where $s_t$ is the state at time $t$, $a$ is action and $\theta$ is some weight to present the correctness of action selection. After maximization, best DQEAF agent is chosen which will carry out optimal actions to perform modifications on a malware file. Workflow of adversarial generation begins by reading original PE malware, followed by modifications using DQEAF and finally correcting the virtual address for sample with integrity ensurance using Cuckoo Sandbox. DQEAF was able to alleviate evasion rate to 70\% in same dataset as used by Gym-Malware.

 Chen et al.~\cite{Chen_2020} proposed a reinforcement learning model based on Gym-Malware~\cite{Anderson2017EvadingML} using Deep Q-Network (DQN) and Advantage Actor Critic (A2C) deep reinforcement algorithm and named the environment as \textit{Gym-malware-mini}. Even though authors claimed to have increased the evasion rate by 18\% than that of Gym-Malware, it could be due to data leakage. Gym-malware-mini is trained and tested using the same data as that of gym-malware. Eleven actions in action space of Gym-Malware scales to uncountable number due to randomness in each action. Gym-malware-mini converts those random actions to 10 deterministic actions making actions space very small. 6 random actions that has been  changed to deterministic are \textit{overlay append}, \textit{imports append}, \textit{section rename}, \textit{section add}, \textit{section append} and \textit{upx pack}. In addition, four actions are directly brought from gym malware including remove signature, remove debug, upx unpack and break optional header checksum. To balance the exploit and exploration, best action are chosen using epsilon-greedy method during the network training. Training workflow begins by initializing network parameters and hyper parameters. Once Gym-malware-mini gets the state of environment in form of feature vector, DQN network calculates action value followed by epsilon greedy algorithm choosing action to execute. The rewards are returned depending on the result of detector and state transitions are stored into replay buffer.  Smaller action space aids in learning policy better. Gym-malware-mini also uses negative reward for punishment which helps to make agent learn faster. 
 
Fang et al.~\cite{Fang2020DeepDetectNetVR} tried to address shortcomings of previous work by proposing own malware detection and adversarial generation method using DRL. MalConv~\cite{raff2017malware}, which has been a standard detector network for Windows PE malware by feeding whole binary bytes has been exploited by various researches. Its vulnerability to gradient based attacks for adversarial motivated authors to build their own malware detection system DeepDetectNet with AUC upto 0.989. For feature extraction, DeepDetecNet uses traditional approach which are based on feature engineering. Static feature extraction mainly includes three categories of features, which are:
\begin{itemize}
    \item Import Functions feature representing common import function
    \item General information feature containing profile or overall attributes of a PE file
    \item Bytes entropy feature is the feature extraction method
\end{itemize}

A total of 2478 features are extracted from \textit{ PE files as \textit{Import Function} features, \textit{General Information} feature and Bytes Entropy} feature. Same action in action space obtains different rewards based on the state of environment, which makes learning by agent confusing. Previous success on adversarial generation using reinforcement learning are found to be UPX packed which are not the actual modifications on PE files. In order to solve this problem, all random modification operations are expanded to 218 specific operations. To combat the problem of overestimation of reward as proposed by DeepMind \cite{DeepMind}, double and dueling DQN algorithm, RLAttackNet is proposed. The architecture is built using DQN and optimized with double DQN method and Dueling DQN method. One of the Q-network is responsible for choosing the most optimal action whereas another one for evaluating the Q-value. Dueling DQN divides Q-values into state values evaluating current environment state and action advantages which evaluates the goodness of actions. The reward is provided in each $turn$ based on constants $k$ and $MAXTURN$ which denotes maximum number of time a file can be modified. Authors could achieve the evasion rate of 19.13\% and retrained DeepDetectNet using adversarial was able to reduce evasion rate to 3.1\%. 

\subsubsection{GAN Based Attacks}
\begin{table*}[t]
    \centering
    \setlength{\tabcolsep}{0.7\tabcolsep}
    \def\arraystretch{1.5}
    \rowcolors{2}{gray!25}{white}
    \caption{{An adversarial attacks based on GAN.}}
    \begin{tabular}{|p{2cm}|p{2cm}|p{2cm}|p{1.7cm}|p{6cm}|p{2cm}|}
    \rowcolor{gray!25}
    \hline
     \textbf{Paper/Year} & \textbf{Key-Motivation} & \textbf{Target Model} & \textbf{Byte/Feature}& \textbf{Approach} & \textbf{Feature Count} \\
     \hline
     Hu. et al. 2017~\cite{hu2017generating} & 
     Need of black-box flexible adversarial attack  &
     ML based (RF, LR, DT, SVM, MLP, VOTE) detectors &
     Feature & 
     \begin{minipage}[t]{\linewidth}
         \begin{itemize}[leftmargin=*]
             \item Feed Forward Neural Networks are used for both generator and substitute detector
             \item Iterative approach, modifying one feature every iteration
         \end{itemize}
    \vspace{1mm}
    \end{minipage}
         & 128 APIs\\
    \hline
    Kawai et al. 2019~\cite{ImprovedMALGAN} &
    Using single malware for realistic attacks &
    ML based (RF, LR, DT, SVM, MLP, VOTE) detectors &
    Feature & 
    \begin{minipage}[t]{\linewidth}
    \begin{itemize}[leftmargin=*]
        \item {Deep Convolutional GAN used for Substitutor({$S$}) and Generator($G$})
        \item {API list from multiple cleanware and single malware}
    \end{itemize} 
    \vspace{1mm}
    \end{minipage}
    & All APIs \\
    \hline
    Castro et al. 2019~\cite{Castro2019PosterTG} &
    Automatic byte level modifications &
    GBDT Model & Byte Level & 
    \begin{minipage}[t]{\linewidth}
    \begin{itemize}[leftmargin=*]
        \item Richer Feature representation
        \item Generates random perturbation sequence with nine different options at each injection
    \end{itemize}
    \vspace{1mm}
    \end{minipage}&
    2350 Features \\
    \hline
    Yuan et al. 2020~\cite{Yuan2020BlackBoxAA} & 
    End-to-End blackbox attacks at byte levels &
    MalConv \cite{raff2017malware} &  Byte Level &
    \begin{minipage}[t]{\linewidth}
    \begin{itemize}[leftmargin=*]
        \item Dynamic thresholding to maintain the effectiveness of payload
        \item Balance in attention of generator to payloads and adversarial samples are brought using automatic weight tuning
    \end{itemize}
    \vspace{1mm}
    \end{minipage}& Raw Bytes\\
    \hline
     
    \end{tabular}\\
     \vspace{1mm}
     \footnotesize\textit{\textbf{Key Motivation}: The major motive behind the published work, \textbf{Target Model}: Target defense for adversarial attack, \textbf{Byte/Feature}: Byte or Feature selected to modify, \textbf{Approach}: Key procedures to carry out adversarial attack, \textbf{Feature Count}: Number of features considered for processing}
    
    \label{tab:GAN_ATTACK}
\end{table*}

\begin{figure}
    \centering
    \includegraphics[scale=0.65]{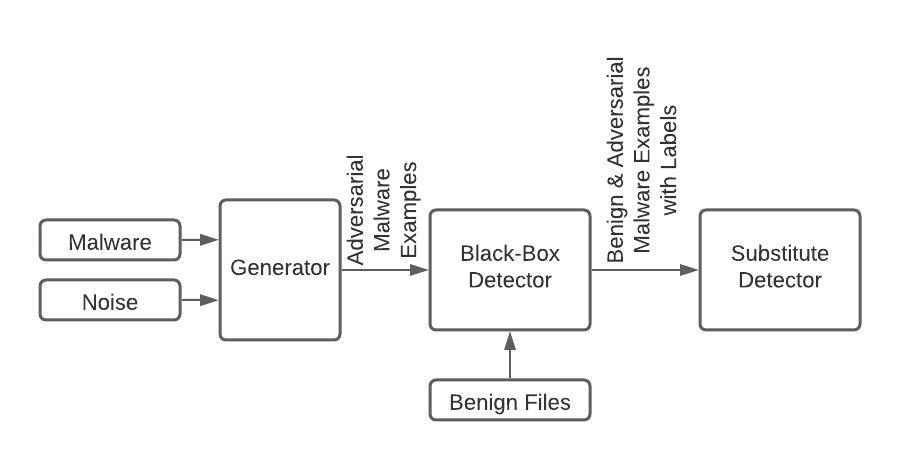}
    \caption{An adversarial malware generation architecture using GAN.}
    \label{fig:MALGAN}
\end{figure}

Most of existing adversarial generation deals with use of gradient information and hand-crafted rules. However, due to constrained representation ability of existing gradient based models, obtaining high true positive rate (TPR) has been the challenge. Generative Adversarial Networks (GAN) originally proposed by Goodfellow et al.~\cite{goodfellow2014generative} has inspired blackbox attack to malware detectors with very high TPR. The common GAN architecture that are used to perform adversarial malware attacks is shown in Figure \ref{fig:MALGAN}. GAN uses discriminative model to distinguish between generated samples and real samples, and a generative model to fool the discriminative model between generated samples and real samples. Table \ref{tab:GAN_ATTACK} summarizes adversarial attacks carried against Windows anti-malware engines.
Hu et al.~\cite{hu2017generating} proposed an adversarial generation technique, MalGAN which is able to bypass black-box machine learning models. Like other GAN architecture, MalGAN is also made up of generator and a substitute detector with both being feed forward neural network. Binary features obtained by presence or absence of API are used as a input to model and number of input feature equals dimension of input. Generator is used to transform malware to its adversarial version by taking the probability distribution of adversarial far away from the detector. Concatenating malware feature vector with noise vector allows generator to produce numerous adversarial examples from a single malware feature vector.  Substitute detector is used to fit the detector model and provide gradient information to train the generator.

MalGAN is trained with 160 system level Application Programming Interfaces (APIs) against different machine learning detector. Experiments were conducted separately with MalGAN and detector model sharing and splitting the training dataset.
True positive rate of zero is obtained on most of machine learning models which shows the ability of substitute detector to fit with very high accuracy. MalGAN's ability to perform complex transformation has resulted in zero TPR for both training and testing set. 

Considering the use of multiple malware to train MalGAN affecting the performance of avoidance, Kawai et al.~\cite{ImprovedMALGAN} proposed improved MalGAN with the use of only one malware for training.
MalGAN imports malware detector for training and predicting which is not convenient for attackers. This improved MalGAN uses python's sub process library to import only detection results to MalGAN. Authors also utilized all APIs used for malware to feature quantities instead of 128 APIs used by original MalGAN. Different dataset is used to prepare API list for training detector and MalGAN. API list are extracted by combining from multiple cleanware and single malware in order to avoid the malware detection process be driven by addition of cleanware's features to malware file. Generator and Substitute model is also changed to Deep Convolutional GAN (DCGAN) originally proposed by Randford et al.~\cite{radford2016unsupervised} in image generation. For malware domain, activation is changed to Parametric Rectified Linear Unit (PReLU) function due to its ability to self-learn negative part of the Leaky ReLU function. Malware detectors are avoided by improved MalGAN through the addition of cleanware features to an initial malware file. Increase in feature quantities has improved the performance of both detectors and MalGAN.

Few assumptions made in designing MalGAN are less realistic and limited to bypass real malware classifier. One of such assumptions is that attackers are assumed to have full access to feature space in detector model. In addition, API features are considered too extended way to represent malware. To overcome these limitations, Castro et al.~\cite{Castro2019PosterTG} published a poster to use GAN approach for generating adversarial examples by injecting byte-level perturbations. Proposed model works with real PE files instead of API feature representations. Automatic byte-level real perturbation is combined with feature representation to produce adversarial examples. Generator takes vector of 2350 features providing extensive overview of each malware and generates random perturbation sequence having 9 different possibilities at each injection. Use of richer feature representation and ability to return valid PE binaries makes system able to bypass not only GBDT detector but also cross-evade different classifiers.

Using API sequences or feature representation demands a lot of manual task to get the training data. Current state of art researches are directed towards end-to-end detection of malware without any feature engineering effort. Yuan et al.~\cite{Yuan2020BlackBoxAA} proposed a GAPGAN framework which performs end-to-end black-box attacks against malware detectors using byte level features. Initial discrete malware binary features are mapped to continuous space before feeding to generator network of GAPGAN which generates adversarial perturbations to be appended at the end of original malware binaries. Dynamic thresholding preserves generated subtle perturbations while mapping back to discrete space from continuous space. The balance on the attention of the generator across payloads and adversarial samples is maintained using an automatic weight tuning strategy. Authors are able to achieve 100\% evasion rate against MalConv~\cite{raff2017malware} with the addition of only 2.5\% perturbations data. Concurrently trained generator and discriminator keeps improving each other and transferability property of adversarial attack enables it to bypass detector. Variable input and output size gives great flexibility to GAPGAN model in contrast to prior research works.

\subsubsection{Recurrent Neural Network Based Attack}

Recent works have focused on use of Recurrent Neural Network (RNN) for malware detection and classification \cite{kimmel2021recurrent, RNN_DETECT_CLASSIF, RNN_DET_KOLS, RNN_DET_PROC}. Sequential malware API is used by RNN to predict whether the program is malware or benign. Papernot et al.~\cite{papernot2016crafting} introduced adversarial sequence for RNN processing sequential data. The authors demonstrated the transferability property of adversarial examples generated from feed-forward neural networks against recurrent neural networks. Table \ref{tab:RNN,XML} summarizes comparison among RNN, explainable ML and malware visualization based adversarial attacks. Hu et al.~\cite{hu2017blackbox} proposed RNN based adversarial attack for RNN malware detector. The approximation of victim RNN model is done by training substitute RNN and generative RNN outputs sequential adversarial examples. Some irrelevant API sequence is generated and inserted in vulnerabilities of original sequence. API sequences, represented as a one-hot vector, are the input for generator network which generates adversarial API sequence. The generative part of RNN generates small API sequence pieces after each API which gets inserted after the API. Small sequence generation is done by sequence decoder where the hidden states are initialized with zero. A benign sequence and the Gumbel-Softmax~\cite{jang2017categorical} output is used to train the substitute network to fit the victim RNN based detector. Due to the use of bidirectional RNN equipped with an attention mechanism, substitute RNN can grasp the complex sequential pattern. Both forward and backward connections are present inside the bidirectional connection to represent the individual unidirectional connection. The attention mechanism helps by spreading the focus on different parts of the sequence. The output label provided by victim RNN for training data is used as training labels for substitute RNN. Gumbel-Softmax~\cite{jang2017categorical} exchanges gradient information across the generative RNN and the substitute RNN while also smoothing API symbols. Generative RNN function tries to minimize the probability of malicious prediction \(p_s\) on \(S_{Gumbel}\) which is given in Equation \ref{eq: RNN_Substitute}.
\begin{equation} \label{eq: RNN_Substitute}
    L_G=log(p_s)-\gamma \mathbb{E}_{t=1\sim T,\tau =1\sim L}\pi ^M_{t\tau}
\end{equation}
where \(\gamma\) is the regularization coefficient used for restricting the number of inserted APIs by maximing null API's expectation probability, M is the index of the null API, $\mathbb{E}$ denotes the expected value, $p_s$ is predicted malicious probability, $\pi$ is any parameter and $\tau$ is learned parameter. By separately tuning hyper-parameters of generative and substitute RNN, proposed architecture was successful in evading LSTM and BiLSTM based RNN malware detectors.

\subsubsection{Explainable Machine Learning based attack}
\begin{table*}[t]
    \centering
    \setlength{\tabcolsep}{0.8\tabcolsep}
    \def\arraystretch{1.5}
    \caption{{RNN, explainable ML and visualization based adversarial.}}
    \rowcolors{2}{gray!25}{white}
    \begin{tabular}{|p{2cm}|p{3cm}|p{2.5cm}|p{2cm}|p{6.5cm}|}
    \rowcolor{gray!25}
     \hline
     \textbf{Paper/Year}  &\textbf{Key-Motivation}&\textbf{Target Model} & \textbf{Algorithms Used}& \textbf{Approach}\\
     \hline
     Hu et al. 2017~\cite{hu2017blackbox} & 
     Attack against RNN processing sequential data &
     LSTM and BiLSTM based detectors &
     Bidirectional RNN with attention mechanism &
     \begin{minipage}[t]{\linewidth}
     \begin{itemize}[leftmargin=*]
         \item Substitute RNN approximates victim RNN 
         \item Generative RNN gives sequential adversarial example
         \item Irrelevant API sequence generated and inserted in vulnerabilities of original sequence
     \end{itemize}
     \vspace{1mm}
     \end{minipage}\\
     \hline
     Rosenberg et al. 2020~\cite{rosenberg2020generating} &
     Use of explainable machine learning for adversarial generation & 
     GBDT Classifer & Integrated Gradient, LRP, DeepLIFT, SHAP &
     \begin{minipage}[t]{\linewidth}
     \begin{itemize}[leftmargin=*]
         \item Unearthing most impactful features using explainability algorithm
         \item Manual selection of easily modifiable features
         \item Feature by feature modification without harming functionality and interdependent features
     \end{itemize} 
     \vspace{1mm}
     \end{minipage}\\
     \hline
     Liu et al. 2019~\cite{liu2019atmpa} & Adversarial malware against visualization based detection & CNN, SVM and RF based malware detectors & ATMPA framework using GoogLeNet, FGSM and C\&W & 
     \begin{minipage}[t]{\linewidth}
     \begin{itemize}[leftmargin=*]
         \item Data transformation to convert code segments into grayscale images
         \item Pre-training module to find function of malware detectors
         \item Optimized FGSM and C\&W attack is used to generate actual AE
     \end{itemize}
     \vspace{1mm}
     \end{minipage}\\
     \hline
     Khormali et al. 2019~\cite{khormali2019copycat} & Targetted and Untargetted misclassification on windows and IoT malware dataset & 
     Convolutional Neural Network & FGSM, C\&W, DeepFool, MIM and PGD &
     \begin{minipage}[t]{\linewidth}
     \begin{itemize}[leftmargin=*]
         \item Adversarial generation using different algorithms
         \item Conversion of adversarial dimension, same as of original image
         \item Appending pixels at the end or injecting
     \end{itemize}
     \vspace{1mm}
     \end{minipage}\\
     \hline
     Benkraouda et al. 2021 ~\cite{benkraouda2021attacks} & Attack against visualization based detection with ability to evade pre-processing filtering without losing functionality & Convolutional Neural Network & 
     Modified version of CW attack\cite{carlini2017evaluating}, Euclidean distance &
     \begin{minipage}[t]{\linewidth}
     \begin{itemize}[leftmargin=*]
         \item Mask generator to flag the locations for perturbation
         \item Modified version of CW attack to generate optimal perturbation 
         \item NOP generator to replace the perturbation from CW attack by semantic NOPs
         \item AE optimizer to choose optimal viable NOPs
     \end{itemize}
     \vspace{1mm}
     \end{minipage}\\
     \hline
    \end{tabular}\\
    \vspace{1mm}
     \footnotesize\textit{\textbf{Key-Motivation}: The major motive behind the published work, \textbf{Target Model}: Target defense for adversarial attack, \textbf{Algorithms Used}: Algorithm used for crafting adversarial example, \textbf{Approach}: Key procedures to carry out adversarial attack}
    \label{tab:RNN,XML}
\end{table*}
Malware detection is one of the most relevant domain for adversarial crafting as attackers are continuously attempting to evade detection networks. However, one of the biggest challenge of machine learning is the lack of explainability or reasoning behind such intelligent decisions. Recent researches have been able to bypass malware detectors using concept of explainable machine learning. Explainability approach involves finding the significance of each features and then conducting feature specific modification based on their importance.

Rosenberg et al.~\cite{rosenberg2020generating} proposed explainable ML approach to generate adversaries against multi-feature type malware classifiers. Authors not only performed feature addition like existing approaches but also modified feature-by-feature. Adversarial attackers first evaluate for the most effective list of features, and the features that are easy to modify are selected. Transferability of explainability allows the proposed attack achieve a very high impact on target classifier even in black-box attack. This approach assumes that the malware classifier and the substitute model possess similar feature importance, leading to modification in feature to impact the target malware classifier. Four different explainability algorithms on white-box \cite{sundararajan2017axiomatic,Bach2015OnPE,shrikumar2019learning} and black-box \cite{lundberg2017unified} are evaluated to make comparisons between substitute model and victim model. Brief introduction of each explainability algorithms is:
\begin{itemize}
    \item Integrated gradient~\cite{sundararajan2017axiomatic} satisfies the completeness property by computing the average of the gradient on varying input while moving in a linear path.
    \item  Layer-wise Relevance Propagation (LRP) \cite{Bach2015OnPE} works on backward pass by starting from the output layer. The relevance of each target neuron is given corresponding to the output of the neuron.
    \item DeepLIFT~\cite{shrikumar2019learning} works in similar fashion to that of LRP but in backward order. 
    \item SHAP (SHapley Additive exPlanation)~\cite{lundberg2017unified} 
    attributes the classifier output to the totality of individual feature effects. SHAP can work without any knowledge about the architecture of the network to explain. 
\end{itemize}
The proposed end-to-end PE adversarial examples performs feature modification without harming the malware functionality as well as interdependent features. Using naive and engineered features of EMBER dataset, explainable ML approach is successful in bypassing GBDT classifier. Rosenberg et al.'s work presents explainability as a duel edged sword that can be used by adversaries to make more explainable models as well as to carry out more robust adversarial attacks.

\subsubsection{Malware Visualization based attack}
\begin{figure}
    \centering
    \includegraphics[scale=0.7]{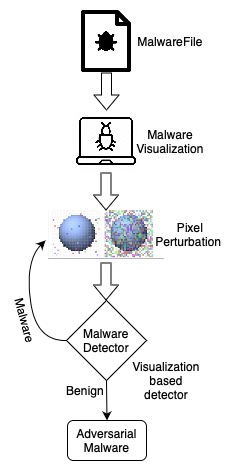}
    \caption{An adversarial generation against malware visualization detection.}
    \label{fig:malware_visualize}
\end{figure}
Machine learning-based visualization detection has been popular due to its ability to prevent zero-day attacks and make detection without extracting pre-selected features~\cite{Han2014MalwareAU,6597204,10.1145/2016904.2016908}. These approaches convert binary code into image data and visualizes the features of sample, improving the detection speed for malicious program. Visualization based techniques are similar to adversarial generation in image domain where pixel perturbations are introduced as shown in Figure \ref{fig:malware_visualize}. Liu et al.~\cite{liu2019atmpa} introduced Adversarial Texture Malware Perturbation Attack (ATMPA) against visualization based malware detection using rectifier in neural-network hidden layers. The framework allows an attacker to probe with the malware image while visualizing and also hiding them from malware detectors. The ATMPA framework is made up of three functional modules: data transformation, adversarial pre-training and malware detection. Code segments are converted into grayscale images during data transformation module. In the adversarial pre-training module, an attacker uses a machine learning approach to train an adversarial example generation model which produces a noise signal \(\delta\). The training process is divided into two modules: one to capture the normal malware behaviors and other to generate AEs. ATMPA framework uses GoogLeNet~\cite{szegedy2014going} for generating adversarial examples in the pre-training process where malware samples are represented using binary indicator vector without any structural properties or inter-dependencies. Adversarial example training uses a feedforward neural network having a rectifier as an activation function.

For generation of AEs, optimized FGSM and C\&W attack is used. Using FGSM approach, perturbation with distortion \(\epsilon\) = 0.35 was able to convert malware with 88.5\% confidence into benign files with 99.8\% confidence. ATMPA method also used \(L_p\)-based C\&W attack to generate adversarial, including $l_0, l_2$ and  $l_{\infty}$ attack. Kaggle Microsoft Malware Classification Challenge\footnote{https://www.kaggle.com/c/malware-classification/overview} dataset is used against malware detection model built on commonly used CNN, SVM and RF-based detectors. Inception V3 architecture followed by FGSM and C\&W attack methods generated pseudo-benign samples, successfully deceiving the detector. Authors also carried out transferability attack comparision on CNN, SVM and wavelet-combined algorithms with transfer rate as high as 88.7\%. 

COPYCAT approach proposed by Khormali et al.~\cite{khormali2019copycat} produced both targeted and untargetted misclassification on Windows and IoT malware dataset. Author used two approaches: AE padding and sample injection to produce adversarial malware for visualization based detector. For padding method, COPYCAT generated adversarial \(x^{'}\) using five different attack methods namely: FGSM~\cite{goodfellow2015explaining}, C\&W~\cite{carlini2017evaluating}, DeepFool~\cite{moosavidezfooli2016deepfool}, Momentum Iterative Method (MIM)~\cite{dong2018boosting} and Projection Gradient Descent (PGD)~\cite{madry2019deep}. The generated adversarial needs to be converted to the dimensions same as that of original image before appending at the end of image. The binary samples from the targeted class are injected into an unreachable section of the target sample. The approach of this work is inspired by Cha et al.~\cite{10.1145/1866307.1866369},  where binaries of an application having different architecture are padded to alter the behavior based on the underlying system.

In order to provide adversarial attack that can evade visualization based detection system in presence of pre-processing filtering, Benkraouda et al.\cite{benkraouda2021attacks} proposed  a binary rewriting based attack on malware files. A mask generator creates the space in the instruction boundary to insert the perturbations. Once the perturbation mask is created, the modified version of CW attack~\cite{carlini2017evaluating} is used to generated adversarial example in image space. The modified version is in the sense that, the perturbation mask is imposed while carrying out attack to restrict the positions of perturbations. NOP generator will replace the perturbation introduced by CW attack with the corresponding binaries, that preserve the malware functionality. And finally AE optimizer will use Euclidean distance metric to choose semantic NOPs that are closed to sequences in the allowed perturbation space. The approach is highly successfully in generating adversarial examples but are constrained by perturbation generation time as time is found to increase drastically with increase in size of malware. 

 \subsection{Android Malware Adversarial}

 \begin{table*}[htbp]
    \centering
    \setlength{\tabcolsep}{0.8\tabcolsep}
    \def\arraystretch{1.5}
    \caption{{An Android adversarial attacks.}}
    \rowcolors{2}{gray!25}{white}
    \begin{tabular}{|p{2cm}|p{2cm}|p{6cm}|p{3cm}|p{3cm}|}
    \rowcolor{gray!25}
    \hline
    \textbf{Paper/Year} & \textbf{Target/Dataset} & \textbf{Approach} & \textbf{Modification} & \textbf{Limitations}\\
    \hline
    Grosse et al. 2017~\cite{grosse_adversarial17} & Feed forward neural network based detector / DREBIN &
    \begin{minipage}[t]{\linewidth}
    \begin{itemize}[leftmargin=*]
        \item Binary feature vector extraction using static evaluation
        \item Jacobian matrix of neural network for adversarial generation
        \item Direction for generated perturbation is given by gradient of the given function with respect to the input
        \item Selection of perturbation with maximal positive gradient towards target class
    \end{itemize}
    \vspace{1mm}
    \end{minipage}&
    \begin{minipage}[t]{\linewidth}
    \begin{itemize}[leftmargin=*]
        \item Feature addition to AndroidManifest.xml
        \item Changing features leading to only one line of code
    \end{itemize}
    \vspace{1mm}
    \end{minipage}&
    \begin{minipage}[t]{\linewidth}
    \begin{itemize}[leftmargin=*]
        \item Constrained on maximum feature space perturbation
        \item Feature modifications confined inside AndroidManifest.xml
    \end{itemize}
    \vspace{1mm}
    \end{minipage}\\
    \hline
    Yang et al. 2017~\cite{Feature_evol_conf_Yang} &
    KNN, DT, SVM, RF/ DREBIN, VirusShare, Genome &
    \begin{minipage}[t]{\linewidth}
    \begin{itemize}[leftmargin=*]
        \item Malware Recomposition Variation conducting semantic analysis
        \item Feature mutation analysis and phylogenetic analysis to perform automatic program transplantation
        \item Malware evolution attack focusing on mimicking and automating the evolution of malware
        \item Conufsion attack making features less differentiable to malware detection
    \end{itemize}
    \vspace{1mm}
    \end{minipage}&
    \begin{minipage}[t]{\linewidth}
    \begin{itemize}[leftmargin=*]
        \item Resource, temporal, locale and dependency features used
        \item Mutation following feature pattern of existing malware
    \end{itemize}
    \vspace{1mm}
    \end{minipage}&
    \begin{minipage}[t]{\linewidth}
    \begin{itemize}[leftmargin=*]
        \item Significant alteration of semantic leading to higher failure rate of app
    \end{itemize}
    \vspace{1mm}
    \end{minipage}\\
    \hline
    Rosenberg et al. 2018~\cite{rosenberg2018generic} & RNN variaant and Feed forward neural networks/ VirusTotal &
    \begin{minipage}[t]{\linewidth}
    \begin{itemize}[leftmargin=*]
        \item Mimicry attacks against surrogate model 
        \item Surrogate model by querying black-box detectors with synthetic inputs selected by Jacobian based heuristics in prioritizing directions
        \item Closest API call in direction indicated by Jacobian are selected
    \end{itemize}
    \vspace{1mm}
    \end{minipage}&
    \begin{minipage}[t]{\linewidth}
    \begin{itemize}[leftmargin=*]
        \item No-op attack by adding API call with valid parameters
        \item Functionality verification using sandbox after modification
    \end{itemize}
    \vspace{1mm}
    \end{minipage}&
    \begin{minipage}[t]{\linewidth}
    \begin{itemize}[leftmargin=*]
        \item Detectable Residual artifacts during app transformation
    \end{itemize}
    \vspace{1mm}
    \end{minipage}\\
    \hline
    Liu et al. 2019~\cite{Liu_2019} &
    Neural network, logistic regression, DT and RF based detectors/DREBIN &
    \begin{minipage}[t]{\linewidth}
    \begin{itemize}[leftmargin=*]
        \item Random forest to filter most significant features
        \item Disturbance randomly generated and disturbance size calculated using genetic algorithm
        \item Mutation using fitness function till fit and evading individual is produced
    \end{itemize}
    \vspace{1mm}
    \end{minipage}&
    \begin{minipage}[t]{\linewidth}
    \begin{itemize}[leftmargin=*]
        \item Restricted permission modification on AndroidManifest file
        \item Functionality changing modifications are deemed unfit
    \end{itemize}
    \vspace{1mm}
    \end{minipage}&
    \begin{minipage}[t]{\linewidth}
    \begin{itemize}[leftmargin=*]
        \item Increased constraint on perturbation
        \item Random perturbation affecting convergence
    \end{itemize}
    \vspace{1mm}
    \end{minipage}\\
    \hline
    Shahpasand et al. 2019~\cite{android_adversarial_Shahpasand} &
    SVM, Neural network, RF and LR / DREBIN &
    \begin{minipage}[t]{\linewidth}
    \begin{itemize}[leftmargin=*]
        \item GAN architecture with threshold on generated distortion
        \item Different loss function to generate benign like adversarial and to produce high mis-classification
    \end{itemize}
    \vspace{1mm}
    \end{minipage}&
    \begin{minipage}[t]{\linewidth}
    \begin{itemize}[leftmargin=*]
        \item Perturbation addition limited by threshold distortion amount
    \end{itemize}
    \vspace{1mm}
    \end{minipage}&
    \begin{minipage}[t]{\linewidth}
    \begin{itemize}[leftmargin=*]
        \item Highly unstable learning of GAN architecture
    \end{itemize}
    \vspace{1mm}
    \end{minipage}\\
    \hline
     Li et al. 2020~\cite{LI_ANDROID_GAN} &
     AdaBoost, CNN, SVM / Tencent Myapp, AndroZoo, VirusShare and Contagio &
     \begin{minipage}[t]{\linewidth}
     \begin{itemize}[leftmargin=*]
         \item Bi-objective GAN with two discriminator and one generator
         \item First discriminator to distinguish malware and benign sample
         \item Second discriminator to distinguish original and adversarial sample
     \end{itemize}
     \vspace{1mm}
     \end{minipage}&
     \begin{minipage}[t]{\linewidth}
     \begin{itemize}[leftmargin=*]
         \item Iterative perturbation addition till evasion
         \item Perturbation evading both malware and adversarial detection
     \end{itemize}
     \vspace{1mm}
     \end{minipage}&
     \begin{minipage}[t]{\linewidth}
     \begin{itemize}[leftmargin=*]
         \item Very limited feature vectors ( Permission, action and API calls) are considered
     \end{itemize}
     \vspace{1mm}
     \end{minipage}\\
     \hline
     Pierazzi et al. 2020~\cite{pierazzi2020intriguing} &
     Linear SVM, Sec-SVM/DREBIN & 
     \begin{minipage}[t]{\linewidth}
     \begin{itemize}[leftmargin=*]
        \item Formalization of adversarial evasion attacks in the problem feature space including transformations, semantics, robustness and plausibility
        \item Automated software transplantation to extract benign slices from donor
        \item Side effect features to find projections that maps perturbation to feasible problem-space regions
        \item Gradient based strategy based on greedy algorithm to choose perturbation
     \end{itemize}
     \vspace{1mm}
     \end{minipage}&
     \begin{minipage}[t]{\linewidth}
     \begin{itemize}[leftmargin=*]
         \item Perturbations appended at the end
         \item Restricted addition of permissions
         \item Cyclomatic Complexity to take heuristic approach maintaining existing homogeneity
     \end{itemize}
     \vspace{1mm}
     \end{minipage}&
     \begin{minipage}[t]{\linewidth}
     \begin{itemize}[leftmargin=*]
         \item Heuristic based approaches are time and resource consuming
     \end{itemize}
     \vspace{1mm}
     \end{minipage}\\
     \hline
     Bostani et al. 2021 ~\cite{bostani2021evadedroid}&
     DREBIN~\cite{Arp2014DREBINEA}, Sec-SVM~\cite{SECSVM_Demontis2019YesML}, MaMaDroid ~\cite{mariconti2016mamadroid} / AndroZoo~\cite{allix2016androzoo} &
     \begin{minipage}[t]{\linewidth}
     \begin{itemize}[leftmargin=*]
        \item  Automated Software Transplantation Technique to prepare action set which includes gadgets extracted from benign Android apps
        \item n-gram-based similarity method to identify benign APKs, closely similar to malware files
        \item Applying extracted gadgets from benign samples into malicious files
        \item Iterative and incremental manipulation 
     \end{itemize}
     \vspace{1mm}
     \end{minipage}&
     \begin{minipage}[t]{\linewidth}
     \begin{itemize}[leftmargin=*]
         \item Random Search(RS) for moving malware sample in problem space applying sequence of transformation in action set
         \item New contents injected inside an IF statement 
     \end{itemize}
     \vspace{1mm}
     \end{minipage}&
     \begin{minipage}[t]{\linewidth}
     \begin{itemize}[leftmargin=*]
         \item In Random Search (RS) algorithm, actions from action space are random
         \item Increase in adversarial size, increasing chances of adversarial detection
     \end{itemize}
     \vspace{1mm}
     \end{minipage}\\
     \hline
    
    \end{tabular}\\
    \vspace{1mm}
    \footnotesize\textit{\textbf{Target/Dataset}: Target defense for adversarial attack/Dataset used, \textbf{Approach}: Key procedures to carry out adversarial attack, \textbf{Modification}: Changes on file to craft the adversarial perturbation, \textbf{Limitations}: Shortcomings of proposed approach}
    \label{tab:Android}
\end{table*}

Android has over 2.8 billion active users and owns 75\% market share in mobile phone industry~\cite{business_of_apps_2021}. The wide usage of Android platform has attracted security threats in numerous forms and adversarial evasion attack is one of them. Table \ref{tab:Android} provides brief comparison among different adversarial attacks crafted against Android files.
Grosse et al.~\cite{Grosse2016AdversarialPA}~\cite{grosse_adversarial17} generated adversarial examples for state-of-art Android malware detection trained on DREBIN dataset~\cite{Arp2014DREBINEA} with more than 63\% accuracy. Authors migrated the method proposed by Papernot et al.~\cite{papernot2015limitations} to handle binary features of Android malware while preserving the malicious functionality. Binary features are derived by statically evaluating code based on system call and usage of specific hardware. These derived features are also known as binary indicator vectors. The final goal of adversarial attack is to find perturbation/noise \(\delta\) such that the prediction results of \(y^{'}\) of \(F(X+\delta)\) is different from the original result, i.e, \(y^{'}\neq y\). \(y^{'}\) gives target class while crafting adversarial. Authors adopted Jacobian matrix of neural network \(F\) for adversarial generation. To get adversarial, the gradient of function \(F\) with respect to $X$ is calculated to get the direction of perturbation such that output of classification function will change. Perturbation \(\delta\) with highest positive gradient in direction of target is selected and is kept small enough to prevent negative change due to intermediary alterations of gradient. Functionality is preserved in this approach by changing features resulting in addition of only single line of code. Research also confine the modifications to manifest features related to \texttt{AndroidManifest.xml} file contained within Android application. With permissions, intents and activities being the most frequently modified features, authors successfully evaded DREBIN classifier\cite{Arp2014DREBINEA} preserving the semantics of malware.

To overcome the white box attack issues, Rosenberg et al.~\cite{rosenberg2018generic} implemented GADGET framework to convert malware binary to an adversarial binary without access to malware source code. Proposed end-to-end black-box method is extended to bypass the multi-feature based malware classifiers relying on the transferability in RNN variants. For target RNN detector, malicious API call sequence is the adversarial example to be generated. Authors perform mimicry attacks where malicious code mimics the system calls of benign code to evade~\cite{mimicry_attack}. Adversaries train surrogate model having same decision boundaries as that of detector and then execute white-box attack on surrogate model. To build the surrogate model, black-box detector is queried with synthetic input values chosen by a Jacobian based heuristics in the prioritizing directions where model output varies. API calls which are nearest to the direction given by Jacobian are inserted to generate adversarial sequence. The label assigned to the input value by a black-box model gives the sign of the Jacobian matrix dimension. Jacobian matrix of the surrogate model is used for evaluation and after each iteration, synthetic example is added to each existing sample. However, finite set of legitimate API embeddings may not be enough for adversarial insertion, causing insertion of most impactful API call in direction indicated by Jacobian. Adversarial examples that are able to fool surrogate model has high likelihood of fooling original model as well~\cite{papernot2017practical}. Adversarial generation showed same success against the substitute and blackbox model with short API sequences, making adversarial generation faster. Framework also uses Cuckoo Sandbox to verify the malicious functionality of generated adversarial malware. GADGET framework wraps malware binary with proxy code and increases the risk even higher providing malware-as-a-service.

Adversarial attack on malware domain has not considered to manipulate the feature vector to see impact of mutation due to strict functionality preserving requirements of malware. Malware Recomposition Variation (MRV) based approach proposed by Yang et al.~\cite{Feature_evol_conf_Yang} performed an analysis of malware file semantically and construct a new variant of malware. Mutation strategies synthesized by conducting semantic-feature mutation analysis and phylogenetic analysis are used to perform automatic program transplantation~\cite{Auto_soft_transplant}. Changing the traditional belief over mutation, authors followed feature patterns of existing malware to preserve the functionality. The proposed framework performs inter-component, inter-app, and inter-method transplantation. More comprehensive attack is performed on both the manifest file as well as dex code. Use of RTLD features allows substitute model to approximate targeted detector and also helps in separation of essential features and contextual features. 

\begin{itemize}
    \item Resource features: These features provide the security-sensitive resources that are impacted by the malware. Resource features are extracted by forming call graphs and pinpointing call-graph nodes of those security-sensitive methods.
    \item Temporal features: These features provide the environmental context at the trigger point of the malicious property and the context is inferred from the attributes at entry points.
    \item Locale features: These features describe the location of programs where malicious activities are observed and can be either of the Android components of concurrency constructs.
    \item Dependency features: These features are provided by constructing an inter-procedure control-flow graph, indicating the control dependencies when malicious activities are invoked. 
\end{itemize}

 The goal of malware evolution attack is to imitate and automate the evolution of malware using phylogenetic evolutionary tree~\cite{phylogenetic_andreas}. A phylogenetic evolutionary tree shows the inferred relations between different samples based on the similarities and differences of feature representation. A pairwise distance between samples is fed to phylogenetic tree generation algorithm, Unweighted Pair Group Method Average (UPGMA), to generate phylogenetic tree which reflects the structure present in a similarity matrix. Another approach proposed by authors, confusion attack, tries to complement malware evolution attack against robust malware detectors~\cite{concept_drift_anshuman}. The feature values modified by confusion attacks can be shared by both the malware sample and benign apps. This approach is more complete due to flexibility of mutation by different number of means but introduces challenges on keeping functionality intact due to higher volume of modification~\cite{pierazzi2020intriguing}. 
 
Several adversarial generation approaches have been conducted by making minor perturbations on existing attacks. Liu et al.~\cite{Liu_2019} proposed a Testing framework for Learning-based Android Malware Detection systems (TLAMD). Framework uses genetic algorithm to perform black-box attack against Android malware detection system. Android files are modified by adding the request permission code to AndroidManifest.xml file which was originally proposed by Grosse et al.~\cite{grosse_adversarial17}. The restriction was imposed on the types and magnitude of permissions that can be added to AndroidManifest file. A random population is generated giving the characteristics of permission to add and followed by calculating the disturbance size for the sample malware. Using the evaluated perturbation size, adversarial is generated and tested against the detection model. Based on the result of detection either new disturbance size is calculated using genetic algorithms or perturbation is successfully added on Android application. Another important aspect of genetic algorithm is to model the fitness function which is this framework has been defined in Equation \ref{eq:fitness_tlamd}.

\begin{equation} \label{eq:fitness_tlamd}
    S(\delta)=\mathrm{min} ~w_{1}.F(X+\delta)+w_{2}.num(\delta)
\end{equation}
where \(w_{1}\) and \(w_{2}\) are the two weights, \(\delta\) is the added disturbance, \(num(\delta)\) is number of permission features added and \(F(X+\delta)\) gives the probability of malicious sample being detected as malware.
Fitness function searches for optimal solution to perform mutation leading to a new fit individual able to evade detection. Random forest approach is used to filter out insignificant features during feature extraction. Disturbance generated by genetic algorithm are able to bypass malware detectors trained on neural networks, logistic regression, decision trees and random forest.

Shahpasand et al.~\cite{android_adversarial_Shahpasand} implemented GAN to generate adversarial by keeping threshold on the distortion values of generated samples. The generated optimum perturbation $\delta$ is added to existing malware to produce adversarial. Like every other GAN architecture, generator can learn the distribution of benign samples, generating perturbations which are able to bypass learning based detector. The discriminator implicitly enhances the perturbation by escalating the loss of generator while the adversarial samples are identifiable with benign files. Loss function for GAN is similar to that of Goodfellow et al.'s~\cite{goodfellow2014generative} work given in Equation \ref{eq:GAN}.
\begin{equation} 
\label{eq:GAN}
    \begin{aligned}
       L_{GAN}=\; _{x \sim P_{Benign}}logD(x) \; +\;  _{x \sim P_{Malware},z \sim P_z(Z)}log(1-\\D(x+G(z))) ; \; s.t.|G(z)| < c
    \end{aligned}
\end{equation}
where $D(x)$ gives probability of sample \(x\) coming from benign software distribution, \(L_{GAN}\) helps maximizing resemblance of adversarial malware with benign sample while limiting distortion less than \textit{c}. 
Loss function of adversarial malware \((L_{adv.Mal})\) is given as:
\begin{equation}
    L_{adv.Mal}=_{x,z}l_f(x+G(z),0)
\end{equation}
where benign class is targeted for an adversarial sample \(x+G(z)\) against classifier $f$ trained on \(l_f\) loss function. Finally, model loss is defined as:

\begin{equation}
    L=\alpha L_{GAN}+(1-\alpha)L_{adv.Mal}
\end{equation}
In above equation, first loss forces the GAN to generate adversarial similar to benign samples while the second loss inclines adversarial samples to have higher miss-classification rate.

The goal of adversarial generation has been to bypass malware detector without losing functionality. However, due to growth in adversarial malware in recent time, defenders are employing firewalls to stop adversarial sample. Li et al.~\cite{LI_ANDROID_GAN} extended the work of MalGAN~\cite{hu2017generating} to make it robust against detection system equipped with firewall. Despite its high evasion rate against malware detectors, MalGAN is found to be less effective against detection systems using firewall. Bi-objective GAN with two discriminator having different objectives are used. One of the discriminator helps distinguish between malware and benign whereas another discriminator helps to find out whether the samples are adversarial or normal ones. Due to this feature, adversarial generated by generator can successfully bypass through the firewall as well as the malware detection. Authors used permissions, actions and application programming interface calls as a features to generate adversarial. In every round of training, gradient descent is used to update the parameters of discriminators and generator, represented by \(\theta_{d_{1}}\), \(\theta_{d_{2}}\) and \(\theta_g\) respectively. $G$, $D_1$ and $D_2$ are considered functions implemented by generator, discriminator 1 and discriminator 2. First discriminator, used to separate benign and malicious class is trained to update \(\theta_{d_{1}}\), by minimizing its loss
\begin{equation}
    L_{d1}=\mathbb{E}_{x\in N_B}log(1-D_1(x))+\mathbb{E}_{x\in N_M}log(D_1(x))
\end{equation}
where \(N_B\) and \(N_M\) denote the distributions of benign and malicious samples.
Second discriminator used to separate between normal and adversarial samples is trained by updating \(\theta_{d_{2}}\) by minimizing its loss
\begin{equation}
    L_{d2}=\mathbb{E}_{x\in N}log(1-D_2(x))+\mathbb{E}_{x\in M}log(D_2(x))
\end{equation}
where \(N\) is the distribution of normal samples and M is the distribution of generated samples detected as malicious. 
Generator is updated twice each round. After updating \(\theta_{d_{1}}\), \(\theta_g\) is updated by minimizing
\begin{equation}
    L_g^{1}=\mathbb{E}_{m\in M, z\in P}log(1-D_1(G(m,z)))
\end{equation}
where \(M\) is distribution of malicious samples fed to generator and \(P\) is the distribution of noises fed to generator. After \(\theta_{d_{2}}\) is updated, \(\theta_g\) is updated by minimizing
\begin{equation}
    L_g^{2}=\mathbb{E}_{m\in M, z\in P}log(1-D_2(G(m,z)))
\end{equation}

 \begin{table*}[t]
    \centering
    \setlength{\tabcolsep}{0.8\tabcolsep}
    \def\arraystretch{1.5}
    \caption{{Adversarial attacks on PDF malware.}}
    \rowcolors{2}{gray!25}{white}
    \begin{tabular}{|p{2cm}|p{1.5cm}|p{6.4cm}|p{4cm}|p{2.5cm}|}
    \rowcolor{gray!25}
     \hline
     \textbf{Paper/Year} & \textbf{Target} & \textbf{Approach} & \textbf{Modification} & \textbf{Limitations}\\
     \hline
      Maiorca et al. 2013 ~\cite{Looking_bag} &
      PJScan, Malware Slayer and PDFRate &
      \begin{minipage}[t]{\linewidth}
      \begin{itemize}[leftmargin=*]
          \item Reverse mimicry attack by manipulating binary files to make it malicious
          \item Malicious embedded EXE payload insertion
          \item Malicious PDF file insertion inside a benign one
          \item Encapsulating malicious JavaScript code
      \end{itemize}
      \vspace{1mm}
      \end{minipage}&
      \begin{minipage}[t]{\linewidth}
      \begin{itemize}[leftmargin=*]
          \item Malicious EXE payload as a new version after trailer
          \item Unrestrained embedded PDF structure insertion
          \item JavaScript code without reference to any other object to minimize structure variation
      \end{itemize}
      \vspace{1mm}
      \end{minipage}&
      \begin{minipage}[t]{\linewidth}
      \begin{itemize}[leftmargin=*]
          \item Less control on malicious goal
      \end{itemize}
      \vspace{1mm}
      \end{minipage}\\
      \hline
      Biggio et al. 2014 ~\cite{Biggio_2013_pdf,biggio2014security} &
      SVM and neural network based detectors &
      \begin{minipage}[t]{\linewidth}
      \begin{itemize}[leftmargin=*]
          \item Gradient based optimization inspired by Golland's discriminative directions technique
          \item Additional panalizing term to reshape objective function, biasing gradient descent towards region of negative class concentration
      \end{itemize}
      \vspace{1mm}
      \end{minipage}&
      \begin{minipage}[t]{\linewidth}
      \begin{itemize}[leftmargin=*]
          \item Insertion of objects creating new PDF files
      \end{itemize}
      \vspace{1mm}
      \end{minipage} &
      \begin{minipage}[t]{\linewidth}
      \begin{itemize}[leftmargin=*]
          \item Feature mapping issues
          \item Non-differential discriminating functions can not be evaded
      \end{itemize}
      \vspace{1mm}
      \end{minipage}\\
      \hline
      Srndic et al. 2014 ~\cite{Practical_Evasion_Srndic}&
      PDFrate employed on Random Forest &
      \begin{minipage}[t]{\linewidth}
      \begin{itemize}[leftmargin=*]
      \item Taking advantage of discrepancy between operation of PDF reader and PDFrate
          \item Mimicry attack to mimic 30 different benign files
          \item GD-KDE attack to defeat classifier with differentiable decision function
      \end{itemize}
      \vspace{1mm}
      \end{minipage}&
      \begin{minipage}[t]{\linewidth}
      \begin{itemize}[leftmargin=*]
          \item Insertion of dummy contents, ignored by PDF readers but affect detector
          \item Trailer section moved away from cross reference table for file injection space
      \end{itemize}
      \vspace{1mm}
      \end{minipage}&
      \begin{minipage}[t]{\linewidth}
      \begin{itemize}[leftmargin=*]
          \item Feature mappings are assumed to be perfect which is unrealistic
      \end{itemize}
      \vspace{1mm}
      \end{minipage}\\
      \hline
       Carmony et al. 2016 ~\cite{Carmony2016ExtractMI} &
       PDFrate and PJScan &
       \begin{minipage}[t]{\linewidth}
       \begin{itemize}[leftmargin=*]
           \item Reference JavaScript extractor by directly tapping into a Adobe reader at locations identified by dynamic binary analysis
           \item Parser confusion attack combined with reverse mimicry attack
       \end{itemize}
       \vspace{1mm}
       \end{minipage}&
       \begin{minipage}[t]{\linewidth}
       \begin{itemize}[leftmargin=*]
           \item Obfuscation based on output of reference extractor
       \end{itemize}
       \vspace{1mm}
       \end{minipage}&
       \begin{minipage}[t]{\linewidth}
       \begin{itemize}[leftmargin=*]
           \item Useful only for JavaScript based detector
           \item Dependent on versions of Adobe Reader
       \end{itemize}
       \vspace{1mm}
       \end{minipage}\\
       \hline
       Xu et al. 2016 ~\cite{autom_evad_clasf} &
       PDFrate and Hidost &
       \begin{minipage}[t]{\linewidth}
       \begin{itemize}[leftmargin=*]
           \item Stochastic manipulations using genetic algorithm to generate population
           \item Iterative population generation till evasion
           \item Successful mutation traces reused for initialization efficiency
           \item Fitness score based on maliciousness detected by oracle
       \end{itemize}
       \vspace{1mm}
       \end{minipage}&
       \begin{minipage}[t]{\linewidth}
       \begin{itemize}[leftmargin=*]
           \item Inserting new, removing and modifying existing contents
           \item Oracle confirming the maliciousness of file
       \end{itemize}
       \vspace{1mm}
       \end{minipage}&
       \begin{minipage}[t]{\linewidth}
       \begin{itemize}[leftmargin=*]
           \item Stochastic approaches are resource intensive
           \item No exact way to choose best fitness function
       \end{itemize}
       \vspace{1mm}
       \end{minipage}\\
       \hline
    \end{tabular}\\
    \vspace{2mm}
    \footnotesize\textit{\textbf{Target}: Target defense for adversarial attack, \textbf{Approach}: Key procedures to carry out adversarial attack, \textbf{Modification}: Changes on file to craft the adversarial perturbation, \textbf{Limitations}: Shortcomings of proposed approach }
    \label{tab:PDF}
\end{table*}

Pierazzi et al.~\cite{pierazzi2020intriguing} formalized the adversarial ML evasion attacks in the problem space and proposed a problem space attack on Android malware. The proposed approach formalizes the set of restriction in transformations available, semantics preserved, robustness of preprocessing approaches, and veracity. Research work is focused on evasion attacks at test time by modifying the objects in real input space corresponding to feature vector. With a goal of overcoming inverse feature-mapping problem from previous researches, author presents the idea of side-effect features. Side effect feature defines and proves the necessary as well as sufficient conditions behind the problem space attacks. An attack on a feature space is projected towards a feasibility region satisfying the problem space constraints to obtain the side effect features. To formally demonstrate the side-effect features, an object \(x\in X\)  is initialized within a feasible region. A gradient-based attack takes an object \(x\) in feature space to \(x+\delta^*\), with $\delta^*$ being the perturbation. The addition of perturbation misclassifies malware files as benign with high confidence. However, the new point with perturbed feature on feature vector may not be inside the feasibility region. Side effect features in perturbation help to map \(x+\delta^*\) to the region of feasible problem-space. 

Since side effect features contribute towards preserving validity of malware due to impact of original gradient based perturbation, side effect features alone can have both positive or negative influence on the classification score. Authors use automated software transplantation~\cite{Auto_soft_transplant} to extract byte-codes from benign donor applications to inject into a malicious host, also known as organ harvesting. Insertion into a host is carried out between statements in non-system package to preserve the functionality and Cyclomatic Complexity is used to take a heuristic approach maintaining existing homogeneity and preventing violation of plausibility. Prior research works relied heavily on addition of permissions to the Android Manifest which is considered dangerous in Android documentation~\cite{PermissionsonAndroidAndroidDev}. Authors bind the modifications to inject single permission to the host app. Gradient based strategy using greedy algorithm proposed in this approach overcomes previous limitations of preserving semantics and pre-processing robustness. 

To overcome the challenges of limited access to target classifiers while circumventing black-box Android malware detectors, Bostani et al.~\cite{bostani2021evadedroid} proposed a novel iterative and incremental manipulation strategy. The attack is carried out in two-step: preparation and manipulation. In preparation phase, automated software transplantation is employed to prepare action sets from Android apps. The n-gram-based similarity method is used to identify benign apps that closely matches to malware files. Insertion of extracted gadgets of closely matching benign files force malware sample towards the blind spots of the classifier. In the manipulation stage, perturbation on malware samples are applied incrementally, choosing from the collected action set. The search method randomly chooses suitable transformation and applies them to malware samples. This approach shows a high success rate in query efficient approach but increases the size of adversarial perturbation which in turn increases the risk of perturbation being easily detected.

\subsection{PDF Malware Adversarial}
Along with widespread applications and adoption, PDF documents have been one of the most exploited avenues for adversarial malware attacks. Initially, JavaScript based and structural properties detection were prominent for recognising malware in PDF. But freedom to distribute chunks of Javascript code and assemble together at run-time and high degree expressiveness in JavaScript language led to failure of Javascript based detection. Despite significant growth in PDF malware detection from JavaScript using deep learning techniques, the challenges posed by adversarial examples still exist. Early evasion attempts on PDF documents were crafted by Smutz et al.~\cite{Smutz2012MaliciousPD} and Šrndić et al.~\cite{Srndic2013DetectionOM} using heuristic approaches. 
The authors proposed approach to build more robust PDF malware detection techniques, showcased the adversarial ability to mislead linear classification algorithm successfully. 

Flexible logical structure of PDF has allowed to craft adversarial by carefully analyzing its structure. Maiorca et al.~\cite{Looking_bag} demonstrated evasion technique called reverse mimicry attack against popular state-of-art malware detectors ~\cite{PjScan,Malware_Slayer,Smutz2012MaliciousPD}. Traditionally, malicious PDF files are believed to be structurally different from benign PDF files. Taking advantage of this structural difference, most of malware detectors were able to discriminate PDF files with very high accuracy. However, malware files which can imitate the benign file structure or vice-versa can easily fool the detector. Reverse mimicry attacks can make benign files malicious with minimal changes in their structure. Malicious payloads poison the samples, initially classified as benign. Three kinds of malicious payloads introduced to benign files take the sample across the decision boundary of malware detector. First one is EXE payload with malicious embedding, which is introduced using Social Engineering Toolkit\footnote{https://www.secmaniac.com/} as a new version after its trailer. On the addition of a new root object, the new trailer will point to a new object. In this payload, authors embedded malicious PDF files inside another benign PDF files using embedded function of PeePDF\cite{esparza2015peepdf} tool. The embedded PDF file automatically opens without user interaction, allowing malicious PDF to be embed inside a benign one without any restriction on embedding file. PDF file injection enabled an attacker to have fine-grained control of structural features in the carrier file. A final kind of payload insertion is carried out by encapsulating a malicious JavaScript code without reference to other object. It helps to minimize variation in the benign-file structure by adding only one object to the tree. Such attacks based on structure are even capable of evading detectors using non-structural features. Table \ref{tab:PDF} provides overview of adversarial attacks carried out on PDF files.

Optimization based evasion attack against PDF malware detection was introduced by Biggio et al.~\cite{Biggio_2013_pdf,biggio2014security}. The attack was carried out using a gradient based optimization procedure inspired by Golland's discriminative directions technique~\cite{Golland_discriminative} to evade linear as well as non-linear classifiers. The proposed work was able to carry out complete knowledge and constrained knowledge attacks on non-linear models like Support Vector Machine(SVM) and neural networks. Work relied on easiness to insert new objects than to remove an embedded object to prevent from corrupting the PDF's file structure. This approach used a gradient descent procedure with special consideration to avoid getting stuck on local optima. To increase the probability of successful evasion, an attacker needs to reach attack points that are legitimate, and to reach this, the additional penalizer term is introduced using a density estimator. The extra component helps imitate features of known legitimate samples, reshaping the objective function by biasing the gradient descent towards the negative class concentration region. This optimization-based approach dates before the realization of adversarial examples against deep learning architectures~\cite{Maiorca_2019}.

\begin{figure}
    \centering
    \includegraphics[scale=0.8]{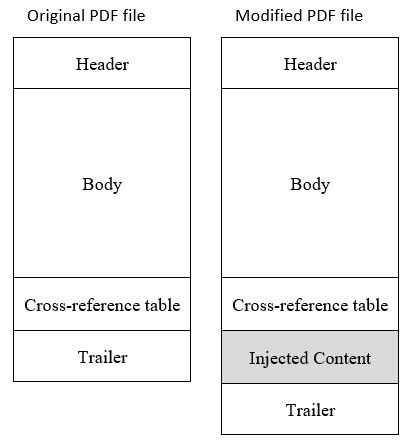}
    \caption{A PDF file modification approach.}
    \label{fig:pdf_modif_srn}
\end{figure}

Srndic et al.~\cite{Practical_Evasion_Srndic} further enhanced optimization based attack against deployed system PDFrate~\cite{Smutz2012MaliciousPD} using mimicry attack, and Gradient Descent and Kennel Density Estimation (GD-KDE) attack. The attack takes advantage of discrepancy between functioning of PDF readers and PDFrate in terms of interpretation of semantic gaps as explained in \cite{abusing_file_proces}. The dummy contents to insert should be ignored by PDF readers but affect the feature computation in PDFrate. PDFrate evaluates sets of regular expressions from raw bytes, reading from beginning of PDF files while PDF readers parse PDF files using PDF format authorized by ISO 32000-1. PDF reader looks at the end of PDF for cross-reference table and goes to locate the object directly. Among 135 features of PDF rate, MIMICUS\footnote{https://github.com/srndic/mimicus}, 35 features are modified while incrementing values for 33 features with preserved functionality. Trailer section of PDF files were moved arbitrarily far away from cross-reference table, generating an empty space for file injection without affecting functionality of PDF document. A string pattern that is separated by whitespace is injected into the gap between CRT and the trailer of targeted PDF files as demonstrated in  Figure \ref{fig:pdf_modif_srn}. To match with specific PDFrate regular expression, patterns are also crafted. Two attack algorithms are implemented as explained:
\begin{itemize}
    \item \textbf{Mimicry attack} independent of underlying classifier, mimics a benign file by changing the modifiable features of malicious file. To increase effectiveness of the approach, each malware is trained to mimic 30 different benign files.
    \item \textbf{Gradient Descent and Kernel Density Estimation (GD-KDE) attack} defeats classifier with a known and differentiable decision function~\cite{Biggio_2013_pdf}. GD-KDE algorithm follows the gradients of the classifier's decision function and the estimated density function of benign samples.
\end{itemize}

\begin{figure}
    \centering
    \includegraphics[scale=0.7]{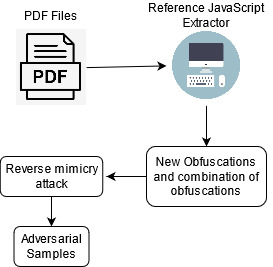}
    \caption{A parser confusion attack.}
    \label{fig:carmony_confusion}
\end{figure}

PDF detection techniques are mostly reliant on PDF parser to extract features for classification~\cite{PDF_DETECT_LASKOV,PDF_DETECT_LIU,pdf_Detect_Tzermias}. These parser are unable to extract all JavaScript of PDF file. Carmony et al.~\cite{Carmony2016ExtractMI} created a reference JavaScript extractor that measured the difference between the parser and Adobe Reader by tapping Adobe reader on locations given by binary analysis. Manual analysis refines the few candidate tap points provided by dynamic binary analysis. JavaScript extraction tap points are a function from which Adobe Reader extracts and executes JavaScript code from PDF documents. The memory accessed by Adobe Readers when reading PDF files using automatically executable JavaScript is analyzed to determine the raw JavaScript extraction tapping points. Proposed PDF parser confusion attack apply obfuscation on malicious PDF sample by analyzing the weaknesses of extractors in approach as shown in Figure \ref{fig:carmony_confusion}. Reference extractor enables several new obfuscation in comparison to existing extractors and combination of these obfuscation were able to bypass all JavaScript extractor based detector. Metadata based detection systems require parser confusion attacks to be combined with reverse mimicry attack as core content of sample is not changed by confusion attacks.

\begin{figure}[!t]
    \centering
    \includegraphics[scale=0.7]{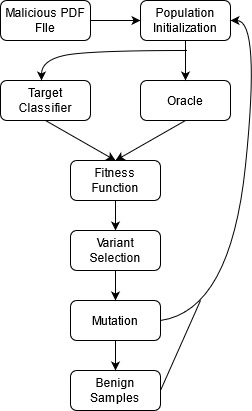}
    \caption{A PDF adversarial generation based on genetic algorithm~\cite{autom_evad_clasf}.}
    \label{fig:pdf_xu}
\end{figure}

In order to preserve maliciousness, most of research works take conservative approach by only inserting new contents and refraining from modification or removal of existing contents. Xu et al.~\cite{autom_evad_clasf} proposed a black-box generic method to evade the classifier as shown in Figure \ref{fig:pdf_xu}. As in figure, first the population is initialized by performing random modifications on malicious file. Then, each member of populations are passed through target classifier to measure maliciousness and through oracle to confirm the functionality. If no any samples are able to evade target classifier with functionality intact, subset of initialized population are chosen for next generation based on fitness score which indicates the progress towards evasive sample. Now, the population generation is repeated and this process is continued till the evasive sample is found or threshold iterations is met. Efficiency of search is enhanced by collecting traces of used mutation operation and reuse the effective operations. These effective traces are used for population initialization to generate variants for other malware. The author uses genetic programming (GP) to bring off stochastic modifications in an iterative manner till evasion. The oracle output and results of prediction from the target classifier need to be fed to the fitness function. Non malicious samples determined by oracle are assigned with low fitness score while malicious samples are provided with high score. Tools used as an oracle for maliciousness verification of adversarial samples, is an open computational challenge along with the selection of appropriate fitness function. The generic method is capable of automatically finding evasive variants irrespective of detection algorithms.

\subsection{Hardware Based Malware Adversarial}
\begin{table*}[t]
    \centering
    \setlength{\tabcolsep}{0.8\tabcolsep}
    \def\arraystretch{1.5}
    \caption{{Summary of hardware based malware adversarial.}}
    \rowcolors{2}{gray!25}{white}
    \begin{tabular}{|p{2cm}|p{1.5cm}|p{4.25cm}|p{2.5cm}|p{5.75cm}|}
    \rowcolor{gray!25}
    \hline
     \textbf{Paper/Year}  &\textbf{Target Model} & \textbf{Data Collection} & \textbf{Feature Vectors} & \textbf{Approach}\\
     \hline
     Khasawneh et al. 2017 ~\cite{RHMD_Khasawneh} &
     Logistic Regression and neural network based detectors &
     \begin{minipage}[t]{\linewidth}
     \begin{itemize}[leftmargin=*]
         \item Running malware and cleanware files on a virtual machine operating on Windows 7
         \item Dynamic traces collected using Pin instrumentation tool
         \item Dynamic traces providing run time behaviour of the programs
     \end{itemize}
     \vspace{1mm}
     \end{minipage}&
     \begin{minipage}[t]{\linewidth}
     \begin{itemize}[leftmargin=*]
         \item Instructions Feature
         \item Memory address patterns
         \item Architectural Events
     \end{itemize}
     \vspace{1mm}
     \end{minipage}&
     \begin{minipage}[t]{\linewidth}
     \begin{itemize}[leftmargin=*]
         \item Reverse engineering to create surrogate model of HMD
         \item Dynamically instruction insertion into malware execution through Dynamic Control Flow Graph
         \item Weighted injection strategy with insertion instruction selection proportional to negative weight
     \end{itemize}
     \vspace{1mm}
     \end{minipage}\\
     \hline
     Dinakarrao et al. 2019 ~\cite{8806987} &
     Logistic Regression and neural network based detectors &
     \begin{minipage}[t]{\linewidth}
     \begin{itemize}[leftmargin=*]
         \item Captured using Hardware Performance Counters (HPC)
         \item Perf tool available under Linux used
     \end{itemize}
     \vspace{1mm}
     \end{minipage}&
     \begin{minipage}[t]{\linewidth}
     \begin{itemize}[leftmargin=*]
         \item Low-level micro-architectural events
         \item LLC load misses, branch instructions, branch misses and executed instructions
     \end{itemize}
     \vspace{1mm}
     \end{minipage}&
     \begin{minipage}[t]{\linewidth}
     \begin{itemize}[leftmargin=*]
         \item Reverse engineering of Black-box HMD 
         \item HPC patterns perturbation mechanism determined using FGSM
         \item Perturbation calculated using neural network
         \item Adversarial generators running as separate thread to avoid interference with original source code
     \end{itemize}
     \vspace{1mm}
     \end{minipage}\\
     \hline
     Nozawa et al. 2021 ~\cite{Kohei_Nozawa2021} &
     Neural network architecture &
     \begin{minipage}[t]{\linewidth}
     \begin{itemize}[leftmargin=*]
         \item Structural features analysis
     \end{itemize}
     \vspace{1mm}
     \end{minipage}&
     \begin{minipage}[t]{\linewidth}
     \begin{itemize}[leftmargin=*]
         \item Gate level netlist
     \end{itemize}
     \vspace{1mm}
     \end{minipage}&
     \begin{minipage}[t]{\linewidth}
     \begin{itemize}[leftmargin=*]
         \item Hardware circuits represented in graph structure and converted to feature space
         \item During design step or after logic synthesis
         \item Trojan-net concealment degree to prevent from detection
         \item Modification evaluating value to limit the extent of modification
     \end{itemize}
     \vspace{1mm}
     \end{minipage}\\
     \hline
    \end{tabular}\\
    \vspace{2mm}
    \footnotesize\textit{\textbf{Target Model}: Defense model under adversarial attack, \textbf{Data Collection}: Feature value collection process, \textbf{Feature Vectors}: Types of features considered, \textbf{Approach}: Process of crafting adversarial}
    \label{tab:HMD}
\end{table*}

\begin{figure}
    \centering
    \includegraphics[scale=0.6]{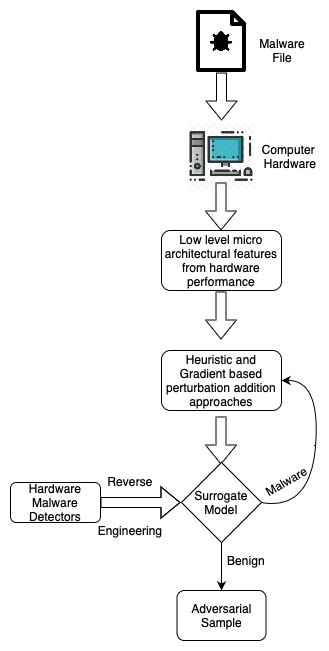}
    \caption{An adversarial generation workflow against HMD.}
    \label{fig:adv_HMD}
\end{figure}

Hardware malware detectors use low-level information of features collected from hardware performance monitoring units available in CPUs. Hardware malware detectors are prone to reverse engineering~\cite{OPTIMAL_RANDOMIZED_Vorobeychik}, allowing mimicry attack~\cite{Wang2006AnagramAC} to reverse-engineered models. Adversarial against such detectors are carried out by generating perturbations in form of low-level hardware features, following the architecture shown in Figure \ref{fig:adv_HMD}. These adversarial generation approaches differ only on type of features used in comparison to previously discussed works. Table \ref{tab:HMD} provides brief comparison of adversarial attacks against hardware malware detectors. Khasawneh et al.~\cite{RHMD_Khasawneh,Khasawneh_resilient_adversarial}
demonstrated evasion of Hardware Malware Detectors(HMD) after being reverse engineered, using low overhead evasion strategies. Data collected by running malware and cleanware programs on a virtual machine operating on Windows 7 are used to train a reverse-engineered model. Execution of malware requires disabling of Windows security services and firewall. Data required for training are dynamic traces while executing the program and are collected by using Pin instrumentation tool~\cite{Luk2005PinBC} by executing 500 system call or 15 millions of committed instructions. These dynamic traces are profiling of a run time behaviour of the programs. This dataset is comprised of three types  of feature vectors: 
\begin{itemize}
    \item \textbf{Instructions feature} giving the frequency of instructions.
    \item \textbf{Memory address} patterns giving distribution of memory references.
    \item \textbf{Architectural events} giving the occurrence of architectural events.
\end{itemize}
Reverse engineering allows to methodically create model similar to HMD, given the attacker has ability to query the target detector. Target detectors are considered to be based on logistic regression and neural networks to generalize most of the classification algorithms. Reverse engineering is carried out in all three types of feature vectors using both algorithms. Attackers have no information about size of instruction window, the specific features used by detector or the classification algorithm, detectors are trained on but have similar detectors to test their hypothesis. Authors\cite{RHMD_Khasawneh} constructed a Dynamic Control Flow Graph (DCFG) of the malware to insert instructions into the executing malware dynamically. Injection of instruction feature increases the weight of the corresponding feature while memory feature injection alters the histogram of memory reference frequencies. Instructions are inserted using two approaches, block level and function level. Khasawneh et al. picked the instructions with negative weights to move the malware away from the decision boundary. Heuristic approach was taken to identify the candidate instructions for insertion. Weighted injection strategy where probability of selecting particular instruction is proportional to negative weight allowed to bypass HMD with around 10\% dynamic overhead. 

Dinakarrao et al.~\cite{8806987} also proposed an adversarial attack on low-level micro-architectural events captured through Hardware Performance Counters (HPC). Victim's defense system (HMD) being black-box needs to be reverse engineered to mimic the behaviour. Number of HPC patterns required to bypass HMD is unknown which leads to need of adversarial sample predictor. The HPC patterns perturbing mechanism are implemented using a lower-complexity gradient approach, Fast Gradient Sign Method (FGSM).
The adversarial perturbations needed to misclassify HPC trace is calculated using gradient based cost function of neural network. With \(\theta\) being hyperparameters of neural network, $x$ being input HPC trace to the model and $y$ as output, cost function \(L(\theta,x,y)\) is defined as:
\begin{equation}
    x^{adv}=x+\epsilon ~sign(\Delta_xL(\theta,x,y))
\end{equation}
where \(\epsilon\) is a scaling constant ranging between 0.0 to 1.0 and used to limit the perturbation to a very small value. 
The LLC load misses and branch misses are the most significant micro-architectural events of malicious application~\cite{8060309}. Linear models are built to find dependency of array sizes(n) and elements flushed(k)  in route to determine LLC load misses generated. The adversarial perturbation generator is executed separately with a sample application to misclassify. The significance of running in a separated thread is to stop adversarial HPC generators from interfering with the application's source code.

Micro-architectural events are useful when the malware program executes on a hardware device causing some changes in the performance of hardware. In addition to malware program execution, threat of injecting a malicious circuit during fabrication of hardware devices has grown~\cite{LIU2016438}. Malicious circuits, also known as hardware Trojans, can be inserted into circuits producing logically equivalent results. Hardware circuits like processors include contribution of number of vendors before reaching to final product, and hence, carries high risk of adversarial circuit insertion anywhere between designing stage to manufacturing stage. Modifications in manufacturing stage are more tedious in comparison to design stage as few changes in hardware description language (HDL) are enough to embed hardware Trojans to the circuit. A trigger circuit allows the payload circuit to trigger malicious behavior such as information leakage and degrading performance after satisfying the trigger condition. A gate level netlist (the lists of nets and circuit elements) is used for malware detection by analyzing its structural features~\cite{7092434}. Several hardware Trojan detection have been recently adopted towards machine learning approaches including neural networks~\cite{7604700,8634823,8252600,8576247}.

Nozawa et al.~\cite{Kohei_Nozawa2021} proposed an architecture to develop the adversarial hardware Trojan using Trojan-net Concealment Degree (TCD) and Modification Evaluating Value (MEV). Feature mapping issues like in all other adversarial attacks in Windows, Android and PDF are prevalent in hardware Trojan as well. Hardware circuits are represented in graphs structure and modifications in feature space does not guarantee the transfer back of modification to graph structure. Two stages in designing period are known for adversarial attack. First one being RTL (Register-transfer level) description design step and second one after logic synthesis. Authors take the assumption of Trojan detector using neural network architecture and the availability of raw output values from the detector to train adversarial model. Goal of adversarial is to maximize the loss function which is given as cross entropy(H) in Equation \ref{Trojan_net_loss}:
\begin{equation} \label{Trojan_net_loss}
    H=-\sum_{i=1}^K p_i(x(e))log\;q_i(x(e))
\end{equation}
where K is the number of units in output layer, \(p_i(x(e))\) is the function to return label of \(x(e)\) and \(q_i(x(e))\) is function that returns the prediction result from classifier. 

Summing up the loss function for each net and calculating the average gives Trojan-net concealment degree (TCD):
\begin{equation}
    TCD=-\frac{1}{|E_t|}\sum_{e_t\epsilon E_t}(\sum_{i=1}^K p_i(x(e_t))log\;q_i(x(e_t))
\end{equation}

Larger TCD indicates the bigger difference in values of the prediction and answer, enabling to achieve the concealment of Trojan nets. However, to monitor the amount of modification authors used modification evaluating value (MEV).
\begin{equation}
    \begin{aligned}
        MEV=-TCD+\sum_{j=1}^N\lambda_jm_j\\
        =\-\frac{1}{|E_t|}\sum_{e_t\epsilon E_t}(\sum_{i=1}^K p_i(x(e_t))log\;q_i(x(e_t))+\sum_{j=1}^N\lambda_jm_j
    \end{aligned}
\end{equation}
where \(m_j(1\leq j\leq N)\) is one of the \(N\) kinds of evaluation indicators and \(\lambda_j(1\leq j \leq N)\) being the corresponding coefficient. The success of research work in evading Trojan nets from Trojan detectors reveal threat of adversarial at a whole new level. 

\subsection{Linux Malware Adversarial}
Distributed edge computing has increased use of IoT devices. With large number of devices using Linux systems, robust malware detection is paramount. Both deep learning networks and Control Flow Graph (CFG) based malware detectors in IoT devices are found to be vulnerable against adversarial samples \cite{Graph_Based_Abusnaina}. 
In off-the-shelf adversarial attack, authors examined different well known adversarial algorithms based on feature extraction. Generic adversarial algorithms are successful in adversarial generation with high evasion rate but limited on applicability of practical changes to feature space. In response to these challenges, adversarial based on control flow graph has been proposed\cite{Graph_Based_Abusnaina}. Programs are structurally analyzed using vertices and edges with help of CFG. Graph embedding and augmentation (GEA) approach combines original graph with target graph, producing mis-classification while preserving the functionality of original program. GNU compiler collection command compiles in a way that only functionality related to original sample is executed. Linux based malware binaries easily evade IoT malware detection from different graph algorithmic constructs. In our search of literature, we found very limited works carried out as adversarial malware attacks in Linux domain.

\section{Challenges and Future Directions}
\label{sec:Future}
Following the introduction of adversarial attacks against deep learning by Szegedy et al.~\cite{szegedy2014intriguing}, machine learning research community have been concerned about its impact in different application domains. The research on adversarial attacks and its countermeasures is gaining momentum. To contribute towards the literature, we conducted a comprehensive research on various adversarial evasion attacks carried out against malware detection domain. Although, our survey highlights several successful adversarial attacks crafted against anti-malware engines, novel attacks are still evolving. In this section, we will discuss potential research open challenges and future direction as the adversarial approaches in malware analysis domain become more prevalent. Our intention is in no way to overlook or understate the contributions of existing adversarial attack researchers in malware domain. 
\subsection{Realistic (Practical) Attacks}
Most of the existing adversarial attacks discussed in the survey are carried out using white-box approaches. In white-box approach, attackers are assumed to have full access to target model providing all required internal information to attackers. White-box approach is considered by most an unrealistic scenario in itself as it is unlikely that any ML based anti-malware engine will reveal information such as algorithms used, gradients of the model and hyper-parameters used to fine tune the model. Getting this information about a target model provides `superpower' to attackers as they can camouflage the data in any way they want. In the future, research is expected to be more inclined towards complete black-box attacks. Few of the existing black-box attacks also depend on the performance of models provided in numeric form. Obtaining numerical performance is also not a realistic approach for attackers. So, we believe that further work should seek for completely black-box approaches for carrying out adversarial attacks. 

In addition, the attacks discussed previously are primarily focused on static malware detection. The main reason behind it is the limited research carried out to test the robustness of dynamic detection. The modern industrial malware detection engines merge both static and dynamic detection techniques. Further, the attacker rarely gets the privilege to work with data at rest. There has been very few successful attempts to craft adversarial examples against data in motion \cite{gong2019realtime,xie2020realtime}. The malware domain can have data which is moving at a very high pace and may require to perform attack on data in motion. Data stored in a storage device or in transit may have enough time to let attackers generate adversarial examples for them while data moving at very high speed might allow smallest of time frames for attacker. Adversarial attacks are not always swift enough to work with data moving across network channels. So, more adversarial attacks are to be experimented for systems deployed with both static and dynamic detection as well as against data at motion.

\subsection{Perturbation Insertion Space}
Smart perturbation insertion plays a key role in determining the success of adversarial attacks. Initial adversarial evasion attacks on malware began by placing perturbations at the end of the malware file~\cite{Kreuk2018AdversarialEO}. Most of the existing attack approaches are concentrated on additive adversarial perturbation. Demetrio et al.~\cite{Demetrio2019} later discovered that perturbations embedded at header sections of file resulted in effective adversarial attacks compared to perturbations appended at the end. However, header perturbation invited the risk of perturbation being detected and also increased the chances of breaking the malware file functionality. Suciu et al.~\cite{Suciu2019Malware} further investigated the possibility of inserting perturbations in slack regions of file which are left behind by the compilers. Experiments demonstrated that perturbations that are inserted in slack region are more effective than perturbations at other locations. These experiments provide inconclusive information about suitable insertion space for perturbation. Hence, further research is needed to determine optimal locations for perturbation that are more effective as well as undetected.

\subsection{Enhancing Efficiency}
Adversarial efficiency can be defined in terms of different parameters. The first and obvious efficiency criteria is the length of the payload to be generated as perturbation. Generating random noise is an inefficient approach while gradient based algorithms are developed to make an efficient adversarial attacks. The significance of the inserted payload determines the efficiency of perturbation. One way to insert efficient features is to first decipher the importance of each feature in the decision making of the machine learning model. Highly influential features can be modified to reach the adversarial goal with minimal perturbation volume. Despite the gradient helping attackers to generate perturbation in right direction, efficiency may be limited due to uncountable iterations to reach the adversarial goal. Applying small perturbations iteratively results in high quality adversarial evasion. However, these approaches will require immense amount of time, making it impossible for real time operation. To challenge this limitation, approaches like Fast Gradient Sign Method are proposed, which produce perturbations at very high pace but are less effective and have a high chance of being detected. Hence, research is needed to ensure that efficiency is looked both in terms of quantity and quality of noise generated to produce adversarial evasion. In addition, trade off between performance and computational complexity should be analysed to evaluate the worth of performing adversarial attacks~\cite{alsmadi2021adversarial}. 


\subsection{Mapping Space Challenge}
Mapping between problem space and feature space is performed by an embedding layer present in between them. The features in problem space can be of any form like n-grams, API names or other non-numeric parameters which can not be directly processed by machine learning models. This causes the problem space vectors to be converted into feature space which are some form of numeric values. The embedding layer however is an approximation mapping table between features in problem space and feature space. One of the biggest challenge of adversarial attacks is to map features in problem space and feature space precisely. Machine learning models require malware features to be converted to feature space from problem space so that adversarial examples can be crafted on them. However, there is no exact mapping between these spaces which results in approximate mapping, leading to slightly altered feature space than original problem space. After adversarial examples are crafted on malware files, mapping features back to problem space also lose few crafted perturbations due to lack of absolute mapping. Therefore, the challenge for defining adversarial space and efficiently searching elements approaching the best replacement has always been there in adversarial domain. Further research is needed to identify and map features in embedding space.

\subsection{Automated Attacks}
All of the discussed adversarial attacks require manual intervention at a few steps of the attack procedure. Human intervention makes the process time-consuming and impractical in many cases. In white-box attacks, the loss function of deep neural network can be used to determine most influential features and the corresponding features can be automatically modified~\cite{zhang2019adversarial}. Current literature relies on human efforts for feature extraction, mapping to adversarial generation and functionality verification. Minimizing human effort while moving towards automated adversarial generation could be the interesting arena to work on the future~\cite{Liang_2018}. Novel research is needed to fully automate the adversarial attack ecosystem.

\subsection{Explainable Adversarial}
Adversarial vulnerabilities have been considered blind spots of machine learning models but current research work fails to assert concrete reasoning behind these blind spots. Having no consensus behind such reasoning leaves explaining the existence of adversarial example an open research domain. Goodfellow et al~\cite{goodfellow2015explaining} first attributed vulnerability to the linear behavior of model in high dimensional space. However, there have been research that contradicts the accountability of adversarial behavior solely to linearity of model as highly non-linear models are also evaded successfully~\cite{li2021verifying}. Explaining the adversarial phenomenon both in terms of models' functionality and features' contribution can pave a path for more robust adversarial attacks. Features can be assigned appropriate weights based on their contribution to alleviate the adversarial effect in the model. With the current state of the literature, explainable adversarial is still at an immature stage and requires concrete efforts from the community.

\subsection{Transferable Attacks}
Transferability refers to generalization property of the attack methods. A machine learning model with transferability property, trained on one particular dataset can generalize well for another different dataset as well. Transferability is a common property for evasion attack and is extensively exploited by black-box attacks. Untargeted attacks are found more transferable than targeted ones due to their generality~\cite{Liang_2018}. Transferability can also take three different forms such as, same architecture with different data, different architecture with same application and different architecture with different data~\cite{yuan2018adversarial}. Although some studies have already been carried out on transferability, there is no any universally accepted postulation. The ability to use same data, model or algorithm to attack all available targets should be one of the goals of future research on adversarial attacks. Thus, attacker having transferability in their models should be able to attack defensive system irrespective of input and context in new targets. 

\subsection{Attacking Adversarial Defense}
The influx of research on adversarial domain during last few years demonstrate the extent and importance of work in performing adversarial attack. The profound activity has not been limited to attack side but considering the threat posed to entire machine learning family, researchers have been equally active on defensive side as well. The cyber war between adversarial attackers and defenders are marching on an extremely high pace to overtake each other. Performing adversarial attacks are turning out to be harder than ever as many systems are designed robustly with adversarial defense in mind. Defensive approaches like adversarial training~\cite{shafahi2019adversarial}, defensive distillation~\cite{papernot2017extending} are proposed to stop adversarial attacks. Some recent techniques are hiding the gradients of target model~\cite{qiu2019review}, which if carried out successfully can completely nullify the threat of gradient based adversarial attacks. Hence, future adversarial attacks are required not only to bypass the machine learning detection but also overcome adversarial defenses. At the same time, novel defense mechanisms and approaches must be designed to make our models resilient against growing adversarial attack ecosystem.

\subsection{Functionality Verification}
Adversarial attacks on image domain carries a longer history than in malware domain. Image classifiers are attacked by modifying the image pixels to create adversarial. While modifying images, the pixels can be abundantly disrupted till it impacts the human perception. However, adversarial attacks is completely a different story in malware domain. The modifications carried out in a malware file should not alter the functionality of malware. The contents in executable file could be very sensitive and modification of a single byte can completely change the functionality of malware or even break the file, making it unexecutable. Most of the adversarial attacks have constrained themselves in perturbation type, volume and insertion techniques, just to preserve the functionality of executable. Malware functionality should not be compromised at the cost of any other adversarial constraints. Despite such gravity, most of the adversarial attacks are still not able to preserve the functionality of a modified files. Moreover, limited mechanisms exist to verify the functionality of malware after perturbing the file. One of the available approach is to run the malware file in an isolated environments like Cuckoo Sandbox~\cite{jamalpur2018dynamic}. But running every individual malware in a sandbox is inefficient and unrealistic. Therefore, further research should be directed to develop tools that can automatically and efficiently verify the functionality of malware post perturbations. 

\subsection{Attacking Federated Learning}
Federated learning~\cite{yang2019federated} is a hot topic as it allows individual nodes in a system to train a shared prediction model by confining all the data on individual devices~\cite{google_ai_blog_2017}. This has gained a great popularity recently as it reduces the computational cost of centralized machine while preserving data privacy of each individual nodes. However, federated learning comes with greater risk of adversarial attacks as central system has no control over the training data. Learning system can easily be poisoned while training, leaving the backdoors in trained model~\cite{bagdasaryan2020backdoor}. The training model with backdoors can be easily evaded using targeted or untargeted attacks. As federated learning is coming into limelight recently, the adversarial risks are yet to be properly quantified before embracing the technology that could be the future. 

\subsection{Benign Files Attack}
Adversarial attacks are performed in malware files by inserting some non-malicious contents which do not tamper with any functionality other than classification decision. Modifying malware files slowly has been a mainstream approach for adversarial. However, no limited or no existing research has studied the possibility of inserting malicious contents to a benign files. This approach works in a reverse way than the established adversarial approaches. Inserting and hiding malicious payload at different locations of file without affecting the classification decision is also a future research topics in adversarial and requires attention. 

\subsection{Targeting Unexplored Algorithms}
Most of the machine learning algorithms have already been victimized by adversarial attackers, including sophisticated deep neural network architectures. However, there are some deep neural networks that haven't yet been compromised by adversarial attackers such as Generative Adversarial Networks (GANs), Deep Reinforcement learning (DRL) and Variational Auto-Encoders (VAEs) \cite{zhang2019adversarial}. These algorithms are itself in development stage, which has capped the adversarial attempt to them till date. Differentiable neural computer~\cite{deepmind_differentiable} are only attacked once ~\cite{chan2018metamorphic}. These new sets of algorithms are yet to be explored by adversarial attackers. 

\subsection{Standardizing Testbed and Metrices}
Adversarial attacks discussed in the survey are carried out in lab environments, taking numerous assumptions which may be unpragmatic for real world challenges. Most of the works have assumed unlimited access to machine learning model, favourable dataset and weak classifiers to bolster their results. The current literature lacks standardized dataset and detection mechanisms to measure the exact performance of adversarial attack. As vast number of research works are performed on different dataset and target model, it is not possible to compare the performance of attacks. Hence, the attack testbed should be standarized to bring the assessment uniformity across the research community and uplift the attack standard out from the lab environment.

The issue is not limited to test environment but also with evaluation metrices. More than often, the performance of adversarial attacker is reflected in terms of evasion accuracy inherited from machine learning models. However, accuracy only provides small fraction of attacker's performance in adversarial domain. To provide the overall quality of attacks, metrics such as transferability, universality and imperceptability need to be studied~\cite{alsmadi2021adversarial}. The metrics should be descriptive, fair, and complete to evaluate the quality of attacks performed across different environments. Some metrics should also be developed to measure the degree of functionality preservation while manipulating the files. Incorporating attack's meaning preservation capability as a quality metrics have shown the benefit in recent works~\cite{michel2019evaluation}. Some distance metrics can be used to determine the differences between original file and adversarial modification. Metrics can also be designed to determine the sensitivity of file structure, helping attacker to determine the level of cautiousness required while modification. These complete and fair metrices will not only help to understand and compare the adversarial quality but also to enhance the performance of attacks. 

\subsection{Adversarial Defense}

The growth in adversarial attacks and novel approaches will also require developing advanced defense mechanisms. Although, our survey is focused on adversarial evasion attacks, we believe it is important to briefly highlight future defense directions to present a comprehensive review paper. Among several defense techniques proposed, defensive distillation~\cite{papernot2016distillation} and adversarial training~\cite{tramer2017ensemble,shafahi2020universal} are found to be the most effective. While talking about the effectiveness of existing works, we cannot undermine the challenges faced by them. Collection of adversarial samples in large amount to perform adversarial training is a tedious task as neural networks require massive volume of adversarial data~\cite{jang2019adversarial}. An adversarial generation approach was proposed by Goodfellow et al.~\cite{goodfellow2014generative}, however is still very far away from being efficient and accurate enough to perform the robust adversarial training. In addition, many defensive approaches that have been tested in an image domain~\cite{liao2018defense,meng2017magnet} are yet to be introduced for a defense in malware adversarial domains. Recent research using robust machine learning architectures like Generative Adversarial Networks (GANs)~\cite{samangouei2018defense} for defending against adversarial attacks require more exploration to thwart or detect sophisticated evasion attacks. Overall, the future research works on adversarial malware should be directed to build more robust, efficient, generalized and reliable defense mechanisms that can protect malware detection models against the adversarial attacks.

\section{Conclusion}
\label{sec:conclusion}
Machine learning and AI solutions are increasingly playing an important role in cyber security domain. However, these data driven systems can be easily manipulated, misled and evaded which can have serious implications. Recent surge and research in adversarial attacks highlight the vulnerability of ML models making them ineffective against even minor perturbations. In this paper, we provide a comprehensive survey of recent work that focuses on adversarial evasion attacks in malware analysis domain. We have summarized the state-of-art adversarial attacks carried out against anti-malware engines in different file domains. The survey demonstrates the flaw of machine learning architectures against minute perturbations in form of adversarial attacks. We taxonomize the adversarial evasion world of malware based on attack domain and approach taken to realize adversarial evasion attacks. Survey briefly discusses approaches taken by researchers, comparing them with other concomitant works. We conclude the survey highlighting current challenges, open issues and future research directions in adversarial malware analysis. This work will provide a definitive guide to researchers and community, to understand the current scenarios of adversarial malware evasion attacks and prompting unexplored research territories in this highly dynamic and evolving domain. 

\ifCLASSOPTIONcaptionsoff
  \newpage
\fi



%
\bibliographystyle{unsrt}
\bibliography{bibliography.bib}

%

\begin{IEEEbiography}[{\includegraphics[width=1.1in,height=1.31in]{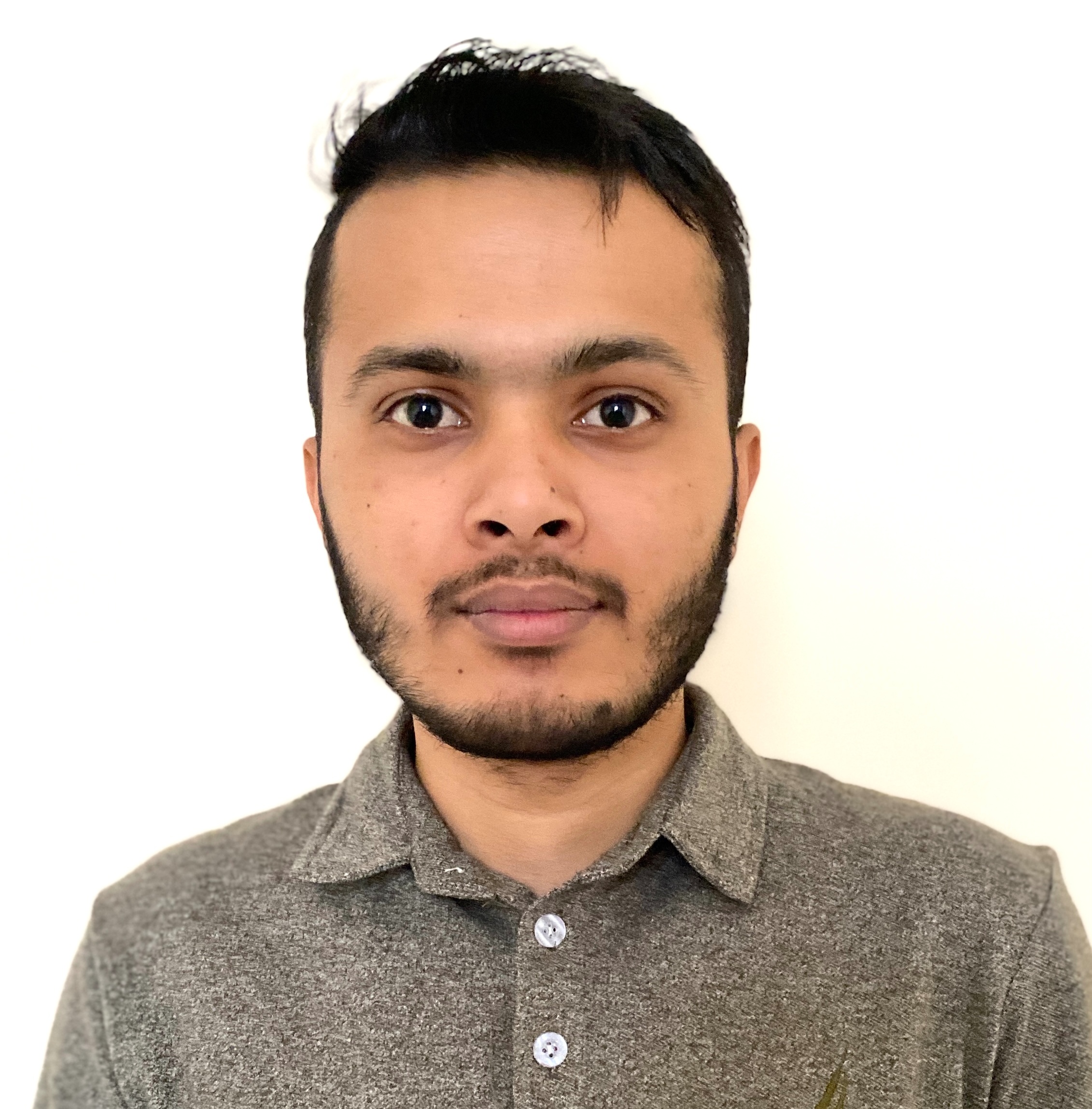}}]{Kshitiz Aryal}
received the B.E. degree in Electronics and Communication Engineering from Paschimanchal Campus, Tribhuvan University, Nepal. He is currently pursuing the PhD degree with the Department of Computer Science, Tennessee Technological University, Cookeville, TN, USA. His current research interests include cybersecurity, adversarial attacks, machine learning, malware analysis, IoT, embedded system and data science.  
\end{IEEEbiography}

\begin{IEEEbiography}[{\includegraphics[width=1.1in,height=1.33in]{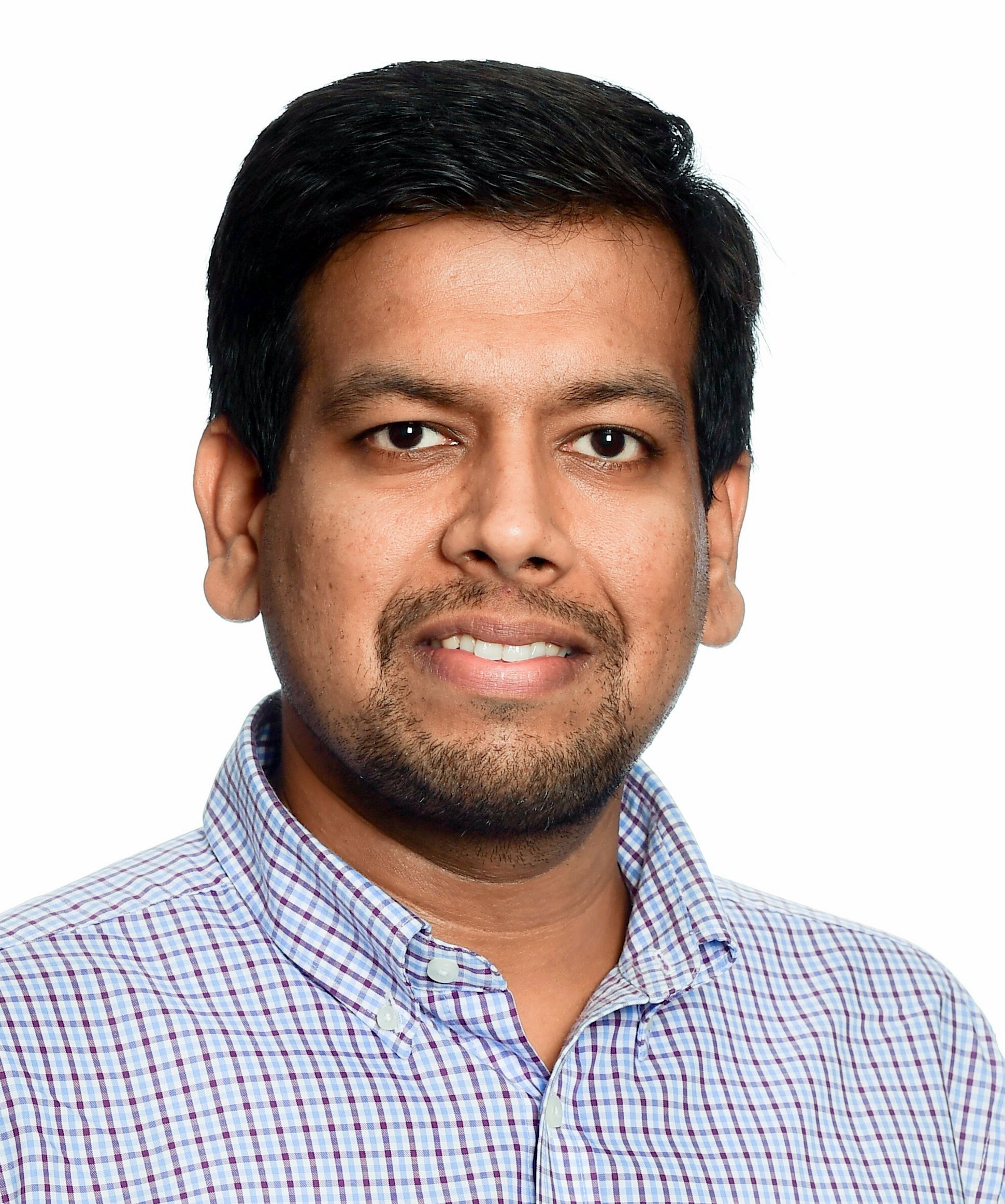}}]{Maanak Gupta} (Member, IEEE) is an Assistant Professor in Computer Science at Tennessee Technological University, Cookeville, USA. He received M.S. and Ph.D. in Computer Science from the University of Texas at San Antonio (UTSA) and has also worked as a postdoctoral fellow at the Institute for Cyber Security (ICS) at UTSA. His primary area of research includes security and privacy in cyber space focused in studying foundational aspects of access control, malware analysis, AI and machine learning assisted cyber security, and their applications in technologies including cyber physical systems, cloud computing, IoT and Big Data. He has worked in developing novel security mechanisms, models and architectures for next generation smart cars, intelligent transportation systems and smart farming. He was awarded the 2019 computer science outstanding doctoral dissertation research award from UT San Antonio. His research has been funded by the US National Science Foundation (NSF), NASA, and US Department of Defense (DoD) among others. He holds a B.Tech degree in Computer Science and Engineering, from India and an M.S. in Information Systems from Northeastern University, Boston, USA.
\end{IEEEbiography}


\begin{IEEEbiography}[{\includegraphics[width=1.1in,height=1.3in,clip,keepaspectratio]{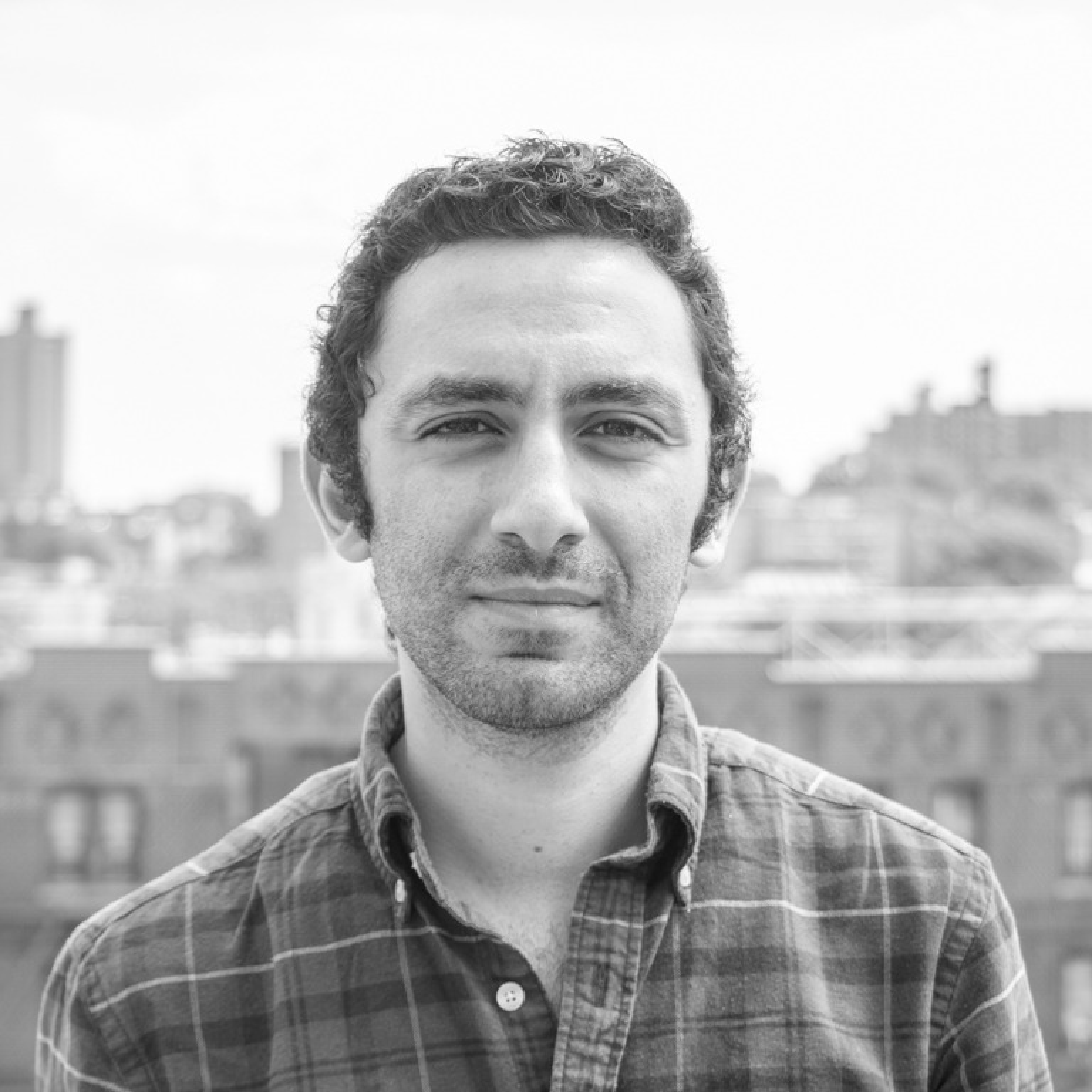}}]{Mahmoud Abdelsalam}
received the B.Sc. degree from the Arab Academy for Science and Technology and Maritime Transportation (AASTMT), in 2013, and the M.Sc. and Ph.D. degrees from the University of Texas at San Antonio (UTSA), in 2017 and 2018, respectively. He was working as a Postdoctoral Research Fellow with the Institute for Cyber Security (ICS), UTSA, and as an Assistant Professor with the Department of Computer Science, Manhattan College. He is currently working as an Assistant Professor with the Department of Computer Science, North Carolina A\&T State University. His research interests include computer systems security, anomaly and malware detection, cloud computing security and monitoring, cyber physical systems security, and applied machine learning.
\end{IEEEbiography}




\end{document}